\documentclass[]{article}
\usepackage{hyperref}
\usepackage[margin=25mm]{geometry}
\usepackage{apacite}
\usepackage{graphicx} 
\usepackage{hyperref}
\hypersetup{
    colorlinks=true,
    linkcolor=blue,
    filecolor=magenta,      
    urlcolor=blue,
    citecolor=blue,
    pdfpagemode=FullScreen,
    }
\usepackage{authblk}
\providecommand{\keywords}[1]
{
  \small	
  \textbf{\textit{Keywords---}} #1
}

\title{\Large The Rise of Artificial Intelligence in Educational Measurement: Opportunities and Ethical Challenges}

\author{Okan Bulut$^{1,†,*}$, Maggie Beiting-Parrish$^{2,†}$, Jodi M. Casabianca$^{3,†}$, Sharon C. Slater$^{3,†}$, Hong Jiao$^{4,†}$, Dan Song$^{5,†}$, Christopher Ormerod$^{6,†}$, Deborah Gbemisola Fabiyi$^{7,†}$, Rodica Ivan$^{8,†}$, Cole Walsh$^{8,†}$, Oscar Rios$^{9,†}$, Joshua Wilson$^{10,†}$, Seyma N. Yildirim-Erbasli$^{11,†}$, Tarid Wongvorachan$^{1}$, Joyce Xinle Liu$^{1}$, Bin Tan$^{1}$, Polina Morilova$^1$}

\date{
 \small $^1$ University of Alberta\\
 \small $^2$ Federation of American Scientists\\
 \small $^3$ Educational Testing Service\\
 \small $^4$ University of Maryland\\
 \small $^5$ University of Iowa\\
 \small $^6$ Cambium Assessment\\
 \small $^7$ Washington State University\\
 \small $^8$ Acuity Insights\\
 \small $^9$ PSI Services\\
 \small $^{10}$ University of Delaware\\
 \small $^{11}$ Concordia University of Edmonton\\
 \small $†$ These authors contributed equally to this work\\
 \small $*$ Corresponding author; \href{mailto:bulut@ualberta.ca}{bulut@ualberta.ca}\\
\normalsize June 2024}

\begin{document}

\maketitle

\section*{Acknowledgement}

\noindent We extend our sincere gratitude to \href{https://www.ncme-aime.org/}{the Special Interest Group on Artificial Intelligence in Measurement and Education (AIME)} and \href{https://www.ncme.org/home}{the National Council of Measurement in Education (NCME)} for initiating this white paper.

\newpage

\begin{abstract}
\noindent The integration of artificial intelligence (AI) in educational measurement has revolutionized assessment methods, enabling automated scoring, rapid content analysis, and personalized feedback through machine learning and natural language processing. These advancements provide timely, consistent feedback and valuable insights into student performance, thereby enhancing the assessment experience. However, the deployment of AI in education also raises significant ethical concerns regarding validity, reliability, transparency, fairness, and equity. Issues such as algorithmic bias and the opacity of AI decision-making processes pose risks of perpetuating inequalities and affecting assessment outcomes. Responding to these concerns, various stakeholders, including educators, policymakers, and organizations, have developed guidelines to ensure ethical AI use in education. The National Council of Measurement in Education's Special Interest Group on AI in Measurement and Education (AIME) also focuses on establishing ethical standards and advancing research in this area. In this paper, a diverse group of AIME members examines the ethical implications of AI-powered tools in educational measurement, explores significant challenges such as automation bias and environmental impact, and proposes solutions to ensure AI's responsible and effective use in education. 
\end{abstract}

\keywords{artificial intelligence, educational measurement, LLM, bias, fairness} 

\newpage
\section{Introduction}

Emerging technologies and applications powered by artificial intelligence (AI) continue to bring significant changes to every scientific field, including educational measurement. The integration of AI into educational measurement has significantly transformed the different methods that are used in practice. For example, using machine learning and deep learning algorithms, AI enables automated scoring (also referred to as automated essay scoring) that can evaluate open-ended responses, essays, and even creative work, providing faster and more consistent feedback to students. This application saves educators time while yielding immediate feedback for learners, allowing for an efficient learning experience. Furthermore, natural language processing (NLP) algorithms can be used to analyze written (or spoken) content rapidly, identify improvement areas, and offer learners personalized feedback. Beyond a conventional assessment setting, AI-driven data analytics tools can also help educators and administrators gather insights into student performance, identify trends, predict future academic outcomes, and recommend interventions to support struggling students.

Although AI-powered innovations have created promising opportunities for more robust, efficient, and personalized assessments, they have also raised significant concerns regarding validity, reliability, transparency, fairness, equity, and test security \cite{hao2024transforming}. For example, advanced AI algorithms (e.g., deep learning) often operate as ``black boxes," making understanding how they arrive at specific decisions challenging. Such algorithms can inadvertently perpetuate or amplify biases present in the data used to train them. In educational contexts, algorithmic bias can affect assessment outcomes (e.g., test scores or grades), exacerbate existing inequalities, and disadvantage certain groups of students. Therefore, the uncontrolled and unregulated development, deployment, and utilization of AI tools in enhancing educational outcomes for learners may lead to unintended consequences, jeopardizing the effectiveness of AI-driven recommendations or assessments.

Ethical concerns have motivated different stakeholders, including educators, researchers, assessment specialists, and policymakers, to regulate the ethical use of AI in educational measurement. Addressing major challenges such as test bias has been a long-standing goal in educational measurement. However, with the rapid development of AI-powered tools in education, resolving ethical challenges has become an urgent priority \cite{zhou_survey_2020}. Various government agencies, non-profit research organizations, and other institutions have developed an active agenda to establish standards of ethical use of AI in education. For example, the European Commission published ethical guidelines on the use of AI and data in teaching and learning for educators \cite{eu2022ethicsai}. The guidelines draw attention to assessment-related topics, such as scoring short-answer items and essays using automated tools, automatic feedback on writing tasks, and algorithms for personalized assessment tools. The Organisation for Economic Co-operation and Development (OECD) also shared a policy brief that discusses the need for explainability and transparency when using digital tools powered by advanced AI technologies in high-stakes settings for students, teachers, or educational establishments \cite{oecd2023}. A recent endeavor that prioritized the ethical utilization of AI for assessment purposes is the implementation of the Duolingo English Test's Responsible AI Standards \shortcite{duolingo}. These standards aim to guide stakeholders on the collaborative and judicious application of AI and human expertise to ensure reliable, secure, and effective assessments.

Recently, the \href{https://www.ncme.org/home}{National Council of Measurement in Education} has established the Special Interest Group on \href{https://www.ncme-aime.org/}{AI in Measurement and Education (AIME)} to advance both theoretical and applied research on the use of AI in educational measurement. AIME consists of a diverse group of members, including data scientists, psychometricians, educational researchers, and other key stakeholders in education. One of AIME's primary objectives is to establish guidelines for the ethical use of AI in educational measurement. In this paper, we aim to contribute to this objective by examining the ethical use of AI-powered applications across various subdomains of educational measurement, such as item generation, scoring, proctoring, and feedback. Also, we discuss other ethical concerns, such as automation bias and the environmental impacts of AI tools in education. Through a detailed exploration of these subdomains, we intend to identify prevailing challenges, highlight ethical issues, and propose viable solutions where appropriate. Considering the rapid evolution of AI algorithms and systems, we anticipate this paper will be an important starting point for researchers and practitioners working on educational measurement, serving as a strong foundation for future research into the new ethical challenges that may emerge from AI-powered assessment tools.

\section{Automated Item Generation}

As the testing industry transitioned from paper-and-pencil to digital formats over the last two decades, the demand for large quantities of high-quality items has increased substantially. Digital assessments have revolutionized how tests are administered and results are analyzed, offering greater flexibility and precision. For example, computerized adaptive testing (CAT) dynamically adjusts the difficulty of questions based on the test-taker's performance in real time. If a student answers a question correctly, the next question is more challenging; if the answer is incorrect, the next question is easier. This approach aims to maintain an optimal difficulty level personalized for each examinee, thereby providing a more accurate measure of their abilities \shortcite{gorgun2023incorporating, weiss1984application}. Another example is multistage adaptive testing (MST), which divides the test into several stages, each consisting of a set of items. Based on the examinee's performance in the initial stage, the system selects the most appropriate set of items for the subsequent stages. This method balances the precision of ability estimation with practical considerations, such as test length and item exposure \shortcite{bulut2023incorporating, zenisky2009multistage}.

Effective item selection in these digital formats requires a large number of high-quality items to control the item exposure rate while finding optimal items from an adequately sized item pool. For instance, adaptive testing programs, such as the Graduate Record Examinations (GRE) and the National Council Licensure Examination (NCLEX), maintain extensive item banks, ensuring that each examinee receives a unique set of questions tailored to their ability level, while also safeguarding the integrity and security of the test content. This process underscores the critical need for a robust item pool to support the sophisticated algorithms that drive CAT and MST systems, ultimately enhancing the accuracy and fairness of digital assessments.

In response to this growing demand, automatic item generation (AIG) was originally proposed over a decade ago by educational measurement researchers as a cost-effective solution to generate a large number of high-quality items (e.g., \shortciteNP{gierl_automatic_2012, holling_automatic_2009, lai2009using}). The most common method of AIG was to use computer algorithms to generate items based on cognitive models and item models developed by human experts, such as subject matter experts, followed by the evaluation of items by experts \shortcite{gierl_role_2012}. Although this method has proven effective in generating a large number of high-quality items, it still heavily relies on human input (i.e., subject matter experts) during the item generation process and yields similar items that may not be diverse enough in terms of content and item structure.

Emerging technologies, such as generative AI and large language models (LLMs), have led researchers to explore their usefulness as an alternative method for item generation. For example, \shortciteA{offerijns2020better} used OpenAI's GPT-2 to generate many items with contextual paragraphs and answers as input text. They also used \shortciteA{devlin_bert_2019}'s Bidirectional Encoder Representations from Transformers (BERT) model to filter out items that were not answerable or not coherent. \shortciteA{kumari2022context} used Google's Text-to-Text Transfer Transformer (T5) to detect answers from source texts, and then they combined the source texts and answers to generate items. In another study, \citeA{bulut_automatic_2022} used OpenAI's GPT-2 for text generation and Google's T5 for item generation associated with the generated text. In a more recent study, \shortciteA{jiao_integrating_2023} compared the model performance of three LLMs for the AIG task by considering the coherence and creativity of the automatically generated items.

A recent scoping review summarized the existing work of leveraging LLMs for AIG \shortcite{tan2024review}. The review identified the commonly used LLMs and their specific usages in the AIG process, as well as the characteristics of the generated items. It concluded that LLMs are a flexible and effective solution for generating various types of items across different languages and subject domains. Although the review suggests that leveraging generative AI tools (i.e., LLMs) is a promising solution for AIG, it also revealed that many of the existing AIG studies lack a solid educational foundation. These findings underscore the need to align item generation with assessment purposes and to integrate measurement and learning theories into the AIG process.

According to \shortciteA{tan2024review}, from both practical and ethical standpoints, AIG should not conclude with merely generating a large number of items but rather with ensuring that the generated items are of high quality for use in educational contexts. However, many existing AIG studies did not involve empirical testing to evaluate the measurement properties of the generated items, unlike traditional item development studies. The measurement properties encompass a variety of item-level or test-level attributes such as item parameters, reliability, validity, and fairness. For instance, it is crucial for a test to include items with varying difficulty levels to accurately measure students' abilities with minimal measurement errors. Neglecting these critical measurement properties could lead to serious ethical issues, potentially resulting in erroneous conclusions about students' abilities, which could unfairly influence their educational paths. For example, in low-stakes assessments, such as practice tests, students might not benefit if they are not properly designed to identify their misconceptions and provide valuable feedback about their learning progress. Therefore, test developers must integrate rigorous pre-testing and validation protocols within the AIG framework. Encouragingly, this gap has been swiftly addressed as more and more researchers have realized the importance of evaluating measurement properties after item generation in AIG (e.g., \shortciteNP{sauberli2024automatic}).

AIG research should adopt a human-in-the-loop framework---a collaborative approach to integrate human expertise into AI-based decision-making. Relying solely on computer-generated items raises ethical concerns, given that assessments play a fundamental role in evaluating students' learning outcomes and monitoring the performance of larger entities such as schools and education systems \shortcite{gierl2022using, sayin2024using}. These assessments can potentially inform educational policies and influence decisions that affect students' lives, such as their education and career paths. Consider a scenario where automated tests contain errors or fail to measure students' abilities accurately; accountability becomes ambiguous. Thus, human oversight is crucial to uphold the reliability of assessments by comprehensively monitoring and supervising AIG systems. Accordingly, we advocate for multidisciplinary teams involving subject matter experts, educators, measurement specialists, and NLP researchers to increase AI's accountability and better harness the utility and potential of LLMs for AIG. For example, NLP researchers can advance the technical aspects of AIG, ensuring the best use of LLMs according to their characteristics and features. Subject matter experts can contribute by providing their expertise and helping to create assessment items that are academically rigorous and aligned with curriculum standards. Measurement specialists can evaluate the functionality of the assessment items (e.g., reliability, validity, and fairness), examining how well they gauge students' learning progress and outcomes.

Furthermore, educators, closely involved with their students, are uniquely positioned to identify their specific needs and learning styles. With educators' observations about students' learning progress and challenges, LLM-based AIG can be adapted to create assessment items that are more effective in diagnosing students' learning gaps, misconceptions, and areas of strength. This can lead to more effective diagnostic items, providing valuable feedback for student learning and fostering targeted instructional strategies \shortcite{drori_neural_2022, rodriguez2022}. An interdisciplinary team can bridge the gap between technical development and educational application, increasing the accountability of using AI and leading to a more appropriate application of AIG in real-world educational and assessment contexts.

Another ethical consideration relates to the bias inherent in the content generated by LLMs. LLMs typically operate in a pre-training phase followed by either a fine-tuning or a prompting-tuning phase \shortcite{radford2018improving}. In the pre-training phase, LLMs adopt an unsupervised learning strategy to learn the conditional probabilities of language tokens from vast text datasets. They can then undergo a subsequent fine-tuning phase to tune the parameters of the LLMs or receive task-relevant prompts without tuning the parameters to perform specific downstream tasks \shortcite{liu2023pre}. Given that LLMs learn the conditional probabilities of texts from the training datasets, they often mirror the biases present in these datasets. Such inherent biases will propagate to downstream tasks, including their manifestation in the generated assessment items in AIG tasks. For instance, if an LLM is trained with datasets that predominantly feature texts associating certain demographic groups with specific characteristics, the model might generate texts that reflect and reinforce these social biases \shortcite{gallegos_bias_2024}.

LLMs trained on data from low-resource languages or underrepresented groups may also generate items containing unfamiliar or inappropriate terms for these groups, raising fairness concerns \shortcite{bender_dangers_2021}. Such biases can manifest in item generation involving texts related to sensitive topics such as gender, race, religion, age, and nationality \shortcite{li2024pre}. To address this ethical issue, several techniques can be implemented during multiple stages of training LLMs, including pre-processing, in-training, and post-processing \shortcite{gallegos_bias_2024}. For instance, an effective method is the inclusion of diverse datasets in the preprocessing phase, enabling the model to learn from a wider array of contexts and reducing the likelihood of generating biased content \cite{buolamwini2018gender}. Moreover, model outputs can be evaluated and modified in the post-processing phase to mitigate bias, such as detecting and replacing harmful or inappropriate words. A more comprehensive introduction to the concepts of fairness and bias and the strategies for debiasing can be found in recent studies conducted by \shortciteA{gallegos_bias_2024} and  \shortciteA{li2024pre}.

\subsection{Prompting for Item Generation}

Inclusiveness stands out as one of the fundamental ethical principles in AI, encompassing the assurance of non-discrimination and the promotion of unbiased algorithms \shortcite{nguyen_ethical_2023}. Bias can be reinforced not only through the utilization of inherently partial data but also by incorporating instructions and cues containing biased language. Apart from the underlying algorithms or training data, the quality of prompts (i.e., specific cues and instructions provided to an AI model to generate desired content) employed for item generation plays a pivotal role in determining the efficacy of AI language models utilized in AIG \cite{bozkurt_generative_2023}. Prompt engineering is ``the process of designing, crafting, and refining contextually appropriate inputs or questions to elicit specific types of responses or behaviors from an AI language model'' \cite{bozkurt_generative_2023}. \shortciteA{heston_prompt_2023} also emphasize the capability of prompt engineering to adjust such aspects of the model's reply as “length, complexity, and style” (p.199). For instance, prompts can enhance item discrimination by eliciting brief, straightforward answers suitable for novice students while prompting more detailed responses tailored towards advanced learners \cite{heston_prompt_2023}. \citeA{zhang2021differentiable} also underscored the benefits of prompts, particularly their effectiveness in leveraging small datasets and their decreased reliance on specialized domain expertise.

Prompts characterized by vagueness or suggestive inclinations toward a specific response can yield biased outputs, and emotionally charged prompts can also impact the objectivity of responses and items generated \cite{heston_prompt_2023}. Such prompts may lead to generating content that perpetuates discrimination and prejudice. Moreover, using inaccurate or misleading prompts can produce false and deceptive information in AI models. To address this issue, \citeA{bozkurt_generative_2023} proposed a set of strategies for designing precise prompts necessary for obtaining the desired output. These recommendations entail setting clear objectives, employing appropriate language and tone, providing context, examples, and references, specifying the expected output format, and incorporating essential details anticipated in the response. Additionally, fine-tuning prompts, experimenting with different variations, and rigorously testing them to analyze results, along with adjusting prompts based on desired responses, can mitigate the generation of irrelevant or erroneous outputs and enhance overall performance \cite{bozkurt_generative_2023}.

Another common issue encountered in prompt engineering and item generation is explainability. The European  \shortciteA{eu2019ethicsai} stresses the significance of providing insights into how a specific AI algorithm operates and makes decisions. A lack of understanding of how prompts are constructed and applied can impede stakeholders' ability to effectively utilize assessments and cause potential misuse of AI capabilities \cite{nguyen_ethical_2023}. One of the solutions entails developing documentation that clearly outlines the purpose, provides guidance using prompt engineering techniques, offers examples of prompts, and showcases generated outcomes. These protocols can assist stakeholders in effectively applying prompts and understanding the mechanism behind item generation. By enhancing the clarity and transparency of prompts, their quality can be enhanced, thereby improving the validity and reliability of generated items \shortcite{lee_few-shot_2023}.

Personalizing prompts with user-specific information can significantly enhance the relevance and quality of generated content, consequently contributing to a more positive learning experience. Strategies such as tailoring content based on previously collected user data or adjusting item tone, style, or complexity can lead to more personalized outputs. However, it is essential to ensure that these strategies are employed only with the consent of all parties whose data is utilized for prompt engineering. Privacy considerations that should be prioritized before implementation include obtaining consent, anonymizing data, ensuring data security, and maintaining transparency throughout the process \cite{nguyen_ethical_2023, vincent-lancrin_trustworthy_2020}. Prompt engineering holds considerable potential and can significantly enhance the quality of generated items. However, optimal design strategies should be developed based on ethical principles to mitigate these risks and ensure the integrity of the generated content \cite{heston_prompt_2023}.

\subsection{Multimodal Stimulus Generation}

Generating multimodal stimuli, such as images or audio, creates content that incorporates multiple modes of communication. Using graphics, audio, video, and interactive elements (e.g., biosensors) enables a more thorough evaluation of unscripted, complex tasks \cite{blikstein_multimodal_2016}. Multimodal assessments advance universal design principles in testing environments through multiple means of engagement, representation, and expression \cite{rao_universal_2015}. These elements of universal design complement multimodal stimuli by ensuring that assessments are not only varied and engaging but also accessible to all learners, reducing barriers and enhancing the opportunity for every student to show their true ability. A study conducted by \shortciteA{smith_emergent_2021} found that providing emergent bilingual students with multimodal assessments supported student identity expression.

By leveraging generative AI, it is possible to generate multimodal stimuli that can complement text-based questions is possible. For instance, in a language comprehension test, an AI could generate a short story as an audio clip alongside visual aids depicting key scenes or concepts. This approach not only aids in comprehension but also engages various cognitive skills, offering a more comprehensive assessment of the learner's abilities compared to traditional multiple-choice test items \shortcite{almond_technology-enabled_2010}. Allowing students to interact with content in more meaningful ways facilitates a deeper measurement of their critical thinking, analysis, and interpretation skills \cite{sharma2020multimodal}. This method can evaluate a broad array of skills and accommodate different learning styles. This ensures a more inclusive and effective assessment process \shortcite{sankey2010engaging}.

The integration of multimodal stimuli in assessments through generative AI raises several ethical considerations that are critical to ensuring the fairness and integrity of the testing process. One primary concern is the potential for bias in the AI algorithms used to generate these stimuli \shortcite{alwahaby_evidence_2022}. It is essential to ensure that these algorithms do not inadvertently favor certain groups of students over others based on cultural, socio-economic, or linguistic backgrounds. To identify and mitigate biases, it is necessary to conduct rigorous testing and maintain continuous oversight of AI systems. As previously stated, using a human-in-the-loop framework is crucial for multiple stages of assessment design, such as item quality review and sensitivity review \shortcite{hao2024transforming}.

Another ethical consideration is the privacy and security of the data used in creating multimodal stimuli \cite{alwahaby_evidence_2022}. As these assessments often incorporate personalized elements to cater to different learning styles, they might also collect sensitive information about students' preferences and abilities. Ensuring the confidentiality and secure handling of this data is paramount to maintaining trust and protecting students' rights. Also, the accessibility of multimodal assessments must be scrutinized to prevent the creation of new barriers to learning. While these assessments aim to be inclusive, there is a risk that the technology required to engage with them might not be equally available to all students, especially in lower-resource environments \shortcite{reiss_use_2021}. Institutions must provide adequate support and resources to ensure that all students have equal access to these innovative testing formats. Overall, while multimodal assessments offer significant advancements in measuring educational achievement, these technologies must be implemented thoughtfully and ethically to truly enhance the learning and assessment landscape.

\section{Automated Scoring}

Automated scoring of constructed-response items is one of the most successful early explorations of AI in assessment. \citeA{page_imminence_1966} and colleagues developed the Project Essay Grader (PEG) system, the first automated scoring system for essays. Automated scoring of short-answer items can be tracked back to the work by \shortciteA{burstein_automated_1998}. AI scoring has also been used to score digitized spoken responses in language assessments \shortcite{higgins_three-stage_2011}. NLP plays a critical role in processing text or speech data so that different machine learning models can be utilized to develop a scoring engine \cite{flor_text_2021}. The methods for automated scoring can be classified into two broad categories: the hand-engineered feature-based approach and the deep learning model-based approach \shortcite{haller_survey_2022, lottridge_psychometric_2023}. A hybrid of both has proven successful in recent years \shortcite{whitmer_results_2023}.

Feature-based models primarily utilize handcrafted features ranging from basic elements like word count to more complex aspects such as argumentation structure and coherence. These features include linguistic features such as syntactic, semantic, lexical, and readability, which are used as input to develop automated scoring models \shortcite{ke_automated_2019,uto_neural_2020}. The major modeling approaches \cite{ramesh_automated_2022} include regression-based, classification models, and neural networks. In the 2021 National Assessment of Educational Progress (NAEP) Automated Scoring Challenge \shortcite{whitmer_results_2023}, the hybrid approach integrating handcrafted features and embeddings from BERT \shortcite{devlin_bert_2019} or extended BERT models performed slightly better compared with the pure ensemble models from LLMs models \cite{lottridge_psychometric_2023, ormerod_short-answer_2022, ormerod2022mapping}. Indeed, recent developments in AI scoring methods show a trend towards integrating LLMs, though use in operational testing is still not widespread, and concerns about validity and fairness persist.

\subsection{Standards for Ethical AI Use in Automated Scoring}

The Standards for Educational and Psychological Testing provide some guidance on AI scoring \shortcite{standards}. Standard 4.19 calls for well-documented support of the engine scores per score level, with theoretical and empirical bases that can link back to the construct. It also states: ``The automated scoring algorithm should have empirical research support, such as agreement rates with human scorers, prior to operational use, as well as evidence that the scoring algorithms do not introduce systematic bias against some subgroups'' \shortcite{standards}. This suggests the need for an evaluation of the automated scoring model as part of the evidence to support the use and interpretation of the scores. The Guidelines for Technology-Based Assessment of the International Test Commission (ITC) and the Association of Test Publishers (ATP) \shortcite{international2022itc} offers a whole chapter on scoring, with an emphasis on constructed-response scoring, offering more detailed advice on AI scoring. In addition, two testing companies have published documents addressing ethical approaches to AI. The Best Practices for Constructed-Response Scoring published by ETS \shortcite{mccaffrey2022best} is focused solely on curating validity evidence for AI scoring and offers very detailed guidance for practitioners (including case studies). Duolingo's Responsible AI Standards \shortcite{duolingo} has a broad focus on the use of AI in assessment and offers high-level principles.

Given the complex nature of developing AI scoring systems, the extent to which the test users are provided with a layman's-level description of the AI scoring engines and scores is vital. Guidance in the Standards \shortcite{standards} predates the introduction of LLMs in educational testing. At that time, the concern with lack of transparency was related to NLP feature definitions and how they are combined to produce a score. Using LLMs creates a true “black box” with thousands or even millions of parameters to interpret. There are several approaches available to improve the interpretability and explainability of the AI scores (see \shortciteNP{boulanger2024explainable, molnar2020interpretable, 
riordan_empirical_2020, 
riordan_probing_2020, zhang_fauss_2024}).

\subsection{Bias in Automated Scoring}

Historically, test bias has been evaluated by comparing scores from different demographic groups or subgroup populations. It may be intuitive to examine mean differences in group means or the overall impact; however, differential item functioning (DIF) analyses became the preferred method for detecting bias because it compares subgroups' item-level performance after conditioning on ability \shortcite{holland_differential_1993, shermis2024ai}. Impact provides a measure of the overall total or scale score differences that are attributable to both actual differences in groups and differences brought about by the use of items with DIF \shortcite{angoff2012perspectives}. AI scoring introduces a different source of potential unfairness into the testing environment that did not exist when DIF analyses and methods were developed.

To ensure that we can measure and address bias in educational testing, we must define the different types of bias and fairness \shortcite{johnson2023evaluating}. The literature on bias in machine learning \shortcite{mehrabi_survey_2021, suresh_framework_2021, suresh_understanding_2021} focuses on seven sources of bias that are particularly relevant in the AI scoring context: historical bias, representation bias, measurement bias, aggregation bias, learning/algorithmic bias, evaluation bias, and deployment bias, and give example sources \shortcite{johnson2023evaluating}. For example, the features being used as part of the model may contribute additional biases if they carry construct irrelevant variance for some subgroups \shortcite{johnson2023evaluating}. Therefore, it is important to thoughtfully choose a model and features that best represent the task for the construct at hand and minimize construct irrelevance for all groups. Importantly, these various sources of bias could affect fairness in different ways. Recent literature \cite{johnson2023evaluating} provides a nuanced conceptualization of fairness: AI scores can have independence (independent of group membership), separation fairness (conditionally independent of group membership, given the true score), sufficiency fairness (true score is conditionally independent of the group membership given the AI score), and/or conditional unbiasedness (i.e., test takers are not negatively affected by the use of machine scores or human ratings). \citeA{johnson2023evaluating} later discussed assessment-level fairness, which examines bias at the level of the reported score.

To minimize unfairness and develop a valid and accurate model, one of the largest considerations is to use a sufficiently large representative sample that includes all of the different demographic attributes of the larger population. Depending on the kind of assessment being performed and the engine used, significantly different sample sizes may be needed. Having more essays in the training data will broaden the set of response types and styles that may appear when the model is used for prediction. When the model contains many features, it will be essential to have larger samples to fully reflect the subgroups' feature distributions. In previous research, as few as 50 features have been found to create sufficiently accurate models \shortcite{burstein_automated_1998}.  However, obtaining sufficiently large sample sizes can be challenging in more contemporary models as LLMs generate thousands or millions of different features representing the test taker response. Features in the underlying language model should also be trained with a wide range of responses from all possible subgroups. See ETS (2021) for more detailed guidance on best practices for minimizing threats to validity and fairness.

\subsection{Detection of Fairness Issues}

Various approaches exist for detecting unfairness or checking to ensure adequate fairness in the context of automated scoring. \shortciteA{williamson_framework_2012} proposed a series of analyses to be conducted by subgroup. In practice, typically, only an analysis of standardized mean difference (SMD) comparing the human and machine scores for each group is performed (or reported). Other analyses may include a comparison of human-human Quadratic Weighted Kappa (QWK) to human-machine QWK by subgroup \shortcite{buzick_comparing_2016} and DIF analyses \shortcite{shermis2024ai, shermis2017use, vo2023human}. Differential feature functioning \shortcite{penfield2016fairness, zhang2017differential} analyses detect whether there are differences in engine feature performance across subgroups, conditioning on overall item score. Many studies report using a mixture of detection methods \shortcite{he2022multilevel, justice2022linear, lottridge2022examining}. \shortciteA{johnson2023evaluating} provided methods to detect different types of fairness, including sufficient and separation fairness. In AI scoring systems based on generative AI with no human ratings, saliency methods may be crucial to understanding what aspects of a response lead to higher or lower scores. Recent work explores the use of saliency methods to detect and understand subgroup differences in short-response scoring \shortcite{zhang_fauss_2024}. Qualitative analyses performed by subject matter experts may also be required to provide evidence for validity in this context.

\subsection{Potential Correction Approaches to Minimize Bias}

To account for bias in feature selection, automated scoring engines can be developed with different sets of features or features weighted by subgroup. For example, in testing programs that use a contributory scoring approach, combining human ratings and machine scores \shortcite{breyer2017implementing} for reporting, penalized best linear predictor models (PBLP) models may be used to minimize subgroup mean score differences between human and machine scores. Error-in-variables regression modeling \cite{johnson2023evaluating} could identify features presenting issues for specific subgroups, and those features could be removed from the model. Another two modeling approaches were proposed to remove differences across groups: constrained optimization and direct penalization \cite{choi_johnson_2024, johnson2023evaluating}. \citeA{liu_fauss_2024} proposed a Bayesian non-parametric model for flexible automated scoring, which captures the potential nonlinear relationship between features and ratings for different subgroups.

Small sample sizes may jeopardize the representation of the characteristics of minority groups in developing automated scoring engines. Some data augmentation methods, such as oversampling \shortcite{chawla_smote_2002}, can be applied to re-balance the small subgroups class in the training data. Synthetic text, audio, and video data can be generated by mimicking the characteristics of minority groups. This may enhance the representation of training data, ultimately minimizing potential sources of bias. However, measures should be in place to ensure that the response data generated are fair and unbiased.

\subsection{Illustrative Example: AP Chinese Language and Culture}

To demonstrate the differences and possible ethical concerns between generative AI and traditional human scoring, we present below the results from a study comparing the scores of a human AP Chinese rater and two versions of ChatGPT (3.5 and 4.0). The essay prompts (scored on a 0 to 6 scale) were sourced from the 2021 AP Chinese Language and Culture Exam. Thirty third-year college students currently enrolled in a third-year Mandarin class participated. Each student was given 15 minutes to complete a story narration writing prompt based on four pictures and asked to imagine writing the story to a friend. In total, 30 essay samples were collected and graded by the human AP Chinese rater, as well as ChatGPT 3.5 and 4.0. Figures~\ref{fig:fig0} shows a selected student writing sample below:

\begin{figure}[!ht]
\includegraphics[scale=0.31]{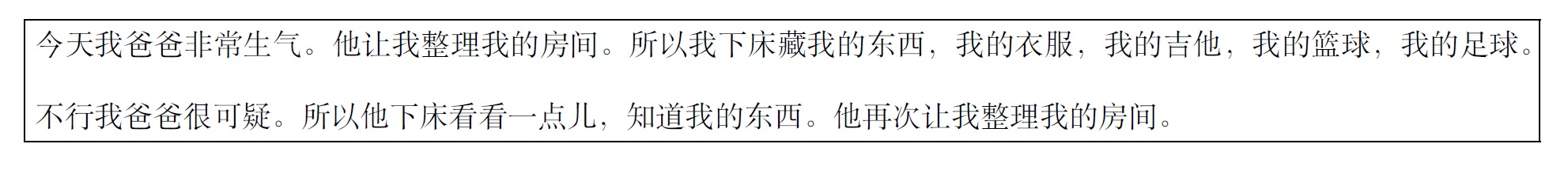}
\caption{A writing sample in Mandarin.}
\label{fig:fig0}
\end{figure}

\begin{figure}[!ht]
\includegraphics[scale=0.9]{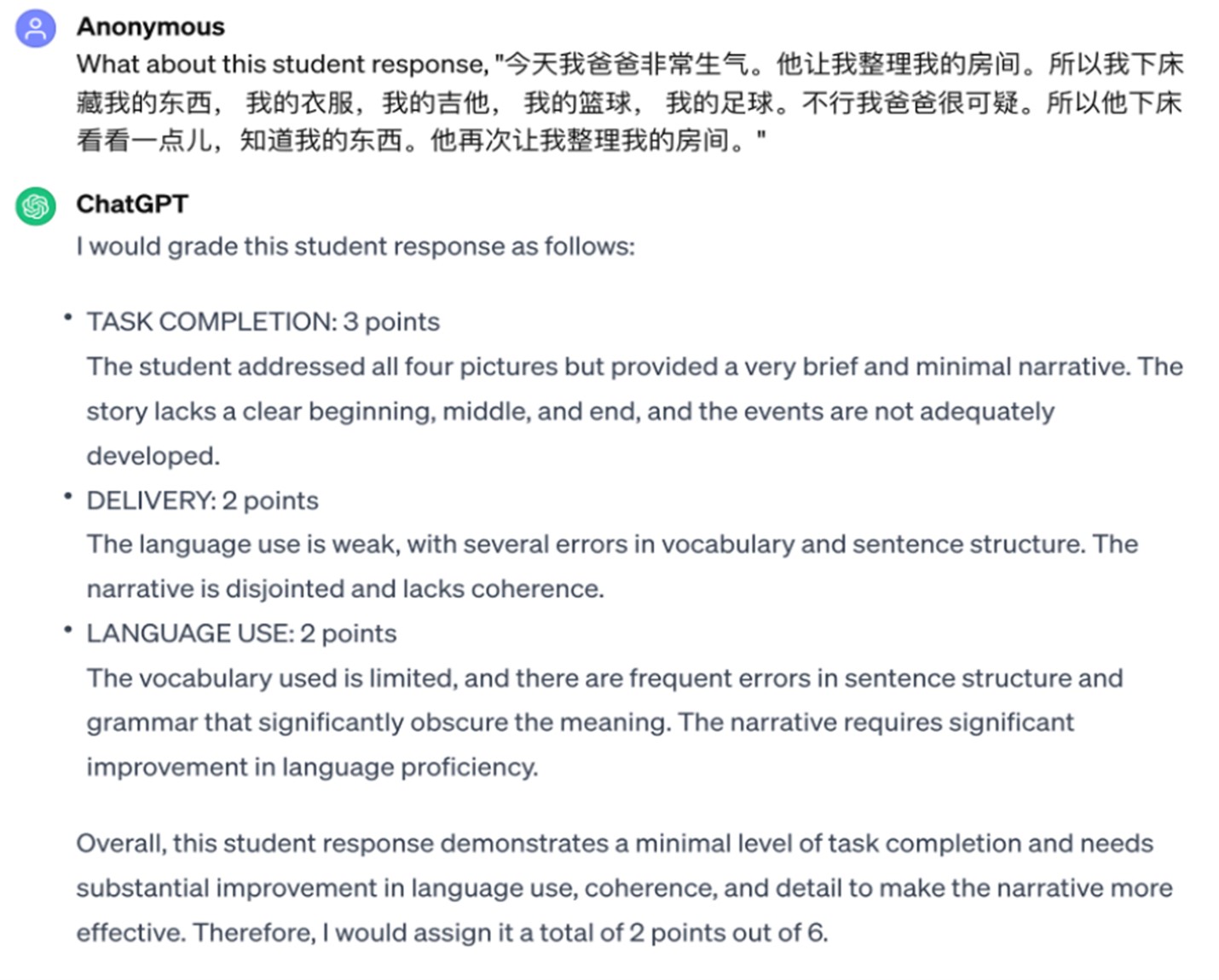}
\caption{Rationales provided by ChatGPT 3.5.}
\label{fig:fig1}
\end{figure}

\begin{figure}[!ht]
\includegraphics[scale=0.9]{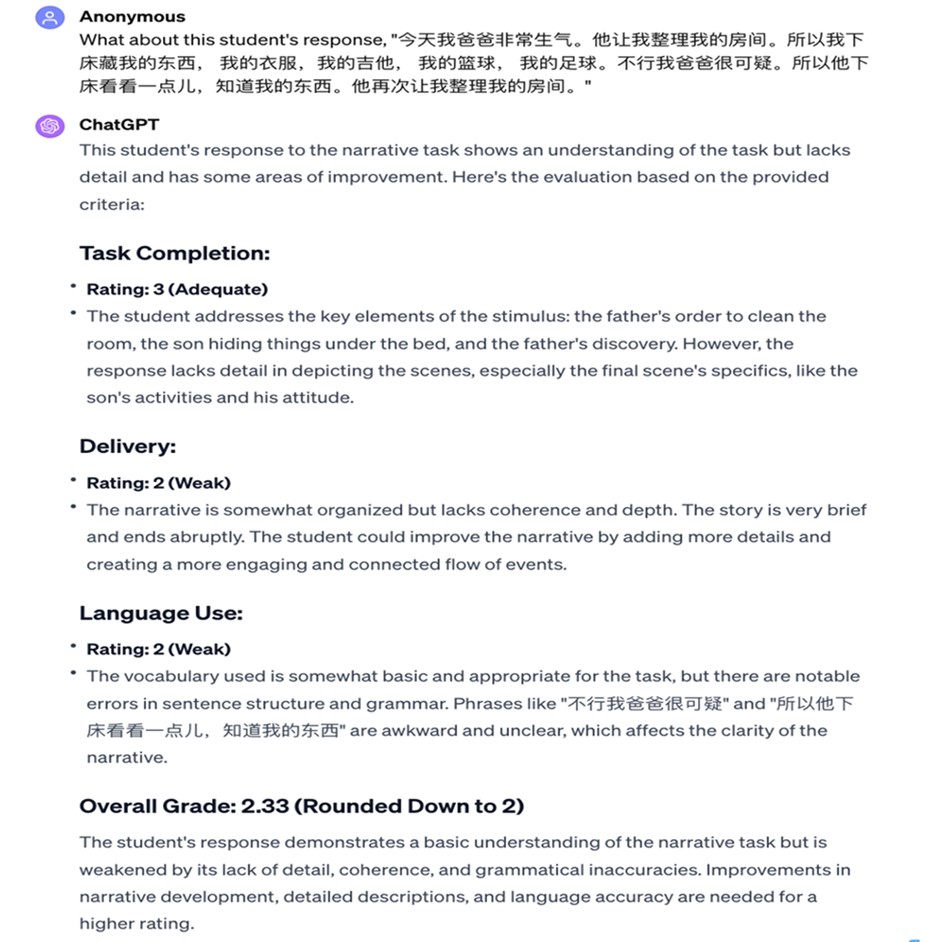}
\caption{Rationales provided by ChatGPT 4.0.}
\label{fig:fig2}
\end{figure}

Before scoring, ChatGPT versions 3.5 and 4.0 were trained by a human rater on using the College Board's scoring rubric, with 10 example samples scored (0-6) by the human rater as references. Furthermore, all human and AI raters provided their rationale for each score based on the rubric. Figures~\ref{fig:fig1} and ~\ref{fig:fig2} display the rationales provided by ChatGPT 3.5 and 4.0 for the selected student writing sample. Additionally, the human rater assigned a score of 3 to the same writing sample and cited the following rationale based on these criteria: ``1) Task Completion: The narrative aligns with the stimulus but contains logical inconsistencies and disconnected sentences; 2) Delivery: The language usage is inconsistent and contains many errors; 3) Language Use: Limited vocabulary and idioms, with frequent errors that sometimes obscure meaning.'' 


After reviewing all the rationales, we found that the scoring rationales provided by the human rater and ChatGPT were internally consistent. The human rater emphasized a holistic assessment, prioritizing task completion, followed by delivery and language use. In contrast, the ChatGPT models evaluated each criterion independently and averaged the scores. Despite their training to follow a similar prioritization as human raters, ChatGPT, in this instance, adhered to a methodology of averaging scores across domains. This methodological difference could potentially lead to a different distribution of scores, raising considerations about the validity evidence for the AI-based scores due to the distinct weighting given to each criterion.

\subsection{Illustrative Example: Comparison of Engines from NAEP Data Challenge}

In 2021, the United States Department of Education's Institute of Education Sciences (IES) and the National Center for Education Statistics (NCES) hosted a data challenge to explore whether students' open-ended responses to the NAEP Reading assessment for fourth and eighth grade could be accurately and fairly scored using AI models. This first data challenge found that the top three challengers could, on average, accurately score all test questions with minimal degradation from the human-human QWK (human-human QWK – human-AI QWK $< 0.05$). The most accurate team overall used proprietary software called PEG, which uses an ensemble model that combines a series of classifier and regression machine models. The next two placing teams used different versions of BERT. However, when the individual subgroup analyses were performed, there was some bias in predicted scores, mostly for English language learners (ELLs) and students with disabilities.

Table 1 shows the mean and standard deviation of the human-AI QWK and SMD in these groups across teams. The mean SMD for ELLs was 0.18 for the “Most Imp. Char.” Task. The mean SMD for students with an individualized education plan (EIP) was 0.17 for the “Imp. Of Fast Deliv.” Task. This level of differences in human and AI scores is typically considered large enough to be concerned, and further investigation is warranted (as per \citeNP{williamson_framework_2012}). Interestingly, the overall QWK for these two tasks was about 0.78 on average, indicating acceptable agreement. However, the tasks that showed higher human and AI agreement had very low mean SMDs. In general, when the overall model evaluation shows high levels of agreement and prediction accuracy \shortcite{mccaffrey2024}, there tend to be no subgroup differences \cite{mccaffrey2022best}. This was found in the next challenge in 2023 when NCES/IES hosted a challenge to explore the accuracy and fairness of scoring open-ended responses to the NAEP Math assessment for fourth and eighth grade. The top three teams were as accurate as the human raters on average across all ten items that were scored—they used completely different methods but yielded a high average QWK ($>0.93$). There was no demonstrated bias in any of the major demographic groups.

\begin{table}
\begin{tabular}{lrrrrrrrrr}
\hline
Item & N Teams & Mean & SD & Mean & SD & Max & Mean & SD & Max \\
Name (Grade) & with Low & QWK & QWK & ELL & ELL & ELL & IEP & IEP & IEP \\
& QWK* &  &  & SMD & SMD & SMD & SMD & SMD & SMD \\
\hline
Most Imp. Char. (8th) & 4 & 0.78 & 0.07 & 0.18 & 0.04 & 0.21 & 0.04 & 0.03 & 0.10 \\
Imp. Of Fast Deliv. (8th) & 6 & 0.78 & 0.04 & 0.16 & 0.09 & 0.25 & 0.18 & 0.07 & 0.28 \\
Text Box Helps (4th) & 5 & 0.79 & 0.06 & 0.19 & 0.08 & 0.26 & 0.03 & 0.03 & 0.08 \\
Thoreau Quotation (8th) & 0 & 0.68 & 0.03 & 0.12 & 0.11 & 0.28 & 0.09 & 0.05 & 0.14 \\
Innkeeper Changes (4th) & 0 & 0.87 & 0.06 & 0.09 & 0.06 & 0.16 & 0.04 & 0.02 & 0.07 \\
Describe Merchant (3th) & 0 & 0.90 & 0.09 & 0.08 & 0.07 & 0.20 & 0.02 & 0.02 & 0.05 \\
Reader Interested (8th) & 1 & 0.88 & 0.13 & 0.09 & 0.09 & 0.24 & 0.05 & 0.05 & 0.11 \\
I'm Ruined (8th) & 5 & 0.79 & 0.10 & 0.04 & 0.03 & 0.09 & 0.11 & 0.05 & 0.18 \\
\hline
\end{tabular}
\caption{The teams with lower QWK had QWK$_{HH}$ -- QWK$_{HL} > 0.05.$}
\end{table}

Automated scoring is not at all a new capability in educational assessment, with the first systems created in the 1960s. These earlier systems were not based on AI, per se, but on supervised models built by NLP scientists using feature sets that were somewhat explainable. With the recent explosion of AI in the mainstream and the availability of open-source advanced LLMs, there has been a resurgence of interest in automated scoring. AI is now more broadly applied to advance and potentially improve automated scoring of constructed-response items. Along with this increase comes the potential responsibility to investigate how the new technology functions, particularly validity and fairness issues. This section aimed to acquaint readers with some of the topics to consider and to encourage more detailed exploration into the current research being done in our field to address validity and fairness in AI scoring of constructed-response items.

\section{Score Reporting and Feedback}

In educational measurement, feedback serves as a crucial component that transforms examinees' performance into actionable insights. Its purpose is to inform individuals about their current ability levels and provide them with guidelines to enhance their knowledge for improved future learning outcomes \cite{boud_rethinking_2013}. Feedback contributes to students' understanding of their performance by operating on multiple levels as follows \shortcite{hattie2007power}:

\begin{enumerate}
\item The \textit{task} level involves guiding students on what specific actions are needed to arrive at the correct answer.
\item The \textit{process} level focuses on informing students about the thought processes required to approach tasks correctly.
\item Self-regulated learning (SRL) level pertains to the ability of students to monitor and guide themselves toward achieving their learning objectives.
\item The \textit{self} level includes reflective feedback of the learners themselves as individuals.
\end{enumerate}

Feedback in its various forms addresses different levels of change. For example, offering scores and specific instructions related to tasks can stimulate modifications at the task level. However, to foster SRL, feedback must be contextualized to the students' circumstances, thereby promoting introspection on their task approach \shortcite{carless_feedback_2019}. An instance of feedback at the SRL level could be, “The strategy you used to approach the task did not do well. What do you think went wrong? What else can you do to reach a different outcome?”

\subsection{AI's Role in Feedback Enhancement}

AI-powered technologies can assist educators in formulating effective feedback through information gained from learning analytics (LA) and NLP \shortcite{wongvorachan_artificial_2022}. LA can provide insights into students' performance and learning by analyzing data about students. Simultaneously, NLP can automatically translate numerical components into actionable verbal feedback. This cutting-edge capability could significantly reduce instructors' workload in the case of large class sizes and the influx of information. Especially in this era where online- and hybrid learning is prevalent, AI can leverage the wealth of information generated from online learning platforms such as Moodle \shortcite{jin_design_2012}. This capability not only informs students of how they did but also informs instructors about patterns of students' learning (e.g., distribution of quiz scores) to inform their instruction planning as well.

The mentioned innovation is driven primarily by students' data, such as students' engagement with course materials through a learning management system and their assessment scores \shortcite{bulutformative, selfregulated2024}. For this reason, instructors and researchers must ensure that all information is given with consent and that no misuse occurs \cite{stahl_student_2016}. Schools and academic institutions should implement data privacy measures, educating students about their data rights and maintaining transparency about how and why their data is being used \shortcite{lin_information_2015, stahl_student_2016}. Potential misuse of students' data in the feedback process includes data misinterpretation, inconsistent data definition (i.e., comparing different kinds of data), poor result visualization (i.e., choice of graphs, texts, and colors) \cite{webber_use_2019}. For example, results about students' performance, including the overall score distribution of the class, should be given with context. Inferring the entire semester's performance from midterm scores, meant for formative assessment, is inappropriate. Such data misuse may inadvertently discourage students from taking initiative in their learning, shifting the focus to test-taking proficiency rather than genuine knowledge acquisition \cite{jones_learning_2007}.

Introducing AI into the feedback process presents an opportunity to enhance and streamline educational practices. AI can automate the feedback process, provide personalized insights, identify learning gaps, and adapt to the learning pace of each student. This can lead to more efficient learning experiences and improved educational outcomes. However, while harnessing the power of AI, it is crucial that ethical considerations are accounted for. The use of AI in education involves handling sensitive student data, which may cause harm to students if misused. Educators and researchers must ensure that all data-driven insights are derived and utilized responsibly. This means being transparent about data collection and usage policies, implementing robust data privacy and security measures, and regularly reviewing and updating these measures in line with evolving ethical standards and regulations \shortcite{gimpel_upside_2018}.

\subsection{Illustrative Examples: Automated Writing Evaluation}

Automated writing evaluation (AWE) has evolved significantly since its initial discussion in the scholarly literature by \citeauthor{warschauer2006automated} in 2006, who set a foundational research agenda for its development. AWE now broadly encompasses tools that provide both quantitative scores and qualitative feedback within classroom-based formative assessments \shortcite{hockly_automated_2019, huawei_systematic_2023}, although some AWE tools solely provide qualitative feedback. Automated scores and feedback are generated using algorithms that rely on NLP and AI—primarily supervised machine learning but increasingly LLMs—to deliver automated feedback aimed at enhancing student writing \shortcite{correnti2024supporting,cotos_automated_2023, deeva_review_2021, fu_review_2024, shi_enhancing_2022}. Linked initially with automated essay scoring (AES) and automated essay evaluation (AEE) for shorter constructed-response items, AWE has expanded to support a broader range of writing activities. It offers utilities that assist with the entire writing process, including planning tools like concept mapping and outlining, alongside feedback mechanisms that focus on grammar, organization, and development \shortcite{burstein_writing_2018, huawei_systematic_2023}. AWE is intended to allow educators and learners to customize feedback, aligning it with individual writing goals and pedagogical objectives.

AWE has demonstrated notable benefits in improving writing quality across various contexts, with studies reporting effect sizes ranging from 0.38 to 0.98 in tertiary and second language (L2) settings \shortcite{fleckenstein2023automated, li_still_2023,ngo_effectiveness_2024, nunes_effectiveness_2022}. Despite these positive outcomes, the effectiveness of AWE varies significantly by educational level, duration of intervention, and the nature of feedback provided. For instance, AWE tends to be less effective for younger students, such as middle schoolers, particularly when feedback is generic or lacks integration with comprehensive writing curricula \cite{ware_feedback_2014, nunes_effectiveness_2022}. Moreover, while some studies indicate that AWE can boost motivation and self-efficacy in writing \shortcite{grimes_utility_2010, moore_student_2016,warschauer_automated_2008, wilson2020automated}, others note mixed effects on students' writing-related beliefs and affect, with variability often tied to the method of AWE implementation and students' prior skills levels \shortcite{chen_beyond_2008, wilson_predictors_2024}. For example, high-achieving students may experience demotivation when using AWE because the system does not provide as positive an evaluation of their work as their teacher does \shortcite{wilson_elementary_2021}.

Given that AWE has differential effects based on how it is implemented, educators, developers, and researchers should carefully consider how AWE is most ethically, equitably, and effectively deployed. These considerations encompass the representativeness of data, the transparency and explainability of algorithms, the assurance of equitable access and benefits across diverse student demographics, and maintaining human oversight. Addressing these factors is essential to ensure that AWE tools support educational objectives ethically, fostering an inclusive and fair learning and assessment environment.

\subsubsection{Representativeness of the Training Data} 

Ensuring data representativeness in AWE systems is a critical ethical consideration that impacts the fairness and effectiveness of these tools. Developers must assemble training data that reflect the diversity of the intended user base, not only in demographic terms but also across the spectrum of achievement levels expected among users (see \shortciteNP{raczynski_appraising_2018}). This involves including a wide range of essays, scores, and feedback that capture both typical and atypical performance outputs—often underrepresented at the extremes of the achievement distribution. In addition, it is crucial to ensure that human ratings and feedback are scrutinized for bias, including rater biases (e.g., \shortciteNP{wind_influence_2018}), as well as human biases, stereotypes, and language ideologies \shortcite{goldshtein2024automating}. Such comprehensive data collection helps train AWE systems that are unbiased and equitable, capable of providing relevant and accurate feedback to all users, regardless of their background or initial skill level. At present, it is not common practice to divulge details about the training data underlying AWE models, but this may be a practice the field pushes to adopt. 

\subsubsection{Transparency and Explainability} 

Transparency and explainability in AWE systems are crucial for building trust and enhancing usability among both students and educators \shortcite{goldshtein2024automating}. These elements ensure that users understand how the AI generates feedback and the rationale behind the specific suggestions offered. As \shortciteA{myers_evaluating_2023} discuss, transparent AWE systems provide detailed explanations of their operational mechanisms, allowing users to see the connection between their input (the student's text) and the output (the feedback provided). This clarity helps demystify the AI processes, which is essential for users to confidently rely on and effectively utilize the feedback. Indeed, a central premise of formative assessment is clarifying and making explicit for learners the criteria for success \shortcite{black_developing_2009}. In writing, knowledge of evaluation criteria serves as the basis for revising in the absence of automated feedback. Moreover, when AWE systems clearly articulate the basis of their feedback, they are more likely to be integrated successfully into educational practices, as both students and educators can better align the automated feedback with instructional objectives and learning outcomes. Finally, ensuring that AWE systems are transparent also aids in accountability, making it easier to facilitate third-party research and evaluation and to identify and correct potential biases or errors in the AI's assessments. This level of transparency not only improves the educational tool's efficacy but also fosters a deeper trust in technology-enhanced learning environments \shortcite{myers_evaluating_2023}.

\subsubsection{Equity and Access} 

Ensuring equity in access and benefits from AWE systems is a pivotal ethical consideration, particularly as these technologies become more integrated into educational environments. Equity challenges in AWE usage stem from the need to ensure that all demographic groups, especially vulnerable or under-served populations such as ELLs, receive comparable benefits from these technologies. This involves more than just making AWE tools available; it requires that these tools are effective across diverse linguistic, cultural, and socioeconomic backgrounds. For instance, the challenge of equitable access and benefit in AWE systems includes ensuring that the feedback provided is linguistically and culturally responsive. This means that the systems must be capable of recognizing and adequately responding to the varied language use among students from different backgrounds. 

Additionally, AWE systems should be tested and proven effective in diverse educational settings to avoid perpetuating existing educational disparities. A recent study by \shortciteA{wilson_elementary_2024} illustrates an approach to conducting such testing. They examined whether elementary-aged ELLs and non-ELLs equally solicited AWE feedback, improved the quality of their first drafts, and productively revised their writing. Findings were promising: language status was unrelated to the degree to which elementary-grade students accessed and benefited from AWE. However, this type of nuanced research is not yet commonplace. Such testing should become the norm, supplementing the rigorous psychometric testing of the underlying scoring and feedback algorithms. 

\subsubsection{Human Oversight}

AWE is not designed to, nor should it, replace the teacher. As articulated in human-centered AI principles \shortcite{shneiderman2022human}, the deployment of AWE should enhance the teacher's (and student's) capabilities. For example, educators using AWE have been shown to focus more on complex writing skills by offloading the evaluation of basic and generic skills to the AWE system. Research supports this approach, indicating that AWE enables teachers to provide richer, more focused feedback on advanced writing elements \cite{wilson_automated_2016}. However, unlike insights from teachers or peers, AWE feedback lacks contextual awareness of a student's developmental progress or curriculum timeline. It analyzes students' texts against the corpus of texts in its training data and the algorithms built thereupon. Thus, AWE may provide feedback on skills students have not yet encountered or are not expected to master at their current educational stage. This situation underscores the importance of teachers' ongoing involvement in correcting any misalignment and ensuring the AWE's feedback supports rather than contradicts their pedagogical objectives. This dynamic was highlighted in focus groups with teachers who have integrated AWE into their classrooms \cite{wilson_elementary_2021}. Moreover, it is crucial for students to understand that AWE systems are tools intended to enhance their ability to communicate effectively with humans rather than serving as the ultimate judge of their writing proficiency. Educators must ensure that AWE is used as an aid in the broader context of developing competent communicators, emphasizing that the technology should complement, not dictate, the learning process.

It is clear that AWE's potential and its challenges are closely tied. The advancements in AWE technology offer opportunities to enrich educational practices by providing timely, consistent feedback and freeing educators to focus on higher-level teaching objectives. However, these benefits hinge on addressing critical ethical considerations, such as ensuring the representativeness of training data, maintaining transparency and explainability, promoting equity in access and benefits across all student demographics, and ensuring that teachers remain in the loop. As AWE systems begin to integrate powerful generative AI technologies––technologies that are more opaque than AI used in legacy AWE systems––these ethical considerations will become all the more salient. 

\section{Other Concerns on AI Use in Education}

\subsection{Aberrant Responses}

In an assessment, aberrant response patterns may occur due to atypical test-taking behaviors such as cheating, careless responding, creative responding, and non-effortful responding \shortcite{bulutLA2024, gorgun2021polytomous, kim_identifying_2016, wan_using_2023}. Such behaviors can arise in both high-stakes and low-stakes assessment contexts \shortcite{liu2020identifying}. For instance, in high-stakes assessments, some test-takers may be motivated to increase their scores by engaging in dishonest behavior \shortcite{ranger2023detecting}. Conversely, non-effortful responding tends to be more prevalent in low-stakes assessments, where students may be less motivated to sustain sufficient effort throughout the test to demonstrate their true ability \shortcite{lindner_onset_2019, wise_response_2005}.

Aberrant responses pose a significant concern for educational measurement because of their impact on data quality and the validity of inferences or predictions made using assessment results \shortcite{gorgun2021polytomous,kim_identifying_2016}. Thus, much research has been devoted to detecting and handling aberrant responses using different techniques. Notably, in recent years, machine learning approaches have been developed to complement psychometric approaches (e.g., person-fit indices, response time models) in this endeavor, making it possible to investigate aberrant response behavior using multiple sources of data \shortcite{kim_identifying_2016, mueller2016have}.

Regarding the detection of non-effortful or disengaged responses, this is usually operationalized by examining response time data collected from digital assessments, seeking to identify instances where test-takers spend unrealistic amounts of time (either too long or too short) on items or tasks \shortcite{gorgun2021polytomous,lindner_onset_2019,liu2020identifying, yildirim-erbasli_designing_2022}. Typically, the data would not include the ground truth about aberrant responses, so the problem is approached from an outlier or anomaly detection perspective. Some approaches include threshold-based methods (e.g., \shortciteNP{soland_comparing_2021}) and mixture modeling (e.g., \shortciteNP{liu2020identifying, wang2018detecting}). Drawing upon the capabilities of machine learning, researchers have also explored the utility of supervised (e.g., \citeNP{yildirim-erbasli_designing_2022}) and unsupervised learning algorithms, often incorporating additional features other than response time (e.g., \shortciteNP{gorgun2022identifying}).

Another area of particular interest is cheating detection. Over the last decade, researchers have applied and tested various machine learning and deep learning algorithms to detect cheating behavior, including supervised and unsupervised learning approaches \shortcite{cizek_handbook_2016,jiao_integrating_2023}. In terms of data input, item-level data (i.e., responses and scores) are most often used to detect aberrant response patterns. \shortciteA{kamalov_machine_2021} presented an algorithm using sequences of grades from students' continuous assessment results to identify cheating on the final exam. More recently, the availability of process data has spurred further advancements. Process data are collected throughout the course of a digital assessment, such as response time, frequency of item revisits, clickstream data, and even sensor data tracking eye movements or head positions \shortcite{alsabhan2023student}. Many researchers (e.g., \shortciteNP{alsabhan2023student,meng_machine_2023, ranger2023detecting}, \shortciteNP{tang2023latent}, \shortciteNP{zhou2023exploration}) have demonstrated how process data could be leveraged to improve the performance of machine learning algorithms.

In the guidelines on quality control in scoring, test analysis, and reporting of test scores set out by the \shortciteA{international2014itc}, it is recommended that aberrant response patterns should be routinely monitored to uphold test security and integrity. With the increasing availability of data from computer-based testing systems, AI no doubt offers a promising solution to enhancing the detection of such patterns. There are several ethical considerations to take into account. From a social perspective, we must be careful about how results from detection algorithms are used in order to avoid unintended consequences. This ties into the Responsibility principle put forth by \shortciteA{taiwo_review_2023}. As \shortciteA{kim_identifying_2016} stated, “The use of data to identify statistically improbable test behaviors is defensible. However, generalizing the results to a testing individual or population to prove cheating is problematic.” (p. 71). In this way, the goal of aberrant response detection should be related to the validity of the scores and not cheater identification \shortcite{kim_identifying_2016}. Suppose results indeed hold consequences for individuals (e.g., score invalidation). In that case, it is paramount that these decisions are explainable and defendable and clear communication channels are set up with test-takers \shortcite{mueller2016have}.

Aberrant response behavior is often of great interest to stakeholders in educational assessment, such as test developers, test sponsors, and educators. A vast body of research has been built around methods to improve its detection. While each detection method has its own merits, \shortciteA{mueller2016have} recommended that it is time for researchers to think more collectively and draw links between different indicators. This means working towards developing a process that identifies aberrant responses using multiple indicators rather than relying on evidence from a single technique \shortcite{mueller2016have}. The development of AI algorithms supports this vision, as multiple sources of data could be incorporated into the same model. However, the challenge lies in its transparency and explainability (another ethical principle outlined by \shortciteNP{taiwo_review_2023}). Fairness and potential algorithmic bias must also be considered, especially when background and demographic variables are included as part of the data input.

\subsection{Predictive Utility of Assessment Results}

Beyond providing information on students' performance, AI-powered innovations can utilize student data, including test results, to predict various educational outcomes. For instance, a student's scores from formative assessments and the time taken to complete tasks have been identified as strong predictors of their summative assessment scores \shortcite{bulut_automatic_2022}. Similarly, a student's Grade Point Average (GPA) in the ninth grade can predict their likelihood of high school dropout \shortcite{bulut_enhancing_2024}. These examples demonstrate that pedagogically grounded predictors, such as assessment-related data, can provide more actionable insights than non-pedagogical predictors like socioeconomic status or complex predictors, such as clickstream data. This application of AI in predictive tasks falls under the domain of Educational Data Mining (EDM), which involves extracting knowledge from educational databases \shortcite{wongvorachan_artificial_2022}. While LA and EDM both use educational database variables to inform students and instructors, EDM distinguishes itself by prioritizing the optimization of predictive models for accurate predictions \shortcite{chen_lets_2020}.

EDM can leverage students' assessment results, among other variables, to predict potential future outcomes. This information can be used for early intervention by informing parents, teachers, and students. For instance, a student's first-generation status and American College Testing (ACT) scores, in conjunction with their GPA, can predict university retention \shortcite{trivedi_improving_2022}. Furthermore, a student's career prospects can be predicted using their GPA and performance in mock interviews, assessing factors like self-confidence, presentation ability, and communication skills \shortcite{casuat_predicting_2019}. These insights can guide students and parents in future preparation. Instructors and administrators can use these results to initiate student support programs, such as remedial classes, writing support centers, or career counseling services.

While EDM is helpful in this regard, it is crucial to consider the explainability and actionability of its predictions. Without these, predictions may not be useful, as we can only anticipate outcomes without the ability to act upon them. To address this, the application of Explainable AI (XAI) is essential. XAI, a branch of AI that focuses on making the output of complex predictive models understandable to humans, helps establish trust between the user (e.g., instructors) and the tool (i.e., the model) \shortcite{biecek2021explanatory}. Specifically, XAI can provide global-level explanations to identify influential predictors through variable importance methods and local-level explanations to elucidate the mechanism behind a case's prediction through methods like Local Interpretable Model-Agnostic Explanations \shortcite{biecek2021explanatory}. By making the results understandable, we enable human validation of the prediction results to confirm their alignment with reality before taking actionable measures \shortcite{bulut_enhancing_2024}.

\subsection{AI-Powered Proctoring}

Another useful application of AI in assessment regards online test administration. Online learning and examinations were on the rise even before the onset of the COVID-19 pandemic when Massive Open Online Courses, also known as MOOCs, and colleges were leveraging online tools to provide more flexible access to resources for students. The COVID-19 pandemic then forced many educational institutions across the globe to rapidly accelerate this investment in online learning and examinations \shortcite{moreno-guerrero_educational_2020, nigam_systematic_2021}. The benefits of conducting exams online as opposed to the traditional in-person format include the ease of exam scheduling and asynchronous exam administration. There is also no need to find a physical space to conduct an exam, allowing exams to be conducted at massive scales without worrying about student capacity \shortcite{arora_is_2021}. Online exams require remote proctoring services, however, which come with one significant drawback: the ratio of human proctors to test takers is typically higher in online formats because online exams provide more opportunities for academic misconduct, which leads to the need for more proctors to effectively invigilate the exams \shortcite{bilen2021online}. One solution to this problem comes in the form of AI-based proctoring systems, which can help alleviate the human proctoring burden.

Remote proctoring may rely on a variety of AI tools, including face detection, eye gaze detection, keystroke analysis, lockdown software, web traffic recording, and others \shortcite{dyer_framework_2024}. Typically, AI-based proctoring starts with identity verification, where the system confirms the test taker's identity through biometrics such as facial recognition and voice recognition or even IP address verification to confirm the test taker's location. Using facial recognition, the test taker's image is taken immediately prior to starting an exam and compared to some verifiable identification, such as a school ID card, to confirm whether the person sitting for an exam is indeed the person who is registered for the exam. This verification procedure can occur periodically throughout an exam to confirm that the registered test taker is taking the exam at any given instance \shortcite{nigam_systematic_2021}. Beyond facial recognition, facial detection can be used to identify how many people are present in an image and flag instances where no faces or more than one face was detected \shortcite{motwani_ai-based_2021}.

Similarly, gaze tracking can be used to identify where a test taker's attention is focused and if they direct their attention away from their screen to other resources such as external notes or a second screen \shortcite{singh_exam_2022}. An AI proctoring system can employ object recognition alongside this functionality to identify what test takers are directing their attention to (if it is within the camera's field of view) and flag whether the test taker uses non-permitted objects (e.g., calculators). When it comes to audio, AI can also be used to identify any background noise, which can be analyzed to determine whether a test taker received unpermitted assistance in completing their exam either from another individual in the room outside of the camera's field of view or from someone on another call with the test taker \shortcite{nigam_systematic_2021}.  

Despite its benefits, AI proctoring also comes with several issues. First, facial recognition and AI audio detection tools are not exempt from producing false positives and can, therefore, mistakenly assign such flags for academic dishonesty \shortcite{slusky_cybersecurity_2020,nigam_systematic_2021}. So, while AI proctoring can minimize human proctoring efforts, human review of these flags is still necessary \cite{dyer_framework_2024} to prevent unfairly penalizing students. Additionally, false negatives are also a concern, and there is little peer-reviewed evidence on the efficacy of remote proctoring in detecting cheating \cite{dawson_remote_2024}. In a controlled study on a particular proctoring service, where six out of 30 students were asked to cheat, \shortciteA{bergmans_efficacy_2021} found that the remote proctoring software did not flag any of the six students. In comparison, a human review of the video recordings led to catching one of the six students who were asked to cheat.

Second, not only can false positives impact the academic standing of test-takers, but they are also more likely to occur for particular demographic groups; as any tool dependent on AI algorithms, remote proctoring is not free from bias. For instance, in a study on automated proctoring software, \shortciteA{yoder2022racial} found evidence of race, skin tone, and gender bias in the facial detection algorithm, which was significantly more likely to flag women with darker skin tones for review than men or women with lighter skin tones. 

Third, in addition to concerns around AI accuracy, reliability, and fairness, there are also key considerations regarding data privacy and security. Since AI proctoring typically involves some form of analyzing biometric data and often even video feed from the test-taker environment, it can also be perceived as an invasion of privacy \shortcite{coghlan_good_2021}, which could sometimes lead to legal action \shortcite{dyer_framework_2024}. With respect to data security, companies that provide remote proctoring services and the institutions that use them must comply with legal guidelines (e.g., GDPR) and have systems in place that can prevent sharing this data with 3rd parties and mitigate the risks of potential cyber-attacks. However, as \shortciteA{coghlan_good_2021} pointed out, and as in the case of any AI tool, these technical controls are necessary \shortcite{slusky_cybersecurity_2020} but not failproof.

Lastly, given that AI proctoring often relies on the test-taker equipment (e.g., the student's laptop webcam), this can lead to different technical issues individual students may face, depending on their equipment and its compatibility with the proctoring and test-taking platform. For this reason, despite the various advances in AI proctoring, online proctoring would also preferably be supplemented by a technical support team that could assist test-takers in the eventuality of technical difficulties during the test. Indeed, even detecting these technical difficulties is made easier with AI. For instance, facial recognition software could flag webcam issues, while AI audio detection could flag whether there is no audio feed for a test that might require the student to speak. Nevertheless, human review of these flags would still be required \shortcite{nigam_systematic_2021}, and ideally, a combination of AI and human live remote proctoring would be employed \shortcite{dyer_framework_2024}.

Some of the risks listed above could be mitigated by employing a hybrid remote proctoring model that uses both AI detection and live human proctors. Alternatively, the needed human review of AI-assigned flags could happen asynchronously, with the recordings being reviewed to determine whether an academic integrity violation occurred. However, research conducted by one company that provides remote proctoring services found that videos flagged by AI systems due to irregular testing behavior underwent human review around 10\% of the time \shortcite{jaschik_proctoru_2021}. This finding led the company to no longer offer remote proctoring solely based on AI and move towards a hybrid remote proctoring model.

Irrespective of the type of AI-assisted proctoring employed, the institutions that rely on these services must establish clear guidelines on the use of such tools and procedures for following up on potential academic integrity violations. Alongside clear guidelines,  best practices for the ethical implementation of remote proctoring also include effective communication and training of faculty and staff \shortcite{dyer_framework_2024}. The goal is for everyone involved in any remote proctoring system to be knowledgeable and able to speak to the services carried out by the system. In this sense, creating an internal remote proctoring implementation team could help by training educators and staff who could then discuss these systems with students. Regular communication between faculty and the remote proctoring team can also help an institution adapt quickly and make changes to the implementation of a remote proctoring system based on ongoing feedback. Crucially, decisions about AI proctoring tools need to be transparently communicated to both educators and students, ensuring that students know what to expect and what will happen during their exams and preparing them for a successful test-taking experience. Lastly, students should also be provided with additional options in case they are unwilling or unable to participate in a remote exam, and this process for requesting an alternative arrangement should be communicated clearly.

\subsection{Automation Bias}

Automation (i.e., the transition from human labor to computerization and mechanization; \shortciteNP{rahm_education_2023}) through AI-based tools plays a critical role in aiding both educators (e.g., grading and analyzing student data) and students (e.g., using automated feedback systems to guide their learning and progress) \shortcite{williamson_re-examining_2023}. Designers and developers of AI-based assessment tools argue that these tools can revolutionize and better educational assessments and student experiences (e.g., \shortciteNP{bulut_automatic_2022, yildirim-erbasli_conversation-based_2023, conversation2023, yildirim-erbasli_designing_2022}). For example, the integration of AI-based tools can offer the promise of minimizing decision errors, such as grading where human error can occur \shortciteA{williamson_framework_2012,zhang_contrasting_2013}. When functioning effectively, this automation can enhance the accuracy and efficiency of assessment processes, benefiting both educators and students. However, this adoption of automation introduces its own possible errors \shortcite{lyell_automation_2017, jones-jang_how_2023}. For instance, automated grading systems may inadvertently misinterpret students' written responses or fail to capture the nuance of complex concepts, leading to inaccuracies in assessment outcomes when they rely on primary factors like essay length \shortcite{andersen_benefits_2021}. Regardless of the accuracy of these tools, the overreliance on automation can lead to errors in decision-making. Automation bias denotes the phenomenon where decisions are influenced by an overreliance or excessive dependence on AI-based systems, even when these systems may be flawed or incorrect \shortcite{bond_human_2019,lyell_automation_2017,parasuraman1997humans}. For example, a student unquestioningly accepts a high grade from an AI-based grading system without considering the validity of the feedback or their own understanding of the material.

One prominent issue about automation bias is blindly accepting AI-based tools' outcomes without critical examination (e.g., \shortciteNP{khera_automation_2023, kupfer2023check}). When educators place too much trust in AI-based tools, they may overlook errors or biases inherent in these systems. This lack of critical scrutiny can result in unjust outcomes for students, as their performance may be inaccurately or unfairly represented. Another concern is to stifle students' skills \shortcite{williamson_re-examining_2023}. If educators become overly reliant on automated assessment tools, they may prioritize conformity to standardized metrics over fostering essential skills such as critical thinking, creativity, and problem-solving. This narrow focus on quantifiable outcomes could limit students' intellectual development and inhibit their ability to thrive in a rapidly changing world. Furthermore, automation bias can raise questions about accountability and transparency \cite{williamson_re-examining_2023}. When decisions about student performance are delegated to automated systems, it can be challenging to hold stakeholders or institutions accountable for errors or biases in the assessment process.

In addition, overreliance on automated assessment systems may reduce human interaction and feedback, depriving students of valuable opportunities for personalized guidance and mentorship: concern about dehumanization \cite{fritts_ai_2021}. Moreover, suppose students become accustomed to receiving automated grading and feedback without understanding the underlying reasoning. In that case, they may become less adept at critically evaluating their work and identifying improvement areas. Second, there might be the issue of unquestioning acceptance of feedback. Automation bias may lead students to accept feedback provided by AI-based tools without critically evaluating its accuracy or considering alternative perspectives. This uncritical acceptance could hinder students' development of essential self-assessment and self-regulation skills. Stakeholders of educational assessments should ensure that automation serves as a valuable tool for enhancing, rather than detracting from, the assessment experience for all students.

Ethical discussions around automation bias extend to considerations of education and training \cite{rahm_education_2023}. It is important to educate students and other stakeholders in educational assessments about the limitations of automated systems and empower them to critically evaluate and supplement automated outputs with human judgment when necessary. In addition, human-centered AI has been discussed as a means to mitigate the automation bias. Human-centered AI places humans at the center of the design process, focusing on creating, developing, and deploying AI-based tools that prioritize human values, needs, and experiences \cite{bond_human_2019}. Following human-centered AI, AI-based assessment tools should have the capability to steer users (e.g., students and educators) through the decision-making process, enabling individuals to make their own micro-decisions (see \shortciteNP{cairns_computer-human_2016}). AI decisions can be reserved until the conclusion, serving to mitigate automation bias and anchoring effects from the outset. The AI tools can also present multiple competing decision statements and explanations to enhance transparency and encourage the user to engage in reasoning and arrive at the final decision, referred to as a differential diagnosis \cite{bond_human_2019}.

\subsection{Evidence-Based Practice in Educational Assessment}

Evidence-based practice refers to the intentional, reliable, and judicious use of empirical evidence to inform real-life decisions. In the context of educational assessment, evidence-based practice serves as the cornerstone for ensuring the efficacy, fairness, and reliability of assessment applications. It involves a deliberate and meticulous approach to decision-making rooted in robust research findings and empirical evidence.

Empirical evidence holds immense significance, especially in the context of AI-powered assessment tools, for several reasons. First, it acts as a safeguard against different sources of bias and discrimination. By rigorously examining data gathered from diverse populations, researchers can identify and mitigate potential biases embedded within algorithms or AI-based assessment frameworks. This process is crucial for ensuring that AI assessments provide equitable opportunities for all individuals, regardless of their background or characteristics.

Second, empirical evidence allows for the continuous refinement and improvement of AI-powered assessment tools. Through ongoing research and data analysis, researchers and assessment experts can identify areas of strength and weakness within AI algorithms, leading to iterative enhancements that optimize accuracy and predictive validity. This iterative process fosters a culture of innovation and continuous improvement, ultimately benefiting both educators and learners.

Third, evidence-based practice fosters transparency and accountability within the educational assessment landscape. By documenting the research methodologies, data sources, and validation procedures used to develop AI-powered assessment tools, researchers and practitioners can enhance the credibility and trustworthiness of their applications. This transparency not only instills confidence in different stakeholders (e.g., learners, educators, parents, and employers) but also encourages collaboration and knowledge-sharing within the broader educational community.

\subsection{Democratizing AI in Education}

AI technologies can immensely benefit education, offering new opportunities for teachers, students, and others involved. However, we must remain aware that these tools have the potential to amplify existing social inequalities. As AI-powered educational solutions become more widespread, it is essential from a fairness perspective to ensure that their advantages are accessible to all, irrespective of ethnicity, gender, disability status, or socioeconomic background. The three measures that ensure that AI benefits everyone include policies of accessibility testing, investing in open-source initiatives, and working more directly with measurement professionals and departments of education as representatives in the governance for the development of AI in the sector.

The divide between those with access to technology and those without access has been termed the "Digital Divide" \cite{cullen2001addressing}. This issue manifests itself in many countries where rural and economically disadvantaged communities lack reliable internet service. As per \shortciteA{katz2021learning}, in the United States of America, only 72 percent of Hispanic-headed households have broadband access compared to 80 percent of white families. This digital divide hinders digital literacy, which has already reduced employment opportunities, exacerbated disparities, and deepened social stratification. As AI-based tools primarily rely upon cloud-based services, the existing digital divide poses a real challenge in providing the benefits of AI-based tools to underprivileged students. The consequence is that the digital divide extends to AI-based skills that are becoming critical to future employment opportunities.

The integration of generative AI is increasingly being heralded as a huge business opportunity. However, it also can widen economic inequalities between those who possess the skills and resources to leverage AI and those who lack them. This divide is gaining recognition as the "AI Divide" \cite{kitsara2022artificial}. In response, there is a growing discourse around the "Democratization of AI" as a possible solution, which broadly calls for greater participation in the utilization, profitability, development, and governance of AI by all sections of society \shortcite{seger2023democratising}). In the book "AI for Everyone?: Critical Perspectives", \shortciteA{verdegem2021ai} argues that for AI to transform society positively, a "radical democratization of AI" is essential, and this can be achieved by adhering to three fundamental principles:
\begin{itemize}
\item AI should be accessible to everyone. Nobody should be excluded from using AI because of differences in race, gender, class, or other distinctions.
\item Developments in AI should contribute to the well-being of everyone in society.
\item In a decent society, all members should have a say about what type of AI is being developed and what services are being offered.
\end{itemize}
From an educational standpoint, the first of these principles calls for developers of AI-based tools to work closely with educators and stakeholders to make AI accessible. As developers, this calls us to work closely to ensure disabilities do not adversely affect the usability of AI-powered tools. Policies concerning accessibility testing ensure that AI-based tools are usable to as many students as possible.

The primary obstacle to making AI accessible to everyone lies in closed-source models and data. As AI becomes more powerful and expensive, companies like OpenAI, Google, and Anthropic have sought to protect their intellectual property by restricting access to the model weights and training data. Not only does relying on closed-source solutions for building AI-based educational applications raise concerns regarding privacy and explainability \shortcite{gimpel_upside_2018}, but it also means that the application would depend on a third party's pricing structure. While current prices are quite reasonable, it is well-known that OpenAI operated at a significant loss of \$540 million last year. If companies like OpenAI were to charge the total cost of running such large models, this move could make AI unaffordable for socioeconomically disadvantaged groups, thereby limiting the benefits of AI-based educational tools to only those who can afford the service.

Fortunately, some companies like Meta and Mistral have taken a positive step towards making AI more accessible by releasing open-source models such as Llama 3 and Mixtral \shortcite{jiang2024mixtral}. These models have been benchmarked and evaluated, and their performance is competitive with closed-source models. Additionally, they have been released under licenses generally suitable for educational applications. Transitioning from closed-source initiatives towards open-source alternatives democratizes the governance and development of AI-based tools, enabling a broader range of people to contribute to the development of AI-based tools. It is also worth noting that libraries like the Transformer Library \shortcite{wolf2019huggingface} play a role in making AI more accessible by lowering the technical barriers to training and deploying AI in production environments. The availability of open-source models and software libraries that simplify the training and inference increases participation in developing AI-based tools, fostering collaboration. It creates a more competitive landscape that potentially reduces the cost of AI-based tools.

Critics who oppose the democratization of AI have raised significant concerns about democratizing the governance of AI systems \shortcite{himmelreich2023against}. They question how ethical standards for open-source models can be effectively imposed and legitimately governed by a majority. While these concerns primarily focus on the process of democratizing AI rather than whether it should be done, they highlight valid points. From an educational perspective, this discussion underscores the practical importance of involving measurement professionals and departments of education to serve as representatives for the interests of students and other stakeholders in the governance and development of AI-based educational tools. Their participation could help ensure that the equity and accessibility concerns of the educational sector are adequately addressed as AI becomes more democratized.

\subsection{Environmental Impact of AI in Education}

We are seeing LLMs being integrated into many aspects of educational technology. As noted in previous sections, these models are favored for their improvements in accuracy in the automated scoring of constructed response items. This accuracy comes at a significant increase in computational requirements, which, in turn, necessitates similar increases in carbon emissions \shortcite{strubell2020energy}. It has been noted that training times for models like BERT are an order of magnitude longer than traditional n-gram approaches. However, the accuracy gains from transformer-based approaches can often be minimal \shortcite{mayfield2020should}.

Regarding educational applications, evaluating whether adopting AI-based tools justifies the significant increase in costs and environmental emissions is crucial. In cases where these gains are significant, we need ways to mitigate the environmental impacts. We emphasize specific strategies discussed in \shortciteA{khowaja2024chatgpt} that can help mitigate the environmental impact associated with the transition to LLMs. The first recommendation is to optimize computations. One way to do this is to pursue efficient model architectures. For example, the MobileBERT architecture is an example that uses bottlenecks to decrease the computational load imposed by the attention mechanism \shortcite{sun2020mobilebert}. Such architectures have been shown to perform comparably to much larger models in an educational context \shortcite{ormerod2021automated}. For larger generative models, the adoption of parameter-efficient methods such as Quantized Low-Rank Adaptation has been estimated to reduce the overall carbon footprint of models by 72\% \shortcite{dettmers2024qlora}. In addition to optimizing architectures, the parallelizable nature of transformer-based models means that they can be run efficiently on GPUs or specialized Tensor Processing Units (TPUs), which can reduce training times and overall electricity consumption \shortcite{khowaja2024chatgpt}.

The second recommendation is to ensure the energy used comes from renewable sources. If we are serving AI on local hardware, the percentage of energy coming from renewable sources depends only on the energy provider, so we only consider cloud computing services. Almost all cloud computing providers have demonstrated some commitment to renewable energy. We highlight the commitments made by the three cloud computing providers with the most significant market share :
\begin{itemize}
\item Amazon Web Services (31\%): Through “The Climate Pledge,” Amazon aims to achieve “Net-Zero Carbon” by 2040.
\item Microsoft Azure (24\%): Microsoft aims to have a negative footprint by 2030.
\item Google Cloud Platform (11\%): Google has been carbon neutral since 2007 and claims to be the first major company to achieve carbon neutrality.
\end{itemize}
This focus on the emissions from AI has prompted \shortciteA{lacoste2019quantifying} to develop a tool quantifying emissions by cloud computing provider, hardware, and region. This tool highlights that services like Google Cloud Provider and Azure generally seek to offset their emissions contribution, whereas others, such as AWS, do not. Ensuring AI-powered applications are powered by renewable energy or mitigated by suitable offsets can simply be a choice of appropriate cloud computing provider.

The third recommendation by \shortciteA{khowaja2024chatgpt} is to encourage collaboration. Open-sourcing models, data, and collaborative research help foster innovation while reducing the duplication of efforts and resources. This being said, we acknowledge that there are practical barriers to some collaboration, such as data privacy and the application of intellectual property to software development in a corporate setting.

Educational measurement aims to leverage AI advancements to improve student outcomes. Despite the potential carbon footprint from AI-enabled educational tools being relatively small compared to overall emissions, no industry should be immune from the moral obligation to address climate change. We have an ethical responsibility to ensure that these tools are utilized responsibly and sustainably, necessitating a heightened awareness of the carbon footprint imposed by AI-based educational tools.

\section{Discussion}

Despite AI's potential to enhance educational measurement, researchers and practitioners must acknowledge and understand the limitations and ethical challenges associated with AI applications. An informed understanding will help prevent unquestioning belief in, over-reliance on, or misuse of AI technologies, especially in high-stakes settings where the implications and consequences of errors or biases can be significant. This paper outlined several ethical challenges common to many AI applications in educational assessment. First, AI technologies mirror and can even amplify biases in the training data; such models are often trained with unrepresentative data and thus inherently hold bias towards certain groups. Second, the transparency and explainability of AI technologies remain significant ethical challenges, particularly for tools that rely on third-party generative models. Third, AI technologies are often not evaluated before use, leading to issues such as creating items without good measurement properties or grading essays and assignments with low consistency. Fourth, the adoption of AI in educational assessments can influence broader social-environmental contexts, such as impacts on students' personal lives, equity of access to educational technologies for students from different backgrounds and cultures, and the environment. 

Addressing these ethical challenges aligns with the principles outlined by \shortciteA{ammanath_trustworthy_2022}, which emphasizes that AI tools for educational purposes must be thoughtfully designed, deployed, and monitored to ensure they are safe, robust, transparent, explainable, and responsible. Additionally, these ethical challenges are particularly important when viewed through the larger lens of the Standards \shortcite{standards}, which already contain a rigorous set of expectations for test takers, psychometricians, and test designers. Given that a new version of the Standards is currently under development, there may need to be an increased focus within the larger measurement community on the ethical use of AI. As the field increasingly adopts AI for educational measurement purposes, it is imperative to recognize when an AI tool may be unsafe, unfair, or unreliable. This vigilance will help harness AI's potential while safeguarding against its pitfalls, thereby advancing educational assessment in a responsible and ethical manner.

To successfully integrate AI technologies into the educational landscape, it is imperative to clearly define and attribute the roles played by human stakeholders and AI systems. By clearly delineating these roles, we can address the ethical challenge of accountability and use AI tools more judiciously. In the current paper, we argue that AI technologies should be applied to augment rather than replace human intelligence. In this framework, regardless of the aspect of educational assessments in which AI is employed (e.g., item generation, essay scoring, feedback generation, or proctoring), humans should maintain a supervisory role to ensure accuracy and responsible use. Therefore, humans must understand the reasons for AI systems' errors and biases and correct them. For humans to correct errors made by AI systems, it is essential to ensure that AI is understandable and transparent to its users \shortcite{shneiderman2022human}. For instance, in AIG tasks, users should understand how the characteristics of their prompts influence the outcomes from AI systems and how their choice of AI systems may impact the generated items, thereby effectively managing the quality and relevance of the generated content \shortcite{tan2024review}. Achieving this level of understanding and transparency may encourage increased collaboration between AI developers, measurement specialists, educators, educational researchers, and other stakeholders. By fostering such collaboration, the educational community can ensure that AI technologies are effectively, responsibly, and ethically integrated into the educational landscape.

As AI-powered tools are increasingly adopted within formal and informal education settings, it will require careful assurance that they are developed and adopted ethically, equitably, and effectively. This necessitates a dual-purpose, evidence-based educational measurement approach that scrutinizes the technical development of AI-powered tools and their real-world applications in educational contexts. The case of AWE systems serves as an excellent illustration of this need. On the development side, researchers must ensure that AWE systems are well-developed, usable, trustworthy, and unbiased. Furthermore, empirical research must analyze potential biases in training data, the outputs, and the validity and reliability of AI scoring \shortcite{goldshtein2024automating, raczynski_appraising_2018, wilson2019generalizability, wind_influence_2018}, while also examining the intended and unintended consequences (i.e., consequential validity) of using these AI measurement tools in authentic educational settings \shortcite{CORRENTI2022100084, correnti2024supporting, wilson_predictors_2024}. Research exploring how AWE feedback is integrated into teaching practices and its effectiveness in enhancing student writing skills is especially important for ensuring that AI-driven AWE tools truly support educational objectives and do not inadvertently undermine critical human skills and relationships \shortcite{wilson_elementary_2021}. Thus, by embracing a holistic research approach that bridges technical and contextual analyses, the educational measurement community can better understand and leverage the potential of AI to enhance teaching and learning while mitigating any adverse or unintended consequences.

Similarly, it is essential to ensure clear, transparent, and effective communication with stakeholders when implementing AI in real-world educational assessments. These stakeholders often include students, parents, educators, policymakers, and other users of assessment scores or feedback. Such communication may include educating stakeholders about how the technology works, where and how it is used in the assessment process, how the results should or should not be interpreted, and addressing concerns and potential misconceptions. In this way, stakeholders can develop a realistic understanding of the AI’s capabilities and limitations, which is crucial for building trust and acceptance. Additionally, open dialogues about the ethical considerations and safeguards in place to protect the integrity of the assessment process could help to further reinforce confidence among stakeholders and encourage test-takers to give their best performance.

Adopting new technologies like AI in the educational sector must also be done through an ethical lens that prioritizes environmental sustainability. No industry can afford to be insulated from the moral imperative to confront climate change head-on. Although the deployment of generative AI carries a larger carbon footprint than traditional approaches, this paper has outlined several pragmatic strategies to mitigate the climate impacts stemming from generative AI usage. Key recommendations include the proliferation of open-source AI models that can be leveraged and improved collaboratively, reducing redundancies. Continued research into developing computationally efficient AI methods that require less energy-intensive training is also crucial. Moreover, educational institutions should aim to harness cloud computing services that are carbon neutral through renewable energy offsets or direct procurement of clean energy.

\subsection{Directions for Future Research}

There are several directions for future AI and educational measurement work, especially around fairness, environmental impact, and explainability. The first of these is creating a set of evidence-based metrics and benchmarks to ensure that AI-based systems treat all students fairly, regardless of demographic background, with respect to educational tasks such as grading, giving feedback, and making predictions around student outcomes. Another option to improve model fairness may be researching and creating a series of vetted training datasets that can be used as additional training material to represent populations who have historically tended to experience AI bias. Additional fairness work may also include research into different human-in-the-loop configurations with respect to high-stakes applications of AI in education such that the AI system alone is not making the final decision when the outcome of these decisions may have a lasting impact on the student. 

Second, there is often a large fiscal and energy burden to training brand-new AI systems as well as the ongoing maintenance and use of these systems \shortciteA{khowaja2024chatgpt}. More research needs to be performed, such as that of \shortciteA{dettmers2024qlora}, to find new ways to design these AI models to be more efficient and thus lower the energy burden of these systems. An additional approach may be to expand research on optimizing and fine-tuning existing general LLMs for specific educational purposes instead of incurring the additional energy burden of training new models. 

Third, to meet the current rigorous expectations of ``Rights and Responsibilities of Test Takers'' found in the Standards \shortcite{standards}, more work is needed to ensure that there are rigorous standards for AI model explainability that would allow any test taker to inquire how their score was calculated. Currently, one of the main concerns over the use of LLM-based tools for scoring purposes is that these models contain thousands, millions, or billions of parameters that make it very difficult for humans to understand all of the aspects of how LLM output was derived; we have referred to this above as the ``black box'' problem. In fact, not even experts and LLM developers can reliably interpret how a specific output was reached \shortcite{bowman2023things}. To encourage deeper trust in using these tools for high-stakes purposes, more research must be performed to foster a deeper understanding of and dissemination about how these scores are generated to support test takers in knowing more about how their scores are being constructed. 

Furthermore, while we provide robust and representative examples of assessments in this paper, it is crucial to recognize the ongoing need to expand our data sources continually. We emphasize the importance of including AI-powered assessment tools designed for culturally and linguistically diverse populations. We encourage future research to incorporate a broader range of these tools from various perspectives and contexts. This strategy will help address potential biases and limitations in data collection, ensuring that AI tools are equitable and effective for all users. By diversifying the data to include culturally and linguistically relevant tools, we can enhance the accuracy and fairness of AI applications in education, leading to more inclusive and reliable outcomes.

\subsection{Conclusion}

This paper has explored broader issues surrounding fairness and equity in AI. We have underscored the critical need to democratize AI technologies, ensuring they remain widely accessible resources that can benefit all segments of society equitably. This democratization entails more than just making AI available--it necessitates proactive efforts to guarantee that the development and implementation of these powerful technologies occur through an inclusive process that empowers diverse voices and perspectives. There is a genuine danger that advanced AI capabilities could become concentrated in the hands of a privileged few nations, corporations, or elite groups. True democratization in education demands that we foster ecosystems where AI development is a participatory endeavor, drawing from the experience of teachers, students, and assessment professionals within government and industry. We can only shape AI to be genuinely universally beneficial through multi-stakeholder collaboration. Seeking the best way to foster such ecosystems requires careful consideration and ensuring that our AI systems align with and abide by human values.

\newpage
\bibliographystyle{apacite}

\begin{thebibliography}{}

\bibitem [\protect \citeauthoryear {%
{AERA, APA, NCME}%
}{%
{AERA, APA, NCME}%
}{%
{\protect \APACyear {2014}}%
}]{%
standards}
\APACinsertmetastar {%
standards}%
\begin{APACrefauthors}%
{AERA, APA, NCME}.%
\end{APACrefauthors}%
\unskip\
\newblock
\APACrefYear{2014}.
\newblock
\APACrefbtitle {Standards for Educational and Psychological Testing} {Standards for educational and psychological testing}.
\newblock
\APACaddressPublisher{Washington, DC}{American Educational Research Association}.
\PrintBackRefs{\CurrentBib}

\bibitem [\protect \citeauthoryear {%
Almond%
\ \protect \BOthers {.}}{%
Almond%
\ \protect \BOthers {.}}{%
{\protect \APACyear {2010}}%
}]{%
almond_technology-enabled_2010}
\APACinsertmetastar {%
almond_technology-enabled_2010}%
\begin{APACrefauthors}%
Almond, P.%
, Winter, P.%
, Cameto, R.%
, Russell, M.%
, Sato, E.%
, Clarke-Midura, J.%
\BDBL {}Lazarus, S.%
\end{APACrefauthors}%
\unskip\
\newblock
\APACrefYearMonthDay{2010}{}{}.
\newblock
{\BBOQ}\APACrefatitle {Technology-Enabled and Universally Designed Assessment: Considering Access in Measuring the Achievement of Students with Disabilities—{A} Foundation for Research} {Technology-enabled and universally designed assessment: Considering access in measuring the achievement of students with disabilities—{A} foundation for research}.{\BBCQ}
\newblock
\APACjournalVolNumPages{The Journal of Technology, Learning and Assessment}{10}{5}{}.
\newblock
\APACrefnote{Number: 5}
\PrintBackRefs{\CurrentBib}

\bibitem [\protect \citeauthoryear {%
Alsabhan%
}{%
Alsabhan%
}{%
{\protect \APACyear {2023}}%
}]{%
alsabhan2023student}
\APACinsertmetastar {%
alsabhan2023student}%
\begin{APACrefauthors}%
Alsabhan, W.%
\end{APACrefauthors}%
\unskip\
\newblock
\APACrefYearMonthDay{2023}{}{}.
\newblock
{\BBOQ}\APACrefatitle {Student cheating detection in higher education by implementing machine learning and LSTM techniques} {Student cheating detection in higher education by implementing machine learning and lstm techniques}.{\BBCQ}
\newblock
\APACjournalVolNumPages{Sensors}{23}{8}{4149}.
\newblock
\begin{APACrefDOI} \doi{10.3390/s23084149} \end{APACrefDOI}
\PrintBackRefs{\CurrentBib}

\bibitem [\protect \citeauthoryear {%
Alwahaby%
, Cukurova%
, Papamitsiou%
\BCBL {}\ \BBA {} Giannakos%
}{%
Alwahaby%
\ \protect \BOthers {.}}{%
{\protect \APACyear {2022}}%
}]{%
alwahaby_evidence_2022}
\APACinsertmetastar {%
alwahaby_evidence_2022}%
\begin{APACrefauthors}%
Alwahaby, H.%
, Cukurova, M.%
, Papamitsiou, Z.%
\BCBL {}\ \BBA {} Giannakos, M.%
\end{APACrefauthors}%
\unskip\
\newblock
\APACrefYearMonthDay{2022}{}{}.
\newblock
{\BBOQ}\APACrefatitle {The Evidence of Impact and Ethical Considerations of Multimodal Learning Analytics: {A} Systematic Literature Review} {The evidence of impact and ethical considerations of multimodal learning analytics: {A} systematic literature review}.{\BBCQ}
\newblock
\BIn{} M.~Giannakos, D.~Spikol, D.~Di~Mitri, K.~Sharma, X.~Ochoa\BCBL {}\ \BBA {} R.~Hammad\ (\BEDS), \APACrefbtitle {The Multimodal Learning Analytics Handbook} {The multimodal learning analytics handbook}\ (\BPGS\ 289--325).
\newblock
\APACaddressPublisher{Cham}{Springer International Publishing}.
\newblock
\begin{APACrefDOI} \doi{10.1007/978-3-031-08076-0\_12} \end{APACrefDOI}
\PrintBackRefs{\CurrentBib}

\bibitem [\protect \citeauthoryear {%
Ammanath%
}{%
Ammanath%
}{%
{\protect \APACyear {2022}}%
}]{%
ammanath_trustworthy_2022}
\APACinsertmetastar {%
ammanath_trustworthy_2022}%
\begin{APACrefauthors}%
Ammanath, B.%
\end{APACrefauthors}%
\unskip\
\newblock
\APACrefYear{2022}.
\newblock
\APACrefbtitle {Trustworthy {AI}: {A} business guide for navigating trust and ethics in {AI}} {Trustworthy {AI}: {A} business guide for navigating trust and ethics in {AI}}.
\newblock
\APACaddressPublisher{}{John Wiley \& Sons}.
\PrintBackRefs{\CurrentBib}

\bibitem [\protect \citeauthoryear {%
Andersen%
, Yuan%
, Watson%
\BCBL {}\ \BBA {} Cheung%
}{%
Andersen%
\ \protect \BOthers {.}}{%
{\protect \APACyear {2021}}%
}]{%
andersen_benefits_2021}
\APACinsertmetastar {%
andersen_benefits_2021}%
\begin{APACrefauthors}%
Andersen, {\O}\BPBI E.%
, Yuan, Z.%
, Watson, R.%
\BCBL {}\ \BBA {} Cheung, K\BPBI Y\BPBI F.%
\end{APACrefauthors}%
\unskip\
\newblock
\APACrefYearMonthDay{2021}{}{}.
\newblock
\APACrefbtitle {Benefits of Alternative Evaluation Methods for Automated Essay Scoring} {Benefits of alternative evaluation methods for automated essay scoring}\ \APACbVolEdTR{}{\BTR{}}.
\newblock
\APACaddressInstitution{Paris, France}{International Educational Data Mining Society}.
\PrintBackRefs{\CurrentBib}

\bibitem [\protect \citeauthoryear {%
Angoff%
}{%
Angoff%
}{%
{\protect \APACyear {2012}}%
}]{%
angoff2012perspectives}
\APACinsertmetastar {%
angoff2012perspectives}%
\begin{APACrefauthors}%
Angoff, W\BPBI H.%
\end{APACrefauthors}%
\unskip\
\newblock
\APACrefYearMonthDay{2012}{}{}.
\newblock
{\BBOQ}\APACrefatitle {Perspectives on differential item functioning methodology} {Perspectives on differential item functioning methodology}.{\BBCQ}
\newblock
\BIn{} \APACrefbtitle {Differential item functioning} {Differential item functioning}\ (\BPGS\ 3--23).
\newblock
\APACaddressPublisher{}{Routledge}.
\PrintBackRefs{\CurrentBib}

\bibitem [\protect \citeauthoryear {%
Arora%
}{%
Arora%
}{%
{\protect \APACyear {2021}}%
}]{%
arora_is_2021}
\APACinsertmetastar {%
arora_is_2021}%
\begin{APACrefauthors}%
Arora, P.%
\end{APACrefauthors}%
\unskip\
\newblock
\APACrefYearMonthDay{2021}{}{}.
\newblock
\APACrefbtitle {Is Remote Proctoring The Future of Academia?} {Is remote proctoring the future of academia?}
\newblock
\begin{APACrefURL} \url{https://elearningindustry.com/is-remote-proctoring-future-academia} \end{APACrefURL}
\PrintBackRefs{\CurrentBib}

\bibitem [\protect \citeauthoryear {%
Bender%
, Gebru%
, McMillan-Major%
\BCBL {}\ \BBA {} Shmitchell%
}{%
Bender%
\ \protect \BOthers {.}}{%
{\protect \APACyear {2021}}%
}]{%
bender_dangers_2021}
\APACinsertmetastar {%
bender_dangers_2021}%
\begin{APACrefauthors}%
Bender, E\BPBI M.%
, Gebru, T.%
, McMillan-Major, A.%
\BCBL {}\ \BBA {} Shmitchell, S.%
\end{APACrefauthors}%
\unskip\
\newblock
\APACrefYearMonthDay{2021}{{\APACmonth{03}}}{}.
\newblock
{\BBOQ}\APACrefatitle {On the Dangers of Stochastic Parrots: {C}an Language Models Be Too Big?} {On the dangers of stochastic parrots: {C}an language models be too big?}{\BBCQ}
\newblock
\BIn{} \APACrefbtitle {Proceedings of the 2021 {ACM} {Conference} on {Fairness}, {Accountability}, and {Transparency}} {Proceedings of the 2021 {ACM} {Conference} on {Fairness}, {Accountability}, and {Transparency}}\ (\BPGS\ 610--623).
\newblock
\APACaddressPublisher{New York, NY, USA}{Association for Computing Machinery}.
\newblock
\begin{APACrefDOI} \doi{10.1145/3442188.3445922} \end{APACrefDOI}
\PrintBackRefs{\CurrentBib}

\bibitem [\protect \citeauthoryear {%
Bergmans%
, Bouali%
, Luttikhuis%
\BCBL {}\ \BBA {} Rensink%
}{%
Bergmans%
\ \protect \BOthers {.}}{%
{\protect \APACyear {2021}}%
}]{%
bergmans_efficacy_2021}
\APACinsertmetastar {%
bergmans_efficacy_2021}%
\begin{APACrefauthors}%
Bergmans, L.%
, Bouali, N.%
, Luttikhuis, M.%
\BCBL {}\ \BBA {} Rensink, A.%
\end{APACrefauthors}%
\unskip\
\newblock
\APACrefYearMonthDay{2021}{}{}.
\newblock
{\BBOQ}\APACrefatitle {On the Efficacy of Online Proctoring using {Proctorio}} {On the efficacy of online proctoring using {Proctorio}}.{\BBCQ}
\newblock
\BIn{} \APACrefbtitle {Proceedings of the 13th {International} {Conference} on {Computer} {Supported} {Education} ({CSEDU} 2021)} {Proceedings of the 13th {International} {Conference} on {Computer} {Supported} {Education} ({CSEDU} 2021)}\ (\BPGS\ 279--290).
\newblock
\APACaddressPublisher{}{SCITEPRESS}.
\newblock
\begin{APACrefDOI} \doi{10.5220/0010399602790290} \end{APACrefDOI}
\PrintBackRefs{\CurrentBib}

\bibitem [\protect \citeauthoryear {%
Biecek%
\ \BBA {} Burzykowski%
}{%
Biecek%
\ \BBA {} Burzykowski%
}{%
{\protect \APACyear {2021}}%
}]{%
biecek2021explanatory}
\APACinsertmetastar {%
biecek2021explanatory}%
\begin{APACrefauthors}%
Biecek, P.%
\BCBT {}\ \BBA {} Burzykowski, T.%
\end{APACrefauthors}%
\unskip\
\newblock
\APACrefYear{2021}.
\newblock
\APACrefbtitle {Explanatory model analysis: Explore, explain, and examine predictive models} {Explanatory model analysis: Explore, explain, and examine predictive models}.
\newblock
\APACaddressPublisher{}{Chapman and Hall/CRC}.
\PrintBackRefs{\CurrentBib}

\bibitem [\protect \citeauthoryear {%
Bilen%
\ \BBA {} Matros%
}{%
Bilen%
\ \BBA {} Matros%
}{%
{\protect \APACyear {2021}}%
}]{%
bilen2021online}
\APACinsertmetastar {%
bilen2021online}%
\begin{APACrefauthors}%
Bilen, E.%
\BCBT {}\ \BBA {} Matros, A.%
\end{APACrefauthors}%
\unskip\
\newblock
\APACrefYearMonthDay{2021}{}{}.
\newblock
{\BBOQ}\APACrefatitle {Online cheating amid {COVID-19}} {Online cheating amid {COVID-19}}.{\BBCQ}
\newblock
\APACjournalVolNumPages{Journal of Economic Behavior \& Organization}{182}{}{196--211}.
\newblock
\begin{APACrefDOI} \doi{10.1016/j.jebo.2020.12.004} \end{APACrefDOI}
\PrintBackRefs{\CurrentBib}

\bibitem [\protect \citeauthoryear {%
Black%
\ \BBA {} Wiliam%
}{%
Black%
\ \BBA {} Wiliam%
}{%
{\protect \APACyear {2009}}%
}]{%
black_developing_2009}
\APACinsertmetastar {%
black_developing_2009}%
\begin{APACrefauthors}%
Black, P.%
\BCBT {}\ \BBA {} Wiliam, D.%
\end{APACrefauthors}%
\unskip\
\newblock
\APACrefYearMonthDay{2009}{}{}.
\newblock
{\BBOQ}\APACrefatitle {Developing the theory of formative assessment} {Developing the theory of formative assessment}.{\BBCQ}
\newblock
\APACjournalVolNumPages{Educational Assessment, Evaluation and Accountability}{21}{1}{5--31}.
\newblock
\begin{APACrefDOI} \doi{10.1007/s11092-008-9068-5} \end{APACrefDOI}
\PrintBackRefs{\CurrentBib}

\bibitem [\protect \citeauthoryear {%
Blikstein%
\ \BBA {} Worsley%
}{%
Blikstein%
\ \BBA {} Worsley%
}{%
{\protect \APACyear {2016}}%
}]{%
blikstein_multimodal_2016}
\APACinsertmetastar {%
blikstein_multimodal_2016}%
\begin{APACrefauthors}%
Blikstein, P.%
\BCBT {}\ \BBA {} Worsley, M.%
\end{APACrefauthors}%
\unskip\
\newblock
\APACrefYearMonthDay{2016}{}{}.
\newblock
{\BBOQ}\APACrefatitle {Multimodal Learning Analytics and Education Data Mining: Using computational technologies to measure complex learning tasks} {Multimodal learning analytics and education data mining: Using computational technologies to measure complex learning tasks}.{\BBCQ}
\newblock
\APACjournalVolNumPages{Journal of Learning Analytics}{3}{2}{220--238}.
\newblock
\begin{APACrefDOI} \doi{10.18608/jla.2016.32.11} \end{APACrefDOI}
\PrintBackRefs{\CurrentBib}

\bibitem [\protect \citeauthoryear {%
Bond%
\ \protect \BOthers {.}}{%
Bond%
\ \protect \BOthers {.}}{%
{\protect \APACyear {2019}}%
}]{%
bond_human_2019}
\APACinsertmetastar {%
bond_human_2019}%
\begin{APACrefauthors}%
Bond, R\BPBI R.%
, Mulvenna, M\BPBI D.%
, Wan, H.%
, Finlay, D\BPBI D.%
, Wong, A.%
, Koene, A.%
\BDBL {}Adel, T.%
\end{APACrefauthors}%
\unskip\
\newblock
\APACrefYearMonthDay{2019}{}{}.
\newblock
{\BBOQ}\APACrefatitle {Human Centered Artificial Intelligence: Weaving {UX} into Algorithmic Decision Making} {Human centered artificial intelligence: Weaving {UX} into algorithmic decision making}.{\BBCQ}
\newblock
\BIn{} \APACrefbtitle {RoCHI} {Rochi}\ (\BPGS\ 2--9).
\PrintBackRefs{\CurrentBib}

\bibitem [\protect \citeauthoryear {%
Boud%
\ \BBA {} Molloy%
}{%
Boud%
\ \BBA {} Molloy%
}{%
{\protect \APACyear {2013}}%
}]{%
boud_rethinking_2013}
\APACinsertmetastar {%
boud_rethinking_2013}%
\begin{APACrefauthors}%
Boud, D.%
\BCBT {}\ \BBA {} Molloy, E.%
\end{APACrefauthors}%
\unskip\
\newblock
\APACrefYearMonthDay{2013}{}{}.
\newblock
{\BBOQ}\APACrefatitle {Rethinking models of feedback for learning: the challenge of design} {Rethinking models of feedback for learning: the challenge of design}.{\BBCQ}
\newblock
\APACjournalVolNumPages{Assessment \& Evaluation in Higher Education}{38}{6}{698--712}.
\newblock
\begin{APACrefDOI} \doi{10.1080/02602938.2012.691462} \end{APACrefDOI}
\PrintBackRefs{\CurrentBib}

\bibitem [\protect \citeauthoryear {%
Boulanger%
\ \BBA {} Kumar%
}{%
Boulanger%
\ \BBA {} Kumar%
}{%
{\protect \APACyear {2024}}%
}]{%
boulanger2024explainable}
\APACinsertmetastar {%
boulanger2024explainable}%
\begin{APACrefauthors}%
Boulanger, D.%
\BCBT {}\ \BBA {} Kumar, V\BPBI S.%
\end{APACrefauthors}%
\unskip\
\newblock
\APACrefYearMonthDay{2024}{}{}.
\newblock
{\BBOQ}\APACrefatitle {Explainable {AI} and {AWE}: {B}alancing tensions between transparency and predictive accuracy} {Explainable {AI} and {AWE}: {B}alancing tensions between transparency and predictive accuracy}.{\BBCQ}
\newblock
\BIn{} \APACrefbtitle {The {R}outledge International Handbook of Automated Essay Evaluation} {The {R}outledge international handbook of automated essay evaluation}\ (\BPGS\ 445--468).
\newblock
\APACaddressPublisher{}{Routledge}.
\PrintBackRefs{\CurrentBib}

\bibitem [\protect \citeauthoryear {%
Bowman%
}{%
Bowman%
}{%
{\protect \APACyear {2023}}%
}]{%
bowman2023things}
\APACinsertmetastar {%
bowman2023things}%
\begin{APACrefauthors}%
Bowman, S\BPBI R.%
\end{APACrefauthors}%
\unskip\
\newblock
\APACrefYearMonthDay{2023}{}{}.
\newblock
\APACrefbtitle {Eight Things to Know about Large Language Models.} {Eight things to know about large language models.}
\PrintBackRefs{\CurrentBib}

\bibitem [\protect \citeauthoryear {%
Bozkurt%
\ \BBA {} Sharma%
}{%
Bozkurt%
\ \BBA {} Sharma%
}{%
{\protect \APACyear {2023}}%
}]{%
bozkurt_generative_2023}
\APACinsertmetastar {%
bozkurt_generative_2023}%
\begin{APACrefauthors}%
Bozkurt, A.%
\BCBT {}\ \BBA {} Sharma, R\BPBI C.%
\end{APACrefauthors}%
\unskip\
\newblock
\APACrefYearMonthDay{2023}{}{}.
\newblock
{\BBOQ}\APACrefatitle {Generative {AI} and Prompt Engineering: {The} Art of Whispering to Let the Genie Out of the Algorithmic World} {Generative {AI} and prompt engineering: {The} art of whispering to let the genie out of the algorithmic world}.{\BBCQ}
\newblock
\APACjournalVolNumPages{Asian Journal of Distance Education}{18}{2}{i--vii}.
\newblock
\begin{APACrefURL} \url{https://www.asianjde.com/ojs/index.php/AsianJDE/article/view/749} \end{APACrefURL}
\PrintBackRefs{\CurrentBib}

\bibitem [\protect \citeauthoryear {%
Breyer%
, Rupp%
\BCBL {}\ \BBA {} Bridgeman%
}{%
Breyer%
\ \protect \BOthers {.}}{%
{\protect \APACyear {2017}}%
}]{%
breyer2017implementing}
\APACinsertmetastar {%
breyer2017implementing}%
\begin{APACrefauthors}%
Breyer, F\BPBI J.%
, Rupp, A\BPBI A.%
\BCBL {}\ \BBA {} Bridgeman, B.%
\end{APACrefauthors}%
\unskip\
\newblock
\APACrefYearMonthDay{2017}{}{}.
\newblock
{\BBOQ}\APACrefatitle {Implementing a contributory scoring approach for the {GRE}{\textregistered} Analytical Writing section: {A} comprehensive empirical investigation} {Implementing a contributory scoring approach for the {GRE}{\textregistered} analytical writing section: {A} comprehensive empirical investigation}.{\BBCQ}
\newblock
\APACjournalVolNumPages{ETS Research Report Series}{2017}{1}{1--28}.
\PrintBackRefs{\CurrentBib}

\bibitem [\protect \citeauthoryear {%
Bulut%
, Gorgun%
\BCBL {}\ \BBA {} Karamese%
}{%
Bulut%
, Gorgun%
\BCBL {}\ \BBA {} Karamese%
}{%
{\protect \APACyear {2023}}%
}]{%
bulut2023incorporating}
\APACinsertmetastar {%
bulut2023incorporating}%
\begin{APACrefauthors}%
Bulut, O.%
, Gorgun, G.%
\BCBL {}\ \BBA {} Karamese, H.%
\end{APACrefauthors}%
\unskip\
\newblock
\APACrefYearMonthDay{2023}{}{}.
\newblock
{\BBOQ}\APACrefatitle {Incorporating Test-Taking Engagement into Multistage Adaptive Testing Design for Large-Scale Assessments} {Incorporating test-taking engagement into multistage adaptive testing design for large-scale assessments}.{\BBCQ}
\newblock
\APACjournalVolNumPages{Journal of Educational Measurement}{}{}{}.
\newblock
\begin{APACrefDOI} \doi{10.1111/jedm.12380} \end{APACrefDOI}
\PrintBackRefs{\CurrentBib}

\bibitem [\protect \citeauthoryear {%
Bulut%
, Gorgun%
, Yildirim-Erbasli%
\BCBL {}\ \protect \BOthers {.}}{%
Bulut%
, Gorgun%
, Yildirim-Erbasli%
\BCBL {}\ \protect \BOthers {.}}{%
{\protect \APACyear {2023}}%
}]{%
bulutformative}
\APACinsertmetastar {%
bulutformative}%
\begin{APACrefauthors}%
Bulut, O.%
, Gorgun, G.%
, Yildirim-Erbasli, S\BPBI N.%
, Wongvorachan, T.%
, Daniels, L\BPBI M.%
, Gao, Y.%
\BDBL {}Shin, J.%
\end{APACrefauthors}%
\unskip\
\newblock
\APACrefYearMonthDay{2023}{}{}.
\newblock
{\BBOQ}\APACrefatitle {Standing on the shoulders of giants: Online formative assessments as the foundation for predictive learning analytics models} {Standing on the shoulders of giants: Online formative assessments as the foundation for predictive learning analytics models}.{\BBCQ}
\newblock
\APACjournalVolNumPages{British Journal of Educational Technology}{54}{1}{19-39}.
\newblock
\begin{APACrefDOI} \doi{https://doi.org/10.1111/bjet.13276} \end{APACrefDOI}
\PrintBackRefs{\CurrentBib}

\bibitem [\protect \citeauthoryear {%
Bulut%
, Wongvorachan%
, He%
\BCBL {}\ \BBA {} Lee%
}{%
Bulut%
, Wongvorachan%
\BCBL {}\ \protect \BOthers {.}}{%
{\protect \APACyear {2024}}%
}]{%
bulut_enhancing_2024}
\APACinsertmetastar {%
bulut_enhancing_2024}%
\begin{APACrefauthors}%
Bulut, O.%
, Wongvorachan, T.%
, He, S.%
\BCBL {}\ \BBA {} Lee, S.%
\end{APACrefauthors}%
\unskip\
\newblock
\APACrefYearMonthDay{2024}{}{}.
\newblock
\APACrefbtitle {Enhancing High-School Dropout Identification: {A} Collaborative Approach Integrating Human and Machine Insights.} {Enhancing high-school dropout identification: {A} collaborative approach integrating human and machine insights.}
\newblock
\begin{APACrefDOI} \doi{10.21203/rs.3.rs-3871667/v1} \end{APACrefDOI}
\PrintBackRefs{\CurrentBib}

\bibitem [\protect \citeauthoryear {%
Bulut%
\ \BBA {} Yildirim-Erbasli%
}{%
Bulut%
\ \BBA {} Yildirim-Erbasli%
}{%
{\protect \APACyear {2022}}%
}]{%
bulut_automatic_2022}
\APACinsertmetastar {%
bulut_automatic_2022}%
\begin{APACrefauthors}%
Bulut, O.%
\BCBT {}\ \BBA {} Yildirim-Erbasli, S\BPBI N.%
\end{APACrefauthors}%
\unskip\
\newblock
\APACrefYearMonthDay{2022}{}{}.
\newblock
{\BBOQ}\APACrefatitle {Automatic story and item generation for reading comprehension assessments with transformers} {Automatic story and item generation for reading comprehension assessments with transformers}.{\BBCQ}
\newblock
\APACjournalVolNumPages{International Journal of Assessment Tools in Education}{9}{Special Issue}{72--87}.
\newblock
\begin{APACrefDOI} \doi{10.21449/ijate.1124382} \end{APACrefDOI}
\PrintBackRefs{\CurrentBib}

\bibitem [\protect \citeauthoryear {%
Bulut%
, Yildirim-Erbasli%
\BCBL {}\ \BBA {} Gorgun%
}{%
Bulut%
, Yildirim-Erbasli%
\BCBL {}\ \BBA {} Gorgun%
}{%
{\protect \APACyear {2024}}%
}]{%
bulutLA2024}
\APACinsertmetastar {%
bulutLA2024}%
\begin{APACrefauthors}%
Bulut, O.%
, Yildirim-Erbasli, S\BPBI N.%
\BCBL {}\ \BBA {} Gorgun, G.%
\end{APACrefauthors}%
\unskip\
\newblock
\APACrefYearMonthDay{2024}{}{}.
\newblock
{\BBOQ}\APACrefatitle {Assessment Analytics for Digital Assessments Identifying, Modeling, and Interpreting Behavioral Engagement} {Assessment analytics for digital assessments identifying, modeling, and interpreting behavioral engagement}.{\BBCQ}
\newblock
\BIn{} M.~Sahin\ \BBA {} D.~Ifenthaler\ (\BEDS), \APACrefbtitle {Assessment Analytics in Education: Designs, Methods and Solutions} {Assessment analytics in education: Designs, methods and solutions}\ (\BPGS\ 35--60).
\newblock
\APACaddressPublisher{Cham}{Springer International Publishing}.
\newblock
\begin{APACrefDOI} \doi{10.1007/978-3-031-56365-2_3} \end{APACrefDOI}
\PrintBackRefs{\CurrentBib}

\bibitem [\protect \citeauthoryear {%
Buolamwini%
\ \BBA {} Gebru%
}{%
Buolamwini%
\ \BBA {} Gebru%
}{%
{\protect \APACyear {2018}}%
}]{%
buolamwini2018gender}
\APACinsertmetastar {%
buolamwini2018gender}%
\begin{APACrefauthors}%
Buolamwini, J.%
\BCBT {}\ \BBA {} Gebru, T.%
\end{APACrefauthors}%
\unskip\
\newblock
\APACrefYearMonthDay{2018}{}{}.
\newblock
{\BBOQ}\APACrefatitle {Gender shades: Intersectional accuracy disparities in commercial gender classification} {Gender shades: Intersectional accuracy disparities in commercial gender classification}.{\BBCQ}
\newblock
\BIn{} \APACrefbtitle {Conference on fairness, accountability and transparency} {Conference on fairness, accountability and transparency}\ (\BPGS\ 77--91).
\PrintBackRefs{\CurrentBib}

\bibitem [\protect \citeauthoryear {%
Burstein%
\ \protect \BOthers {.}}{%
Burstein%
\ \protect \BOthers {.}}{%
{\protect \APACyear {2018}}%
}]{%
burstein_writing_2018}
\APACinsertmetastar {%
burstein_writing_2018}%
\begin{APACrefauthors}%
Burstein, J.%
, Elliot, N.%
, Klebanov, B\BPBI B.%
, Madnani, N.%
, Napolitano, D.%
, Schwartz, M.%
\BDBL {}Molloy, H.%
\end{APACrefauthors}%
\unskip\
\newblock
\APACrefYearMonthDay{2018}{}{}.
\newblock
{\BBOQ}\APACrefatitle {Writing {MentorTM}: {Writing} Progress Using Self-Regulated Writing Support} {Writing {MentorTM}: {Writing} progress using self-regulated writing support}.{\BBCQ}
\newblock
\APACjournalVolNumPages{The Journal of Writing Analytics}{2}{1}{285--313}.
\newblock
\begin{APACrefDOI} \doi{10.37514/JWA-J.2018.2.1.12} \end{APACrefDOI}
\PrintBackRefs{\CurrentBib}

\bibitem [\protect \citeauthoryear {%
Burstein%
\ \protect \BOthers {.}}{%
Burstein%
\ \protect \BOthers {.}}{%
{\protect \APACyear {1998}}%
}]{%
burstein_automated_1998}
\APACinsertmetastar {%
burstein_automated_1998}%
\begin{APACrefauthors}%
Burstein, J.%
, Kukich, K.%
, Wolff, S.%
, Lu, C.%
, Chodorow, M.%
, Braden-Harder, L.%
\BCBL {}\ \BBA {} Harris, M\BPBI D.%
\end{APACrefauthors}%
\unskip\
\newblock
\APACrefYearMonthDay{1998}{}{}.
\newblock
{\BBOQ}\APACrefatitle {Automated scoring using a hybrid feature identification technique} {Automated scoring using a hybrid feature identification technique}.{\BBCQ}
\newblock
\BIn{} \APACrefbtitle {Proceedings of the 36th {Annual} {Meeting} of the {Association} for {Computational} {Linguistics} and 17th {International} {Conference} on {Computational} {Linguistics} - {Volume} 1} {Proceedings of the 36th {Annual} {Meeting} of the {Association} for {Computational} {Linguistics} and 17th {International} {Conference} on {Computational} {Linguistics} - {Volume} 1}\ (\BPGS\ 206--210).
\newblock
\APACaddressPublisher{USA}{Association for Computational Linguistics}.
\newblock
\begin{APACrefDOI} \doi{10.3115/980845.980879} \end{APACrefDOI}
\PrintBackRefs{\CurrentBib}

\bibitem [\protect \citeauthoryear {%
Burstein%
\ \protect \BOthers {.}}{%
Burstein%
\ \protect \BOthers {.}}{%
{\protect \APACyear {2023}}%
}]{%
duolingo}
\APACinsertmetastar {%
duolingo}%
\begin{APACrefauthors}%
Burstein, J.%
, von Davier, A.%
, Yancey, K.%
, Belzak, W.%
, Bicknell, K.%
, Gottlieb, C.%
\BDBL {}Zheng, M.%
\end{APACrefauthors}%
\unskip\
\newblock
\APACrefYearMonthDay{2023}{}{}.
\newblock
\APACrefbtitle {{DET} Responsible {AI} Standards 2024.} {{DET} responsible {AI} standards 2024.}
\newblock
\begin{APACrefURL} \url{https://duolingo-papers.s3.amazonaws.com/other/DET%2BResponsible%2BAI%2BStandards%2B-%2B040824.pdf} \end{APACrefURL}
\PrintBackRefs{\CurrentBib}

\bibitem [\protect \citeauthoryear {%
Buzick%
, Oliveri%
, Attali%
\BCBL {}\ \BBA {} Flor%
}{%
Buzick%
\ \protect \BOthers {.}}{%
{\protect \APACyear {2016}}%
}]{%
buzick_comparing_2016}
\APACinsertmetastar {%
buzick_comparing_2016}%
\begin{APACrefauthors}%
Buzick, H.%
, Oliveri, M\BPBI E.%
, Attali, Y.%
\BCBL {}\ \BBA {} Flor, M.%
\end{APACrefauthors}%
\unskip\
\newblock
\APACrefYearMonthDay{2016}{}{}.
\newblock
{\BBOQ}\APACrefatitle {Comparing Human and Automated Essay Scoring for Prospective Graduate Students With Learning Disabilities and/or {ADHD}} {Comparing human and automated essay scoring for prospective graduate students with learning disabilities and/or {ADHD}}.{\BBCQ}
\newblock
\APACjournalVolNumPages{Applied Measurement in Education}{29}{3}{161--172}.
\newblock
\begin{APACrefDOI} \doi{10.1080/08957347.2016.1171765} \end{APACrefDOI}
\PrintBackRefs{\CurrentBib}

\bibitem [\protect \citeauthoryear {%
Cairns%
\ \protect \BOthers {.}}{%
Cairns%
\ \protect \BOthers {.}}{%
{\protect \APACyear {2016}}%
}]{%
cairns_computer-human_2016}
\APACinsertmetastar {%
cairns_computer-human_2016}%
\begin{APACrefauthors}%
Cairns, A\BPBI W.%
, Bond, R\BPBI R.%
, Finlay, D\BPBI D.%
, Breen, C.%
, Guldenring, D.%
, Gaffney, R.%
\BDBL {}Henn, P.%
\end{APACrefauthors}%
\unskip\
\newblock
\APACrefYearMonthDay{2016}{}{}.
\newblock
{\BBOQ}\APACrefatitle {A computer-human interaction model to improve the diagnostic accuracy and clinical decision-making during 12-lead electrocardiogram interpretation} {A computer-human interaction model to improve the diagnostic accuracy and clinical decision-making during 12-lead electrocardiogram interpretation}.{\BBCQ}
\newblock
\APACjournalVolNumPages{Journal of Biomedical Informatics}{64}{}{93--107}.
\newblock
\begin{APACrefDOI} \doi{10.1016/j.jbi.2016.09.016} \end{APACrefDOI}
\PrintBackRefs{\CurrentBib}

\bibitem [\protect \citeauthoryear {%
Carless%
}{%
Carless%
}{%
{\protect \APACyear {2019}}%
}]{%
carless_feedback_2019}
\APACinsertmetastar {%
carless_feedback_2019}%
\begin{APACrefauthors}%
Carless, D.%
\end{APACrefauthors}%
\unskip\
\newblock
\APACrefYearMonthDay{2019}{}{}.
\newblock
{\BBOQ}\APACrefatitle {Feedback loops and the longer-term: Towards feedback spirals} {Feedback loops and the longer-term: Towards feedback spirals}.{\BBCQ}
\newblock
\APACjournalVolNumPages{Assessment \& Evaluation in Higher Education}{44}{5}{705--714}.
\newblock
\begin{APACrefDOI} \doi{10.1080/02602938.2018.1531108} \end{APACrefDOI}
\PrintBackRefs{\CurrentBib}

\bibitem [\protect \citeauthoryear {%
Casuat%
\ \BBA {} Festijo%
}{%
Casuat%
\ \BBA {} Festijo%
}{%
{\protect \APACyear {2019}}%
}]{%
casuat_predicting_2019}
\APACinsertmetastar {%
casuat_predicting_2019}%
\begin{APACrefauthors}%
Casuat, C\BPBI D.%
\BCBT {}\ \BBA {} Festijo, E\BPBI D.%
\end{APACrefauthors}%
\unskip\
\newblock
\APACrefYearMonthDay{2019}{{\APACmonth{12}}}{}.
\newblock
{\BBOQ}\APACrefatitle {Predicting Students' Employability using Machine Learning Approach} {Predicting students' employability using machine learning approach}.{\BBCQ}
\newblock
\BIn{} \APACrefbtitle {2019 {IEEE} 6th {International} {Conference} on {Engineering} {Technologies} and {Applied} {Sciences} ({ICETAS})} {2019 {IEEE} 6th {International} {Conference} on {Engineering} {Technologies} and {Applied} {Sciences} ({ICETAS})}\ (\BPGS\ 1--5).
\newblock
\begin{APACrefDOI} \doi{10.1109/ICETAS48360.2019.9117338} \end{APACrefDOI}
\PrintBackRefs{\CurrentBib}

\bibitem [\protect \citeauthoryear {%
Chawla%
, Bowyer%
, Hall%
\BCBL {}\ \BBA {} Kegelmeyer%
}{%
Chawla%
\ \protect \BOthers {.}}{%
{\protect \APACyear {2002}}%
}]{%
chawla_smote_2002}
\APACinsertmetastar {%
chawla_smote_2002}%
\begin{APACrefauthors}%
Chawla, N\BPBI V.%
, Bowyer, K\BPBI W.%
, Hall, L\BPBI O.%
\BCBL {}\ \BBA {} Kegelmeyer, W\BPBI P.%
\end{APACrefauthors}%
\unskip\
\newblock
\APACrefYearMonthDay{2002}{}{}.
\newblock
{\BBOQ}\APACrefatitle {{SMOTE}: Synthetic Minority Over-sampling Technique} {{SMOTE}: Synthetic minority over-sampling technique}.{\BBCQ}
\newblock
\APACjournalVolNumPages{Journal of Artificial Intelligence Research}{16}{}{321--357}.
\newblock
\begin{APACrefDOI} \doi{10.1613/jair.953} \end{APACrefDOI}
\PrintBackRefs{\CurrentBib}

\bibitem [\protect \citeauthoryear {%
C\BHBI F\BPBI E.~Chen%
\ \BBA {} Cheng%
}{%
C\BHBI F\BPBI E.~Chen%
\ \BBA {} Cheng%
}{%
{\protect \APACyear {2008}}%
}]{%
chen_beyond_2008}
\APACinsertmetastar {%
chen_beyond_2008}%
\begin{APACrefauthors}%
Chen, C\BHBI F\BPBI E.%
\BCBT {}\ \BBA {} Cheng, W\BHBI Y\BPBI E.%
\end{APACrefauthors}%
\unskip\
\newblock
\APACrefYearMonthDay{2008}{}{}.
\newblock
{\BBOQ}\APACrefatitle {Beyond the Design of Automated Writing Evaluation: {P}edagogical Practices and Perceived Learning Effectiveness in {EFL} Writing Classes} {Beyond the design of automated writing evaluation: {P}edagogical practices and perceived learning effectiveness in {EFL} writing classes}.{\BBCQ}
\newblock
\APACjournalVolNumPages{Language Learning \& Technology}{12}{2}{94--112}.
\newblock
\begin{APACrefDOI} \doi{10125/44145} \end{APACrefDOI}
\PrintBackRefs{\CurrentBib}

\bibitem [\protect \citeauthoryear {%
G.~Chen%
, Rolim%
, Mello%
\BCBL {}\ \BBA {} Gašević%
}{%
G.~Chen%
\ \protect \BOthers {.}}{%
{\protect \APACyear {2020}}%
}]{%
chen_lets_2020}
\APACinsertmetastar {%
chen_lets_2020}%
\begin{APACrefauthors}%
Chen, G.%
, Rolim, V.%
, Mello, R\BPBI F.%
\BCBL {}\ \BBA {} Gašević, D.%
\end{APACrefauthors}%
\unskip\
\newblock
\APACrefYearMonthDay{2020}{March}{}.
\newblock
{\BBOQ}\APACrefatitle {Let's shine together! A comparative study between learning analytics and educational data mining} {Let's shine together! a comparative study between learning analytics and educational data mining}.{\BBCQ}
\newblock
\BIn{} \APACrefbtitle {Proceedings of the {Tenth} {International} {Conference} on {Learning} {Analytics} \& {Knowledge}} {Proceedings of the {Tenth} {International} {Conference} on {Learning} {Analytics} \& {Knowledge}}\ (\BPGS\ 544--553).
\newblock
\APACaddressPublisher{New York, NY, USA}{Association for Computing Machinery}.
\newblock
\begin{APACrefDOI} \doi{10.1145/3375462.3375500} \end{APACrefDOI}
\PrintBackRefs{\CurrentBib}

\bibitem [\protect \citeauthoryear {%
Choi%
\ \BBA {} Johnson%
}{%
Choi%
\ \BBA {} Johnson%
}{%
{\protect \APACyear {2024}}%
}]{%
choi_johnson_2024}
\APACinsertmetastar {%
choi_johnson_2024}%
\begin{APACrefauthors}%
Choi, I.%
\BCBT {}\ \BBA {} Johnson, M\BPBI S.%
\end{APACrefauthors}%
\unskip\
\newblock
\APACrefYearMonthDay{2024}{April}{}.
\newblock
{\BBOQ}\APACrefatitle {Examining Partial Derivatives to Identify Causes of Differential Prediction Bias in Automated Scores} {Examining partial derivatives to identify causes of differential prediction bias in automated scores}.{\BBCQ}
\newblock
\APACjournalVolNumPages{Paper presented at the annual meeting of the National Council for Measurement in Education, Philadelphia, PA}{}{}{}.
\PrintBackRefs{\CurrentBib}

\bibitem [\protect \citeauthoryear {%
Cizek%
\ \BBA {} Wollack%
}{%
Cizek%
\ \BBA {} Wollack%
}{%
{\protect \APACyear {2016}}%
}]{%
cizek_handbook_2016}
\APACinsertmetastar {%
cizek_handbook_2016}%
\begin{APACrefauthors}%
Cizek, G\BPBI J.%
\BCBT {}\ \BBA {} Wollack, J\BPBI A.%
\end{APACrefauthors}%
\unskip\
\newblock
\APACrefYear{2016}.
\newblock
\APACrefbtitle {Handbook of Quantitative Methods for Detecting Cheating on Tests} {Handbook of quantitative methods for detecting cheating on tests}.
\newblock
\APACaddressPublisher{}{Taylor \& Francis}.
\PrintBackRefs{\CurrentBib}

\bibitem [\protect \citeauthoryear {%
Coghlan%
, Miller%
\BCBL {}\ \BBA {} Paterson%
}{%
Coghlan%
\ \protect \BOthers {.}}{%
{\protect \APACyear {2021}}%
}]{%
coghlan_good_2021}
\APACinsertmetastar {%
coghlan_good_2021}%
\begin{APACrefauthors}%
Coghlan, S.%
, Miller, T.%
\BCBL {}\ \BBA {} Paterson, J.%
\end{APACrefauthors}%
\unskip\
\newblock
\APACrefYearMonthDay{2021}{}{}.
\newblock
{\BBOQ}\APACrefatitle {Good Proctor or “Big Brother”? {E}thics of Online Exam Supervision Technologies} {Good proctor or “big brother”? {E}thics of online exam supervision technologies}.{\BBCQ}
\newblock
\APACjournalVolNumPages{Philosophy \& Technology}{34}{4}{1581--1606}.
\newblock
\begin{APACrefDOI} \doi{10.1007/s13347-021-00476-1} \end{APACrefDOI}
\PrintBackRefs{\CurrentBib}

\bibitem [\protect \citeauthoryear {%
Correnti%
, Matsumura%
, Wang%
, Litman%
\BCBL {}\ \BBA {} Zhang%
}{%
Correnti%
\ \protect \BOthers {.}}{%
{\protect \APACyear {2022}}%
}]{%
CORRENTI2022100084}
\APACinsertmetastar {%
CORRENTI2022100084}%
\begin{APACrefauthors}%
Correnti, R.%
, Matsumura, L\BPBI C.%
, Wang, E\BPBI L.%
, Litman, D.%
\BCBL {}\ \BBA {} Zhang, H.%
\end{APACrefauthors}%
\unskip\
\newblock
\APACrefYearMonthDay{2022}{}{}.
\newblock
{\BBOQ}\APACrefatitle {Building a validity argument for an automated writing evaluation system (eRevise) as a formative assessment} {Building a validity argument for an automated writing evaluation system (erevise) as a formative assessment}.{\BBCQ}
\newblock
\APACjournalVolNumPages{Computers and Education Open}{3}{}{100084}.
\newblock
\begin{APACrefDOI} \doi{10.1016/j.caeo.2022.100084} \end{APACrefDOI}
\PrintBackRefs{\CurrentBib}

\bibitem [\protect \citeauthoryear {%
Correnti%
\ \protect \BOthers {.}}{%
Correnti%
\ \protect \BOthers {.}}{%
{\protect \APACyear {2024}}%
}]{%
correnti2024supporting}
\APACinsertmetastar {%
correnti2024supporting}%
\begin{APACrefauthors}%
Correnti, R.%
, Wang, E\BPBI L.%
, Matsumura, L\BPBI C.%
, Litman, D.%
, Liu, Z.%
\BCBL {}\ \BBA {} Li, T.%
\end{APACrefauthors}%
\unskip\
\newblock
\APACrefYearMonthDay{2024}{}{}.
\newblock
{\BBOQ}\APACrefatitle {Supporting Students' Text-Based Evidence Use via Formative Automated Writing and Revision Assessment} {Supporting students' text-based evidence use via formative automated writing and revision assessment}.{\BBCQ}
\newblock
\BIn{} \APACrefbtitle {The {R}outledge {I}nternational {H}andbook of {A}utomated {E}ssay {E}valuation} {The {R}outledge {I}nternational {H}andbook of {A}utomated {E}ssay {E}valuation}\ (\BPGS\ 221--243).
\newblock
\APACaddressPublisher{}{Routledge}.
\PrintBackRefs{\CurrentBib}

\bibitem [\protect \citeauthoryear {%
Cotos%
}{%
Cotos%
}{%
{\protect \APACyear {2023}}%
}]{%
cotos_automated_2023}
\APACinsertmetastar {%
cotos_automated_2023}%
\begin{APACrefauthors}%
Cotos, E.%
\end{APACrefauthors}%
\unskip\
\newblock
\APACrefYearMonthDay{2023}{}{}.
\newblock
{\BBOQ}\APACrefatitle {Automated Feedback on Writing} {Automated feedback on writing}.{\BBCQ}
\newblock
\BIn{} O.~Kruse\ \BOthers {.}\ (\BEDS), \APACrefbtitle {Digital Writing Technologies in Higher Education: Theory, Research, and Practice} {Digital writing technologies in higher education: Theory, research, and practice}\ (\BPGS\ 347--364).
\newblock
\APACaddressPublisher{Cham}{Springer International Publishing}.
\newblock
\begin{APACrefDOI} \doi{10.1007/978-3-031-36033-6_22} \end{APACrefDOI}
\PrintBackRefs{\CurrentBib}

\bibitem [\protect \citeauthoryear {%
Cullen%
}{%
Cullen%
}{%
{\protect \APACyear {2001}}%
}]{%
cullen2001addressing}
\APACinsertmetastar {%
cullen2001addressing}%
\begin{APACrefauthors}%
Cullen, R.%
\end{APACrefauthors}%
\unskip\
\newblock
\APACrefYearMonthDay{2001}{}{}.
\newblock
{\BBOQ}\APACrefatitle {Addressing the digital divide} {Addressing the digital divide}.{\BBCQ}
\newblock
\APACjournalVolNumPages{Online information review}{25}{5}{311--320}.
\PrintBackRefs{\CurrentBib}

\bibitem [\protect \citeauthoryear {%
Dawson%
}{%
Dawson%
}{%
{\protect \APACyear {2024}}%
}]{%
dawson_remote_2024}
\APACinsertmetastar {%
dawson_remote_2024}%
\begin{APACrefauthors}%
Dawson, P.%
\end{APACrefauthors}%
\unskip\
\newblock
\APACrefYearMonthDay{2024}{}{}.
\newblock
{\BBOQ}\APACrefatitle {Remote Proctoring: {U}nderstanding the Debate} {Remote proctoring: {U}nderstanding the debate}.{\BBCQ}
\newblock
\BIn{} S\BPBI E.~Eaton\ (\BED), \APACrefbtitle {Second Handbook of Academic Integrity} {Second handbook of academic integrity}\ (\BPGS\ 1511--1526).
\newblock
\APACaddressPublisher{Cham}{Springer Nature Switzerland}.
\newblock
\begin{APACrefDOI} \doi{10.1007/978-3-031-54144-5\_150} \end{APACrefDOI}
\PrintBackRefs{\CurrentBib}

\bibitem [\protect \citeauthoryear {%
Deeva%
, Bogdanova%
, Serral%
, Snoeck%
\BCBL {}\ \BBA {} De~Weerdt%
}{%
Deeva%
\ \protect \BOthers {.}}{%
{\protect \APACyear {2021}}%
}]{%
deeva_review_2021}
\APACinsertmetastar {%
deeva_review_2021}%
\begin{APACrefauthors}%
Deeva, G.%
, Bogdanova, D.%
, Serral, E.%
, Snoeck, M.%
\BCBL {}\ \BBA {} De~Weerdt, J.%
\end{APACrefauthors}%
\unskip\
\newblock
\APACrefYearMonthDay{2021}{}{}.
\newblock
{\BBOQ}\APACrefatitle {A review of automated feedback systems for learners: {Classification} framework, challenges and opportunities} {A review of automated feedback systems for learners: {Classification} framework, challenges and opportunities}.{\BBCQ}
\newblock
\APACjournalVolNumPages{Computers \& Education}{162}{}{104094}.
\newblock
\begin{APACrefDOI} \doi{10.1016/j.compedu.2020.104094} \end{APACrefDOI}
\PrintBackRefs{\CurrentBib}

\bibitem [\protect \citeauthoryear {%
Dettmers%
, Pagnoni%
, Holtzman%
\BCBL {}\ \BBA {} Zettlemoyer%
}{%
Dettmers%
\ \protect \BOthers {.}}{%
{\protect \APACyear {2024}}%
}]{%
dettmers2024qlora}
\APACinsertmetastar {%
dettmers2024qlora}%
\begin{APACrefauthors}%
Dettmers, T.%
, Pagnoni, A.%
, Holtzman, A.%
\BCBL {}\ \BBA {} Zettlemoyer, L.%
\end{APACrefauthors}%
\unskip\
\newblock
\APACrefYearMonthDay{2024}{}{}.
\newblock
{\BBOQ}\APACrefatitle {Qlora: Efficient finetuning of quantized {LLMs}} {Qlora: Efficient finetuning of quantized {LLMs}}.{\BBCQ}
\newblock
\APACjournalVolNumPages{Advances in Neural Information Processing Systems}{36}{}{}.
\PrintBackRefs{\CurrentBib}

\bibitem [\protect \citeauthoryear {%
Devlin%
, Chang%
, Lee%
\BCBL {}\ \BBA {} Toutanova%
}{%
Devlin%
\ \protect \BOthers {.}}{%
{\protect \APACyear {2019}}%
}]{%
devlin_bert_2019}
\APACinsertmetastar {%
devlin_bert_2019}%
\begin{APACrefauthors}%
Devlin, J.%
, Chang, M\BHBI W.%
, Lee, K.%
\BCBL {}\ \BBA {} Toutanova, K.%
\end{APACrefauthors}%
\unskip\
\newblock
\APACrefYearMonthDay{2019}{}{}.
\newblock
\APACrefbtitle {{BERT}: {Pre}-training of Deep Bidirectional Transformers for Language Understanding.} {{BERT}: {Pre}-training of deep bidirectional transformers for language understanding.}
\newblock
\APACaddressPublisher{}{arXiv}.
\newblock
\APACrefnote{arXiv:1810.04805 [cs]}
\newblock
\begin{APACrefDOI} \doi{10.48550/arXiv.1810.04805} \end{APACrefDOI}
\PrintBackRefs{\CurrentBib}

\bibitem [\protect \citeauthoryear {%
Drori%
\ \protect \BOthers {.}}{%
Drori%
\ \protect \BOthers {.}}{%
{\protect \APACyear {2022}}%
}]{%
drori_neural_2022}
\APACinsertmetastar {%
drori_neural_2022}%
\begin{APACrefauthors}%
Drori, I.%
, Zhang, S.%
, Shuttleworth, R.%
, Tang, L.%
, Lu, A.%
, Ke, E.%
\BDBL {}Strang, G.%
\end{APACrefauthors}%
\unskip\
\newblock
\APACrefYearMonthDay{2022}{}{}.
\newblock
{\BBOQ}\APACrefatitle {A neural network solves, explains, and generates university math problems by program synthesis and few-shot learning at human level} {A neural network solves, explains, and generates university math problems by program synthesis and few-shot learning at human level}.{\BBCQ}
\newblock
\APACjournalVolNumPages{Proceedings of the National Academy of Sciences}{119}{32}{e2123433119}.
\newblock
\begin{APACrefDOI} \doi{10.1073/pnas.2123433119} \end{APACrefDOI}
\PrintBackRefs{\CurrentBib}

\bibitem [\protect \citeauthoryear {%
Dyer%
}{%
Dyer%
}{%
{\protect \APACyear {2024}}%
}]{%
dyer_framework_2024}
\APACinsertmetastar {%
dyer_framework_2024}%
\begin{APACrefauthors}%
Dyer, J.%
\end{APACrefauthors}%
\unskip\
\newblock
\APACrefYearMonthDay{2024}{}{}.
\newblock
{\BBOQ}\APACrefatitle {Framework for Ethical Implementation of Remote Proctoring in Education} {Framework for ethical implementation of remote proctoring in education}.{\BBCQ}
\newblock
\BIn{} S\BPBI E.~Eaton\ (\BED), \APACrefbtitle {Second Handbook of Academic Integrity} {Second handbook of academic integrity}\ (\BPGS\ 1527--1550).
\newblock
\APACaddressPublisher{Cham}{Springer Nature Switzerland}.
\newblock
\begin{APACrefDOI} \doi{10.1007/978-3-031-54144-5\_151} \end{APACrefDOI}
\PrintBackRefs{\CurrentBib}

\bibitem [\protect \citeauthoryear {%
{European Commission}%
}{%
{European Commission}%
}{%
{\protect \APACyear {2019}}%
}]{%
eu2019ethicsai}
\APACinsertmetastar {%
eu2019ethicsai}%
\begin{APACrefauthors}%
{European Commission}.%
\end{APACrefauthors}%
\unskip\
\newblock
\APACrefYearMonthDay{2019}{}{}.
\newblock
\APACrefbtitle {The {E}uropean {C}ommission's high-level expert group on intelligence: {E}thics Guidelines for Trustworthy {AI}} {The {E}uropean {C}ommission's high-level expert group on intelligence: {E}thics guidelines for trustworthy {AI}}\ \APACbVolEdTR{}{\BTR{}}.
\newblock
\APACaddressInstitution{}{Publications Office of the European Union}.
\newblock
\begin{APACrefURL} \url{https://digital-strategy.ec.europa.eu/en/policies/expert-group-ai} \end{APACrefURL}
\PrintBackRefs{\CurrentBib}

\bibitem [\protect \citeauthoryear {%
European Commission{,} Directorate-General~for Education{,}~Youth%
\ \BBA {} Culture%
}{%
European Commission{,} Directorate-General~for Education{,}~Youth%
\ \BBA {} Culture%
}{%
{\protect \APACyear {2022}}%
}]{%
eu2022ethicsai}
\APACinsertmetastar {%
eu2022ethicsai}%
\begin{APACrefauthors}%
European Commission{,} Directorate-General~for Education{,}~Youth, S.%
\BCBT {}\ \BBA {} Culture.%
\end{APACrefauthors}%
\unskip\
\newblock
\APACrefYearMonthDay{2022}{}{}.
\newblock
\APACrefbtitle {Ethical guidelines on the use of artificial intelligence ({AI}) and data in teaching and learning for educators} {Ethical guidelines on the use of artificial intelligence ({AI}) and data in teaching and learning for educators}\ \APACbVolEdTR{}{\BTR{}}.
\newblock
\APACaddressInstitution{}{Publications Office of the European Union}.
\newblock
\begin{APACrefURL} \url{https://data.europa.eu/doi/10.2766/153756} \end{APACrefURL}
\PrintBackRefs{\CurrentBib}

\bibitem [\protect \citeauthoryear {%
Fleckenstein%
, Liebenow%
\BCBL {}\ \BBA {} Meyer%
}{%
Fleckenstein%
\ \protect \BOthers {.}}{%
{\protect \APACyear {2023}}%
}]{%
fleckenstein2023automated}
\APACinsertmetastar {%
fleckenstein2023automated}%
\begin{APACrefauthors}%
Fleckenstein, J.%
, Liebenow, L\BPBI W.%
\BCBL {}\ \BBA {} Meyer, J.%
\end{APACrefauthors}%
\unskip\
\newblock
\APACrefYearMonthDay{2023}{}{}.
\newblock
{\BBOQ}\APACrefatitle {Automated feedback and writing: A multi-level meta-analysis of effects on students' performance} {Automated feedback and writing: A multi-level meta-analysis of effects on students' performance}.{\BBCQ}
\newblock
\APACjournalVolNumPages{Frontiers in Artificial Intelligence}{6}{}{1162454}.
\PrintBackRefs{\CurrentBib}

\bibitem [\protect \citeauthoryear {%
Flor%
\ \BBA {} Hao%
}{%
Flor%
\ \BBA {} Hao%
}{%
{\protect \APACyear {2021}}%
}]{%
flor_text_2021}
\APACinsertmetastar {%
flor_text_2021}%
\begin{APACrefauthors}%
Flor, M.%
\BCBT {}\ \BBA {} Hao, J.%
\end{APACrefauthors}%
\unskip\
\newblock
\APACrefYearMonthDay{2021}{}{}.
\newblock
{\BBOQ}\APACrefatitle {Text Mining and Automated Scoring} {Text mining and automated scoring}.{\BBCQ}
\newblock
\BIn{} A\BPBI A.~von Davier, R\BPBI J.~Mislevy\BCBL {}\ \BBA {} J.~Hao\ (\BEDS), \APACrefbtitle {Computational Psychometrics: New Methodologies for a New Generation of Digital Learning and Assessment: With Examples in {R} and {Python}} {Computational psychometrics: New methodologies for a new generation of digital learning and assessment: With examples in {R} and {Python}}\ (\BPGS\ 245--262).
\newblock
\APACaddressPublisher{Cham}{Springer International Publishing}.
\newblock
\begin{APACrefDOI} \doi{10.1007/978-3-030-74394-9\_14} \end{APACrefDOI}
\PrintBackRefs{\CurrentBib}

\bibitem [\protect \citeauthoryear {%
Fritts%
\ \BBA {} Cabrera%
}{%
Fritts%
\ \BBA {} Cabrera%
}{%
{\protect \APACyear {2021}}%
}]{%
fritts_ai_2021}
\APACinsertmetastar {%
fritts_ai_2021}%
\begin{APACrefauthors}%
Fritts, M.%
\BCBT {}\ \BBA {} Cabrera, F.%
\end{APACrefauthors}%
\unskip\
\newblock
\APACrefYearMonthDay{2021}{}{}.
\newblock
{\BBOQ}\APACrefatitle {{AI} recruitment algorithms and the dehumanization problem} {{AI} recruitment algorithms and the dehumanization problem}.{\BBCQ}
\newblock
\APACjournalVolNumPages{Ethics and Information Technology}{23}{4}{791--801}.
\newblock
\begin{APACrefDOI} \doi{10.1007/s10676-021-09615-w} \end{APACrefDOI}
\PrintBackRefs{\CurrentBib}

\bibitem [\protect \citeauthoryear {%
Fu%
, Zou%
, Xie%
\BCBL {}\ \BBA {} Cheng%
}{%
Fu%
\ \protect \BOthers {.}}{%
{\protect \APACyear {2024}}%
}]{%
fu_review_2024}
\APACinsertmetastar {%
fu_review_2024}%
\begin{APACrefauthors}%
Fu, Q\BHBI K.%
, Zou, D.%
, Xie, H.%
\BCBL {}\ \BBA {} Cheng, G.%
\end{APACrefauthors}%
\unskip\
\newblock
\APACrefYearMonthDay{2024}{}{}.
\newblock
{\BBOQ}\APACrefatitle {A review of {AWE} feedback: types, learning outcomes, and implications} {A review of {AWE} feedback: types, learning outcomes, and implications}.{\BBCQ}
\newblock
\APACjournalVolNumPages{Computer Assisted Language Learning}{37}{1-2}{179--221}.
\newblock
\begin{APACrefDOI} \doi{10.1080/09588221.2022.2033787} \end{APACrefDOI}
\PrintBackRefs{\CurrentBib}

\bibitem [\protect \citeauthoryear {%
Gallegos%
\ \protect \BOthers {.}}{%
Gallegos%
\ \protect \BOthers {.}}{%
{\protect \APACyear {2024}}%
}]{%
gallegos_bias_2024}
\APACinsertmetastar {%
gallegos_bias_2024}%
\begin{APACrefauthors}%
Gallegos, I\BPBI O.%
, Rossi, R\BPBI A.%
, Barrow, J.%
, Tanjim, M\BPBI M.%
, Kim, S.%
, Dernoncourt, F.%
\BDBL {}Ahmed, N\BPBI K.%
\end{APACrefauthors}%
\unskip\
\newblock
\APACrefYearMonthDay{2024}{}{}.
\newblock
\APACrefbtitle {Bias and Fairness in Large Language Models: {A} Survey.} {Bias and fairness in large language models: {A} survey.}
\newblock
\APACaddressPublisher{}{arXiv}.
\newblock
\begin{APACrefDOI} \doi{10.48550/arXiv.2309.00770} \end{APACrefDOI}
\PrintBackRefs{\CurrentBib}

\bibitem [\protect \citeauthoryear {%
Gierl%
\ \BBA {} Haladyna%
}{%
Gierl%
\ \BBA {} Haladyna%
}{%
{\protect \APACyear {2012}}%
}]{%
gierl_automatic_2012}
\APACinsertmetastar {%
gierl_automatic_2012}%
\begin{APACrefauthors}%
Gierl, M\BPBI J.%
\BCBT {}\ \BBA {} Haladyna, T\BPBI M.%
\end{APACrefauthors}%
\unskip\
\newblock
\APACrefYearMonthDay{2012}{}{}.
\newblock
{\BBOQ}\APACrefatitle {Automatic Item Generation: {An} Introduction} {Automatic item generation: {An} introduction}.{\BBCQ}
\newblock
\BIn{} \APACrefbtitle {Automatic Item Generation.} {Automatic item generation.}
\newblock
\APACaddressPublisher{}{Routledge}.
\PrintBackRefs{\CurrentBib}

\bibitem [\protect \citeauthoryear {%
Gierl%
\ \BBA {} Lai%
}{%
Gierl%
\ \BBA {} Lai%
}{%
{\protect \APACyear {2012}}%
}]{%
gierl_role_2012}
\APACinsertmetastar {%
gierl_role_2012}%
\begin{APACrefauthors}%
Gierl, M\BPBI J.%
\BCBT {}\ \BBA {} Lai, H.%
\end{APACrefauthors}%
\unskip\
\newblock
\APACrefYearMonthDay{2012}{}{}.
\newblock
{\BBOQ}\APACrefatitle {The Role of Item Models in Automatic Item Generation} {The role of item models in automatic item generation}.{\BBCQ}
\newblock
\APACjournalVolNumPages{International Journal of Testing}{12}{3}{273--298}.
\newblock
\begin{APACrefDOI} \doi{10.1080/15305058.2011.635830} \end{APACrefDOI}
\PrintBackRefs{\CurrentBib}

\bibitem [\protect \citeauthoryear {%
Gierl%
, Shin%
, Firoozi%
\BCBL {}\ \BBA {} Lai%
}{%
Gierl%
\ \protect \BOthers {.}}{%
{\protect \APACyear {2022}}%
}]{%
gierl2022using}
\APACinsertmetastar {%
gierl2022using}%
\begin{APACrefauthors}%
Gierl, M\BPBI J.%
, Shin, J.%
, Firoozi, T.%
\BCBL {}\ \BBA {} Lai, H.%
\end{APACrefauthors}%
\unskip\
\newblock
\APACrefYearMonthDay{2022}{}{}.
\newblock
{\BBOQ}\APACrefatitle {Using content coding and automatic item generation to improve test security} {Using content coding and automatic item generation to improve test security}.{\BBCQ}
\newblock
\APACjournalVolNumPages{Frontiers in Education}{7}{}{853578}.
\newblock
\begin{APACrefDOI} \doi{10.3389/feduc.2022.853578} \end{APACrefDOI}
\PrintBackRefs{\CurrentBib}

\bibitem [\protect \citeauthoryear {%
Gimpel%
, Kleindienst%
, Nüske%
, Rau%
\BCBL {}\ \BBA {} Schmied%
}{%
Gimpel%
\ \protect \BOthers {.}}{%
{\protect \APACyear {2018}}%
}]{%
gimpel_upside_2018}
\APACinsertmetastar {%
gimpel_upside_2018}%
\begin{APACrefauthors}%
Gimpel, H.%
, Kleindienst, D.%
, Nüske, N.%
, Rau, D.%
\BCBL {}\ \BBA {} Schmied, F.%
\end{APACrefauthors}%
\unskip\
\newblock
\APACrefYearMonthDay{2018}{}{}.
\newblock
{\BBOQ}\APACrefatitle {The upside of data privacy – delighting customers by implementing data privacy measures} {The upside of data privacy – delighting customers by implementing data privacy measures}.{\BBCQ}
\newblock
\APACjournalVolNumPages{Electronic Markets}{28}{4}{437--452}.
\newblock
\begin{APACrefDOI} \doi{10.1007/s12525-018-0296-3} \end{APACrefDOI}
\PrintBackRefs{\CurrentBib}

\bibitem [\protect \citeauthoryear {%
Goldshtein%
, Alhashim%
\BCBL {}\ \BBA {} Roscoe%
}{%
Goldshtein%
\ \protect \BOthers {.}}{%
{\protect \APACyear {2024}}%
}]{%
goldshtein2024automating}
\APACinsertmetastar {%
goldshtein2024automating}%
\begin{APACrefauthors}%
Goldshtein, M.%
, Alhashim, A\BPBI G.%
\BCBL {}\ \BBA {} Roscoe, R\BPBI D.%
\end{APACrefauthors}%
\unskip\
\newblock
\APACrefYearMonthDay{2024}{}{}.
\newblock
{\BBOQ}\APACrefatitle {Automating Bias in Writing Evaluation: Sources, Barriers, and Recommendations} {Automating bias in writing evaluation: Sources, barriers, and recommendations}.{\BBCQ}
\newblock
\BIn{} \APACrefbtitle {The {R}outledge International Handbook of Automated Essay Evaluation} {The {R}outledge international handbook of automated essay evaluation}\ (\BPGS\ 421--444).
\newblock
\APACaddressPublisher{}{Routledge}.
\PrintBackRefs{\CurrentBib}

\bibitem [\protect \citeauthoryear {%
Gorgun%
\ \BBA {} Bulut%
}{%
Gorgun%
\ \BBA {} Bulut%
}{%
{\protect \APACyear {2021}}%
}]{%
gorgun2021polytomous}
\APACinsertmetastar {%
gorgun2021polytomous}%
\begin{APACrefauthors}%
Gorgun, G.%
\BCBT {}\ \BBA {} Bulut, O.%
\end{APACrefauthors}%
\unskip\
\newblock
\APACrefYearMonthDay{2021}{}{}.
\newblock
{\BBOQ}\APACrefatitle {A polytomous scoring approach to handle not-reached items in low-stakes assessments} {A polytomous scoring approach to handle not-reached items in low-stakes assessments}.{\BBCQ}
\newblock
\APACjournalVolNumPages{Educational and Psychological Measurement}{81}{5}{847--871}.
\newblock
\begin{APACrefDOI} \doi{10.1177/0013164421991211} \end{APACrefDOI}
\PrintBackRefs{\CurrentBib}

\bibitem [\protect \citeauthoryear {%
Gorgun%
\ \BBA {} Bulut%
}{%
Gorgun%
\ \BBA {} Bulut%
}{%
{\protect \APACyear {2022}}%
}]{%
gorgun2022identifying}
\APACinsertmetastar {%
gorgun2022identifying}%
\begin{APACrefauthors}%
Gorgun, G.%
\BCBT {}\ \BBA {} Bulut, O.%
\end{APACrefauthors}%
\unskip\
\newblock
\APACrefYearMonthDay{2022}{}{}.
\newblock
{\BBOQ}\APACrefatitle {Identifying aberrant responses in intelligent tutoring systems: an application of anomaly detection methods} {Identifying aberrant responses in intelligent tutoring systems: an application of anomaly detection methods}.{\BBCQ}
\newblock
\APACjournalVolNumPages{Psychological Test and Assessment Modeling}{64}{4}{359--384}.
\PrintBackRefs{\CurrentBib}

\bibitem [\protect \citeauthoryear {%
Gorgun%
\ \BBA {} Bulut%
}{%
Gorgun%
\ \BBA {} Bulut%
}{%
{\protect \APACyear {2023}}%
}]{%
gorgun2023incorporating}
\APACinsertmetastar {%
gorgun2023incorporating}%
\begin{APACrefauthors}%
Gorgun, G.%
\BCBT {}\ \BBA {} Bulut, O.%
\end{APACrefauthors}%
\unskip\
\newblock
\APACrefYearMonthDay{2023}{}{}.
\newblock
{\BBOQ}\APACrefatitle {Incorporating test-taking engagement into the item selection algorithm in low-stakes computerized adaptive tests} {Incorporating test-taking engagement into the item selection algorithm in low-stakes computerized adaptive tests}.{\BBCQ}
\newblock
\APACjournalVolNumPages{Large-scale Assessments in Education}{11}{1}{27}.
\newblock
\begin{APACrefDOI} \doi{10.1186/s40536-023-00177-5} \end{APACrefDOI}
\PrintBackRefs{\CurrentBib}

\bibitem [\protect \citeauthoryear {%
Grimes%
\ \BBA {} Warschauer%
}{%
Grimes%
\ \BBA {} Warschauer%
}{%
{\protect \APACyear {2010}}%
}]{%
grimes_utility_2010}
\APACinsertmetastar {%
grimes_utility_2010}%
\begin{APACrefauthors}%
Grimes, D.%
\BCBT {}\ \BBA {} Warschauer, M.%
\end{APACrefauthors}%
\unskip\
\newblock
\APACrefYearMonthDay{2010}{}{}.
\newblock
{\BBOQ}\APACrefatitle {Utility in a Fallible Tool: {A} Multi-Site Case Study of Automated Writing Evaluation} {Utility in a fallible tool: {A} multi-site case study of automated writing evaluation}.{\BBCQ}
\newblock
\APACjournalVolNumPages{The Journal of Technology, Learning and Assessment}{8}{6}{}.
\PrintBackRefs{\CurrentBib}

\bibitem [\protect \citeauthoryear {%
Haller%
, Aldea%
, Seifert%
\BCBL {}\ \BBA {} Strisciuglio%
}{%
Haller%
\ \protect \BOthers {.}}{%
{\protect \APACyear {2022}}%
}]{%
haller_survey_2022}
\APACinsertmetastar {%
haller_survey_2022}%
\begin{APACrefauthors}%
Haller, S.%
, Aldea, A.%
, Seifert, C.%
\BCBL {}\ \BBA {} Strisciuglio, N.%
\end{APACrefauthors}%
\unskip\
\newblock
\APACrefYearMonthDay{2022}{}{}.
\newblock
\APACrefbtitle {Survey on Automated Short Answer Grading with Deep Learning: {F}rom Word Embeddings to Transformers.} {Survey on automated short answer grading with deep learning: {F}rom word embeddings to transformers.}
\newblock
\APACaddressPublisher{}{arXiv}.
\newblock
\begin{APACrefDOI} \doi{10.48550/arXiv.2204.03503} \end{APACrefDOI}
\PrintBackRefs{\CurrentBib}

\bibitem [\protect \citeauthoryear {%
Hao%
\ \protect \BOthers {.}}{%
Hao%
\ \protect \BOthers {.}}{%
{\protect \APACyear {2024}}%
}]{%
hao2024transforming}
\APACinsertmetastar {%
hao2024transforming}%
\begin{APACrefauthors}%
Hao, J.%
, von Davier, A\BPBI A.%
, Yaneva, V.%
, Lottridge, S.%
, von Davier, M.%
\BCBL {}\ \BBA {} Harris, D\BPBI J.%
\end{APACrefauthors}%
\unskip\
\newblock
\APACrefYearMonthDay{2024}{}{}.
\newblock
{\BBOQ}\APACrefatitle {Transforming assessment: The impacts and implications of large language models and generative {AI}} {Transforming assessment: The impacts and implications of large language models and generative {AI}}.{\BBCQ}
\newblock
\APACjournalVolNumPages{Educational Measurement: Issues and Practice}{43}{2}{16--29}.
\newblock
\begin{APACrefDOI} \doi{10.1111/emip.12602} \end{APACrefDOI}
\PrintBackRefs{\CurrentBib}

\bibitem [\protect \citeauthoryear {%
Hattie%
\ \BBA {} Timperley%
}{%
Hattie%
\ \BBA {} Timperley%
}{%
{\protect \APACyear {2007}}%
}]{%
hattie2007power}
\APACinsertmetastar {%
hattie2007power}%
\begin{APACrefauthors}%
Hattie, J.%
\BCBT {}\ \BBA {} Timperley, H.%
\end{APACrefauthors}%
\unskip\
\newblock
\APACrefYearMonthDay{2007}{}{}.
\newblock
{\BBOQ}\APACrefatitle {The power of feedback} {The power of feedback}.{\BBCQ}
\newblock
\APACjournalVolNumPages{Review of Educational Research}{77}{1}{81--112}.
\newblock
\begin{APACrefDOI} \doi{10.3102/003465430298487} \end{APACrefDOI}
\PrintBackRefs{\CurrentBib}

\bibitem [\protect \citeauthoryear {%
He%
, Jing%
\BCBL {}\ \BBA {} Lu%
}{%
He%
\ \protect \BOthers {.}}{%
{\protect \APACyear {2022}}%
}]{%
he2022multilevel}
\APACinsertmetastar {%
he2022multilevel}%
\begin{APACrefauthors}%
He, Y.%
, Jing, S.%
\BCBL {}\ \BBA {} Lu, Y.%
\end{APACrefauthors}%
\unskip\
\newblock
\APACrefYearMonthDay{2022}{April}{}.
\newblock
{\BBOQ}\APACrefatitle {A multilevel multinomial logit approach to bias detection} {A multilevel multinomial logit approach to bias detection}.{\BBCQ}
\newblock
\BIn{} \APACrefbtitle {the annual meeting of the {N}ational {C}ouncil on {M}easurement in {E}ducation. {S}an {D}iego, {CA}.} {the annual meeting of the {N}ational {C}ouncil on {M}easurement in {E}ducation. {S}an {D}iego, {CA}.}
\PrintBackRefs{\CurrentBib}

\bibitem [\protect \citeauthoryear {%
Heston%
\ \BBA {} Khun%
}{%
Heston%
\ \BBA {} Khun%
}{%
{\protect \APACyear {2023}}%
}]{%
heston_prompt_2023}
\APACinsertmetastar {%
heston_prompt_2023}%
\begin{APACrefauthors}%
Heston, T\BPBI F.%
\BCBT {}\ \BBA {} Khun, C.%
\end{APACrefauthors}%
\unskip\
\newblock
\APACrefYearMonthDay{2023}{}{}.
\newblock
{\BBOQ}\APACrefatitle {Prompt Engineering in Medical Education} {Prompt engineering in medical education}.{\BBCQ}
\newblock
\APACjournalVolNumPages{International Medical Education}{2}{3}{198--205}.
\newblock
\begin{APACrefDOI} \doi{10.3390/ime2030019} \end{APACrefDOI}
\PrintBackRefs{\CurrentBib}

\bibitem [\protect \citeauthoryear {%
Higgins%
, Xi%
, Zechner%
\BCBL {}\ \BBA {} Williamson%
}{%
Higgins%
\ \protect \BOthers {.}}{%
{\protect \APACyear {2011}}%
}]{%
higgins_three-stage_2011}
\APACinsertmetastar {%
higgins_three-stage_2011}%
\begin{APACrefauthors}%
Higgins, D.%
, Xi, X.%
, Zechner, K.%
\BCBL {}\ \BBA {} Williamson, D.%
\end{APACrefauthors}%
\unskip\
\newblock
\APACrefYearMonthDay{2011}{}{}.
\newblock
{\BBOQ}\APACrefatitle {A three-stage approach to the automated scoring of spontaneous spoken responses} {A three-stage approach to the automated scoring of spontaneous spoken responses}.{\BBCQ}
\newblock
\APACjournalVolNumPages{Computer Speech \& Language}{25}{2}{282--306}.
\newblock
\begin{APACrefDOI} \doi{10.1016/j.csl.2010.06.001} \end{APACrefDOI}
\PrintBackRefs{\CurrentBib}

\bibitem [\protect \citeauthoryear {%
Himmelreich%
}{%
Himmelreich%
}{%
{\protect \APACyear {2023}}%
}]{%
himmelreich2023against}
\APACinsertmetastar {%
himmelreich2023against}%
\begin{APACrefauthors}%
Himmelreich, J.%
\end{APACrefauthors}%
\unskip\
\newblock
\APACrefYearMonthDay{2023}{}{}.
\newblock
{\BBOQ}\APACrefatitle {Against “democratizing {AI}”} {Against “democratizing {AI}”}.{\BBCQ}
\newblock
\APACjournalVolNumPages{AI \& Society}{38}{4}{1333--1346}.
\newblock
\begin{APACrefDOI} \doi{10.1007/s00146-021-01357-z} \end{APACrefDOI}
\PrintBackRefs{\CurrentBib}

\bibitem [\protect \citeauthoryear {%
Hockly%
}{%
Hockly%
}{%
{\protect \APACyear {2019}}%
}]{%
hockly_automated_2019}
\APACinsertmetastar {%
hockly_automated_2019}%
\begin{APACrefauthors}%
Hockly, N.%
\end{APACrefauthors}%
\unskip\
\newblock
\APACrefYearMonthDay{2019}{}{}.
\newblock
{\BBOQ}\APACrefatitle {Automated writing evaluation} {Automated writing evaluation}.{\BBCQ}
\newblock
\APACjournalVolNumPages{ELT Journal}{73}{1}{82--88}.
\newblock
\begin{APACrefDOI} \doi{10.1093/elt/ccy044} \end{APACrefDOI}
\PrintBackRefs{\CurrentBib}

\bibitem [\protect \citeauthoryear {%
Holland%
\ \BBA {} Wainer%
}{%
Holland%
\ \BBA {} Wainer%
}{%
{\protect \APACyear {1993}}%
}]{%
holland_differential_1993}
\APACinsertmetastar {%
holland_differential_1993}%
\begin{APACrefauthors}%
Holland, P\BPBI W.%
\BCBT {}\ \BBA {} Wainer, H.%
\end{APACrefauthors}%
\ (\BEDS).
\unskip\
\newblock
\APACrefYear{1993}.
\newblock
\APACrefbtitle {Differential item functioning} {Differential item functioning}.
\newblock
\APACaddressPublisher{Hillsdale, NJ, US}{Lawrence Erlbaum Associates, Inc}.
\PrintBackRefs{\CurrentBib}

\bibitem [\protect \citeauthoryear {%
Holling%
, Bertling%
\BCBL {}\ \BBA {} Zeuch%
}{%
Holling%
\ \protect \BOthers {.}}{%
{\protect \APACyear {2009}}%
}]{%
holling_automatic_2009}
\APACinsertmetastar {%
holling_automatic_2009}%
\begin{APACrefauthors}%
Holling, H.%
, Bertling, J\BPBI P.%
\BCBL {}\ \BBA {} Zeuch, N.%
\end{APACrefauthors}%
\unskip\
\newblock
\APACrefYearMonthDay{2009}{}{}.
\newblock
{\BBOQ}\APACrefatitle {Automatic item generation of probability word problems} {Automatic item generation of probability word problems}.{\BBCQ}
\newblock
\APACjournalVolNumPages{Studies in Educational Evaluation}{35}{2}{71--76}.
\newblock
\begin{APACrefDOI} \doi{10.1016/j.stueduc.2009.10.004} \end{APACrefDOI}
\PrintBackRefs{\CurrentBib}

\bibitem [\protect \citeauthoryear {%
Huawei%
\ \BBA {} Aryadoust%
}{%
Huawei%
\ \BBA {} Aryadoust%
}{%
{\protect \APACyear {2023}}%
}]{%
huawei_systematic_2023}
\APACinsertmetastar {%
huawei_systematic_2023}%
\begin{APACrefauthors}%
Huawei, S.%
\BCBT {}\ \BBA {} Aryadoust, V.%
\end{APACrefauthors}%
\unskip\
\newblock
\APACrefYearMonthDay{2023}{}{}.
\newblock
{\BBOQ}\APACrefatitle {A systematic review of automated writing evaluation systems} {A systematic review of automated writing evaluation systems}.{\BBCQ}
\newblock
\APACjournalVolNumPages{Education and Information Technologies}{28}{1}{771--795}.
\newblock
\begin{APACrefDOI} \doi{10.1007/s10639-022-11200-7} \end{APACrefDOI}
\PrintBackRefs{\CurrentBib}

\bibitem [\protect \citeauthoryear {%
{International Test Commission}%
}{%
{International Test Commission}%
}{%
{\protect \APACyear {2014}}%
}]{%
international2014itc}
\APACinsertmetastar {%
international2014itc}%
\begin{APACrefauthors}%
{International Test Commission}.%
\end{APACrefauthors}%
\unskip\
\newblock
\APACrefYearMonthDay{2014}{}{}.
\newblock
{\BBOQ}\APACrefatitle {{ITC} guidelines on quality control in scoring, test analysis, and reporting of test scores} {{ITC} guidelines on quality control in scoring, test analysis, and reporting of test scores}.{\BBCQ}
\newblock
\APACjournalVolNumPages{International Journal of Testing}{14}{3}{195--217}.
\newblock
\begin{APACrefDOI} \doi{10.1080/15305058.2014.918040} \end{APACrefDOI}
\PrintBackRefs{\CurrentBib}

\bibitem [\protect \citeauthoryear {%
{ITC}%
\ \BBA {} {ATP}%
}{%
{ITC}%
\ \BBA {} {ATP}%
}{%
{\protect \APACyear {2022}}%
}]{%
international2022itc}
\APACinsertmetastar {%
international2022itc}%
\begin{APACrefauthors}%
{ITC}%
\BCBT {}\ \BBA {} {ATP}.%
\end{APACrefauthors}%
\unskip\
\newblock
\APACrefYear{2022}.
\newblock
\APACrefbtitle {The Guidelines for Technology-Based Assessment} {The guidelines for technology-based assessment}.
\newblock
\APACaddressPublisher{}{Association of Test Publishers, International Test Commission}.
\newblock
\begin{APACrefURL} \url{https://www.intestcom.org/page/16} \end{APACrefURL}
\PrintBackRefs{\CurrentBib}

\bibitem [\protect \citeauthoryear {%
Jaschik%
}{%
Jaschik%
}{%
{\protect \APACyear {2021}}%
}]{%
jaschik_proctoru_2021}
\APACinsertmetastar {%
jaschik_proctoru_2021}%
\begin{APACrefauthors}%
Jaschik, S.%
\end{APACrefauthors}%
\unskip\
\newblock
\APACrefYearMonthDay{2021}{}{}.
\newblock
\APACrefbtitle {{ProctorU} Abandons Business Based Solely on {AI}.} {{ProctorU} abandons business based solely on {AI}.}
\newblock
\begin{APACrefURL} \url{https://www.insidehighered.com/news/2021/05/24/proctoru-abandons-business-based-solely-ai} \end{APACrefURL}
\PrintBackRefs{\CurrentBib}

\bibitem [\protect \citeauthoryear {%
Jiang%
\ \protect \BOthers {.}}{%
Jiang%
\ \protect \BOthers {.}}{%
{\protect \APACyear {2024}}%
}]{%
jiang2024mixtral}
\APACinsertmetastar {%
jiang2024mixtral}%
\begin{APACrefauthors}%
Jiang, A\BPBI Q.%
, Sablayrolles, A.%
, Roux, A.%
, Mensch, A.%
, Savary, B.%
, Bamford, C.%
\BDBL {}others%
\end{APACrefauthors}%
\unskip\
\newblock
\APACrefYearMonthDay{2024}{}{}.
\newblock
{\BBOQ}\APACrefatitle {Mixtral of experts} {Mixtral of experts}.{\BBCQ}
\newblock
\APACjournalVolNumPages{arXiv Preprint}{}{}{}.
\newblock
\begin{APACrefDOI} \doi{10.48550/arXiv.2401.04088} \end{APACrefDOI}
\PrintBackRefs{\CurrentBib}

\bibitem [\protect \citeauthoryear {%
Jiao%
, Yadav%
\BCBL {}\ \BBA {} Li%
}{%
Jiao%
\ \protect \BOthers {.}}{%
{\protect \APACyear {2023}}%
}]{%
jiao_integrating_2023}
\APACinsertmetastar {%
jiao_integrating_2023}%
\begin{APACrefauthors}%
Jiao, H.%
, Yadav, C.%
\BCBL {}\ \BBA {} Li, G.%
\end{APACrefauthors}%
\unskip\
\newblock
\APACrefYearMonthDay{2023}{}{}.
\newblock
\APACrefbtitle {Integrating Psychometric Analysis and Machine Learning to Augment Data for Cheating Detection in Large-Scale Assessment.} {Integrating psychometric analysis and machine learning to augment data for cheating detection in large-scale assessment.}
\newblock
\APACaddressPublisher{}{OSF}.
\newblock
\begin{APACrefDOI} \doi{10.31234/osf.io/fjz2c} \end{APACrefDOI}
\PrintBackRefs{\CurrentBib}

\bibitem [\protect \citeauthoryear {%
Jin%
}{%
Jin%
}{%
{\protect \APACyear {2012}}%
}]{%
jin_design_2012}
\APACinsertmetastar {%
jin_design_2012}%
\begin{APACrefauthors}%
Jin, S.%
\end{APACrefauthors}%
\unskip\
\newblock
\APACrefYearMonthDay{2012}{}{}.
\newblock
{\BBOQ}\APACrefatitle {Design of an online learning platform with {Moodle}} {Design of an online learning platform with {Moodle}}.{\BBCQ}
\newblock
\BIn{} \APACrefbtitle {the 7th {International} {Conference} on {Computer} {Science} \& {Education} ({ICCSE})} {the 7th {International} {Conference} on {Computer} {Science} \& {Education} ({ICCSE})}\ (\BPGS\ 1710--1714).
\newblock
\begin{APACrefDOI} \doi{10.1109/ICCSE.2012.6295395} \end{APACrefDOI}
\PrintBackRefs{\CurrentBib}

\bibitem [\protect \citeauthoryear {%
Johnson%
\ \BBA {} McCaffrey%
}{%
Johnson%
\ \BBA {} McCaffrey%
}{%
{\protect \APACyear {2023}}%
}]{%
johnson2023evaluating}
\APACinsertmetastar {%
johnson2023evaluating}%
\begin{APACrefauthors}%
Johnson, M\BPBI S.%
\BCBT {}\ \BBA {} McCaffrey, D\BPBI F.%
\end{APACrefauthors}%
\unskip\
\newblock
\APACrefYearMonthDay{2023}{}{}.
\newblock
{\BBOQ}\APACrefatitle {Evaluating Fairness of Automated Scoring in Educational Measurement} {Evaluating fairness of automated scoring in educational measurement}.{\BBCQ}
\newblock
\APACjournalVolNumPages{Advancing Natural Language Processing in Educational Assessment}{}{}{142}.
\PrintBackRefs{\CurrentBib}

\bibitem [\protect \citeauthoryear {%
Jones%
\ \BBA {} Egley%
}{%
Jones%
\ \BBA {} Egley%
}{%
{\protect \APACyear {2007}}%
}]{%
jones_learning_2007}
\APACinsertmetastar {%
jones_learning_2007}%
\begin{APACrefauthors}%
Jones, B\BPBI D.%
\BCBT {}\ \BBA {} Egley, R\BPBI J.%
\end{APACrefauthors}%
\unskip\
\newblock
\APACrefYearMonthDay{2007}{}{}.
\newblock
{\BBOQ}\APACrefatitle {Learning to Take Tests or Learning for Understanding? Teachers' Beliefs about Test-Based Accountability} {Learning to take tests or learning for understanding? teachers' beliefs about test-based accountability}.{\BBCQ}
\newblock
\APACjournalVolNumPages{The Educational Forum}{71}{3}{232--248}.
\newblock
\begin{APACrefDOI} \doi{10.1080/00131720709335008} \end{APACrefDOI}
\PrintBackRefs{\CurrentBib}

\bibitem [\protect \citeauthoryear {%
Jones-Jang%
\ \BBA {} Park%
}{%
Jones-Jang%
\ \BBA {} Park%
}{%
{\protect \APACyear {2023}}%
}]{%
jones-jang_how_2023}
\APACinsertmetastar {%
jones-jang_how_2023}%
\begin{APACrefauthors}%
Jones-Jang, S\BPBI M.%
\BCBT {}\ \BBA {} Park, Y\BPBI J.%
\end{APACrefauthors}%
\unskip\
\newblock
\APACrefYearMonthDay{2023}{}{}.
\newblock
{\BBOQ}\APACrefatitle {How do people react to {AI} failure? Automation bias, algorithmic aversion, and perceived controllability} {How do people react to {AI} failure? automation bias, algorithmic aversion, and perceived controllability}.{\BBCQ}
\newblock
\APACjournalVolNumPages{Journal of Computer-Mediated Communication}{28}{1}{}.
\newblock
\begin{APACrefDOI} \doi{10.1093/jcmc/zmac029} \end{APACrefDOI}
\PrintBackRefs{\CurrentBib}

\bibitem [\protect \citeauthoryear {%
Justice%
}{%
Justice%
}{%
{\protect \APACyear {2022}}%
}]{%
justice2022linear}
\APACinsertmetastar {%
justice2022linear}%
\begin{APACrefauthors}%
Justice, D.%
\end{APACrefauthors}%
\unskip\
\newblock
\APACrefYearMonthDay{2022}{April}{}.
\newblock
{\BBOQ}\APACrefatitle {A linear model approach to bias detection} {A linear model approach to bias detection}.{\BBCQ}
\newblock
\BIn{} \APACrefbtitle {the annual meeting of the {National Council on Measurement in Education}.} {the annual meeting of the {National Council on Measurement in Education}.}
\newblock
\APACaddressPublisher{San Diego, CA}{}.
\PrintBackRefs{\CurrentBib}

\bibitem [\protect \citeauthoryear {%
Kamalov%
, Sulieman%
\BCBL {}\ \BBA {} Calonge%
}{%
Kamalov%
\ \protect \BOthers {.}}{%
{\protect \APACyear {2021}}%
}]{%
kamalov_machine_2021}
\APACinsertmetastar {%
kamalov_machine_2021}%
\begin{APACrefauthors}%
Kamalov, F.%
, Sulieman, H.%
\BCBL {}\ \BBA {} Calonge, D\BPBI S.%
\end{APACrefauthors}%
\unskip\
\newblock
\APACrefYearMonthDay{2021}{}{}.
\newblock
{\BBOQ}\APACrefatitle {Machine learning based approach to exam cheating detection} {Machine learning based approach to exam cheating detection}.{\BBCQ}
\newblock
\APACjournalVolNumPages{PLOS ONE}{16}{8}{e0254340}.
\newblock
\begin{APACrefDOI} \doi{10.1371/journal.pone.0254340} \end{APACrefDOI}
\PrintBackRefs{\CurrentBib}

\bibitem [\protect \citeauthoryear {%
Katz%
\ \BBA {} Rideout%
}{%
Katz%
\ \BBA {} Rideout%
}{%
{\protect \APACyear {2021}}%
}]{%
katz2021learning}
\APACinsertmetastar {%
katz2021learning}%
\begin{APACrefauthors}%
Katz, V.%
\BCBT {}\ \BBA {} Rideout, V.%
\end{APACrefauthors}%
\unskip\
\newblock
\APACrefYearMonthDay{2021}{}{}.
\newblock
{\BBOQ}\APACrefatitle {Learning at Home While Under-Connected: Lower-Income Families during the {COVID-19} Pandemic.} {Learning at home while under-connected: Lower-income families during the {COVID-19} pandemic.}{\BBCQ}
\newblock
\APACjournalVolNumPages{New America}{}{}{}.
\newblock
\begin{APACrefURL} \url{https://www.carnegie.org/publications/learning-home-while-under-connected-lower-income-families-during-covid-19-pandemic/} \end{APACrefURL}
\PrintBackRefs{\CurrentBib}

\bibitem [\protect \citeauthoryear {%
Ke%
\ \BBA {} Ng%
}{%
Ke%
\ \BBA {} Ng%
}{%
{\protect \APACyear {2019}}%
}]{%
ke_automated_2019}
\APACinsertmetastar {%
ke_automated_2019}%
\begin{APACrefauthors}%
Ke, Z.%
\BCBT {}\ \BBA {} Ng, V.%
\end{APACrefauthors}%
\unskip\
\newblock
\APACrefYearMonthDay{2019}{August}{}.
\newblock
{\BBOQ}\APACrefatitle {Automated Essay Scoring: {A} Survey of the State of the Art} {Automated essay scoring: {A} survey of the state of the art}.{\BBCQ}
\newblock
\BIn{} \APACrefbtitle {{Proceedings of the Twenty-Eighth International Joint Conference on Artificial Intelligence}} {{Proceedings of the Twenty-Eighth International Joint Conference on Artificial Intelligence}}\ (\BPGS\ 6300--6308).
\newblock
\APACaddressPublisher{Macao, China}{}.
\newblock
\begin{APACrefDOI} \doi{10.24963/ijcai.2019/879} \end{APACrefDOI}
\PrintBackRefs{\CurrentBib}

\bibitem [\protect \citeauthoryear {%
Khera%
, Simon%
\BCBL {}\ \BBA {} Ross%
}{%
Khera%
\ \protect \BOthers {.}}{%
{\protect \APACyear {2023}}%
}]{%
khera_automation_2023}
\APACinsertmetastar {%
khera_automation_2023}%
\begin{APACrefauthors}%
Khera, R.%
, Simon, M\BPBI A.%
\BCBL {}\ \BBA {} Ross, J\BPBI S.%
\end{APACrefauthors}%
\unskip\
\newblock
\APACrefYearMonthDay{2023}{}{}.
\newblock
{\BBOQ}\APACrefatitle {Automation Bias and Assistive {AI}: {Risk} of Harm From {AI}-Driven Clinical Decision Support} {Automation bias and assistive {AI}: {Risk} of harm from {AI}-driven clinical decision support}.{\BBCQ}
\newblock
\APACjournalVolNumPages{JAMA}{330}{23}{2255--2257}.
\newblock
\begin{APACrefDOI} \doi{10.1001/jama.2023.22557} \end{APACrefDOI}
\PrintBackRefs{\CurrentBib}

\bibitem [\protect \citeauthoryear {%
Khowaja%
, Khuwaja%
, Dev%
, Wang%
\BCBL {}\ \BBA {} Nkenyereye%
}{%
Khowaja%
\ \protect \BOthers {.}}{%
{\protect \APACyear {2024}}%
}]{%
khowaja2024chatgpt}
\APACinsertmetastar {%
khowaja2024chatgpt}%
\begin{APACrefauthors}%
Khowaja, S\BPBI A.%
, Khuwaja, P.%
, Dev, K.%
, Wang, W.%
\BCBL {}\ \BBA {} Nkenyereye, L.%
\end{APACrefauthors}%
\unskip\
\newblock
\APACrefYearMonthDay{2024}{}{}.
\newblock
{\BBOQ}\APACrefatitle {ChatGPT needs spade (sustainability, privacy, digital divide, and ethics) evaluation: A review} {Chatgpt needs spade (sustainability, privacy, digital divide, and ethics) evaluation: A review}.{\BBCQ}
\newblock
\APACjournalVolNumPages{Cognitive Computation}{}{}{1--23}.
\newblock
\begin{APACrefDOI} \doi{10.1007/s12559-024-10285-1} \end{APACrefDOI}
\PrintBackRefs{\CurrentBib}

\bibitem [\protect \citeauthoryear {%
Kim%
, Woo%
\BCBL {}\ \BBA {} Dickison%
}{%
Kim%
\ \protect \BOthers {.}}{%
{\protect \APACyear {2016}}%
}]{%
kim_identifying_2016}
\APACinsertmetastar {%
kim_identifying_2016}%
\begin{APACrefauthors}%
Kim, D.%
, Woo, A.%
\BCBL {}\ \BBA {} Dickison, P.%
\end{APACrefauthors}%
\unskip\
\newblock
\APACrefYearMonthDay{2016}{}{}.
\newblock
{\BBOQ}\APACrefatitle {Identifying and Investigating Aberrant Responses using Psychometrics-Based and Machine Learning-Based Approaches} {Identifying and investigating aberrant responses using psychometrics-based and machine learning-based approaches}.{\BBCQ}
\newblock
\BIn{} G\BPBI J.~Cizek\ \BBA {} J\BPBI A.~Wollack\ (\BEDS), \APACrefbtitle {Handbook of Quantitative Methods for Detecting Cheating on Tests} {Handbook of quantitative methods for detecting cheating on tests}\ (\BPGS\ 70--97).
\newblock
\APACaddressPublisher{}{Routledge}.
\PrintBackRefs{\CurrentBib}

\bibitem [\protect \citeauthoryear {%
Kitsara%
}{%
Kitsara%
}{%
{\protect \APACyear {2022}}%
}]{%
kitsara2022artificial}
\APACinsertmetastar {%
kitsara2022artificial}%
\begin{APACrefauthors}%
Kitsara, I.%
\end{APACrefauthors}%
\unskip\
\newblock
\APACrefYearMonthDay{2022}{}{}.
\newblock
{\BBOQ}\APACrefatitle {Artificial intelligence and the digital divide: From an innovation perspective} {Artificial intelligence and the digital divide: From an innovation perspective}.{\BBCQ}
\newblock
\BIn{} \APACrefbtitle {Platforms and Artificial Intelligence: The Next Generation of Competences} {Platforms and artificial intelligence: The next generation of competences}\ (\BPGS\ 245--265).
\newblock
\APACaddressPublisher{}{Springer}.
\PrintBackRefs{\CurrentBib}

\bibitem [\protect \citeauthoryear {%
Kumari%
, Keshari%
, Sharma%
\BCBL {}\ \BBA {} Goel%
}{%
Kumari%
\ \protect \BOthers {.}}{%
{\protect \APACyear {2022}}%
}]{%
kumari2022context}
\APACinsertmetastar {%
kumari2022context}%
\begin{APACrefauthors}%
Kumari, V.%
, Keshari, S.%
, Sharma, Y.%
\BCBL {}\ \BBA {} Goel, L.%
\end{APACrefauthors}%
\unskip\
\newblock
\APACrefYearMonthDay{2022}{}{}.
\newblock
{\BBOQ}\APACrefatitle {Context-based question answering system with suggested questions} {Context-based question answering system with suggested questions}.{\BBCQ}
\newblock
\BIn{} \APACrefbtitle {{the 12th International Conference on Cloud Computing, Data Science \& Engineering (Confluence)}} {{the 12th International Conference on Cloud Computing, Data Science \& Engineering (Confluence)}}\ (\BPGS\ 368--373).
\newblock
\APACaddressPublisher{Noida, India}{}.
\PrintBackRefs{\CurrentBib}

\bibitem [\protect \citeauthoryear {%
Kupfer%
\ \protect \BOthers {.}}{%
Kupfer%
\ \protect \BOthers {.}}{%
{\protect \APACyear {2023}}%
}]{%
kupfer2023check}
\APACinsertmetastar {%
kupfer2023check}%
\begin{APACrefauthors}%
Kupfer, C.%
, Prassl, R.%
, Flei{\ss}, J.%
, Malin, C.%
, Thalmann, S.%
\BCBL {}\ \BBA {} Kubicek, B.%
\end{APACrefauthors}%
\unskip\
\newblock
\APACrefYearMonthDay{2023}{}{}.
\newblock
{\BBOQ}\APACrefatitle {Check the box! How to deal with automation bias in AI-based personnel selection} {Check the box! how to deal with automation bias in ai-based personnel selection}.{\BBCQ}
\newblock
\APACjournalVolNumPages{Frontiers in Psychology}{14}{}{1118723}.
\newblock
\begin{APACrefDOI} \doi{10.3389/fpsyg.2023.1118723} \end{APACrefDOI}
\PrintBackRefs{\CurrentBib}

\bibitem [\protect \citeauthoryear {%
Lacoste%
, Luccioni%
, Schmidt%
\BCBL {}\ \BBA {} Dandres%
}{%
Lacoste%
\ \protect \BOthers {.}}{%
{\protect \APACyear {2019}}%
}]{%
lacoste2019quantifying}
\APACinsertmetastar {%
lacoste2019quantifying}%
\begin{APACrefauthors}%
Lacoste, A.%
, Luccioni, A.%
, Schmidt, V.%
\BCBL {}\ \BBA {} Dandres, T.%
\end{APACrefauthors}%
\unskip\
\newblock
\APACrefYearMonthDay{2019}{}{}.
\newblock
{\BBOQ}\APACrefatitle {Quantifying the Carbon Emissions of Machine Learning} {Quantifying the carbon emissions of machine learning}.{\BBCQ}
\newblock
\APACjournalVolNumPages{arXiv preprint arXiv:1910.09700}{}{}{}.
\newblock
\begin{APACrefDOI} \doi{10.48550/arXiv.1910.09700} \end{APACrefDOI}
\PrintBackRefs{\CurrentBib}

\bibitem [\protect \citeauthoryear {%
Lai%
, Alves%
\BCBL {}\ \BBA {} Gierl%
}{%
Lai%
\ \protect \BOthers {.}}{%
{\protect \APACyear {2009}}%
}]{%
lai2009using}
\APACinsertmetastar {%
lai2009using}%
\begin{APACrefauthors}%
Lai, H.%
, Alves, C.%
\BCBL {}\ \BBA {} Gierl, M.%
\end{APACrefauthors}%
\unskip\
\newblock
\APACrefYearMonthDay{2009}{June}{}.
\newblock
{\BBOQ}\APACrefatitle {Using automatic item generation to address item demands for {CAT}} {Using automatic item generation to address item demands for {CAT}}.{\BBCQ}
\newblock
\BIn{} \APACrefbtitle {{the 2009 GMAC Conference on Computerized Adaptive Testing}.} {{the 2009 GMAC Conference on Computerized Adaptive Testing}.}
\newblock
\begin{APACrefURL} \url{http://iacat.org/sites/default/files/biblio/cat09lai.pdf} \end{APACrefURL}
\PrintBackRefs{\CurrentBib}

\bibitem [\protect \citeauthoryear {%
Lee%
\ \protect \BOthers {.}}{%
Lee%
\ \protect \BOthers {.}}{%
{\protect \APACyear {2023}}%
}]{%
lee_few-shot_2023}
\APACinsertmetastar {%
lee_few-shot_2023}%
\begin{APACrefauthors}%
Lee, U.%
, Jung, H.%
, Jeon, Y.%
, Sohn, Y.%
, Hwang, W.%
, Moon, J.%
\BCBL {}\ \BBA {} Kim, H.%
\end{APACrefauthors}%
\unskip\
\newblock
\APACrefYearMonthDay{2023}{}{}.
\newblock
{\BBOQ}\APACrefatitle {Few-shot is enough: {E}xploring {ChatGPT} prompt engineering method for automatic question generation in English education} {Few-shot is enough: {E}xploring {ChatGPT} prompt engineering method for automatic question generation in english education}.{\BBCQ}
\newblock
\APACjournalVolNumPages{Education and Information Technologies}{}{}{}.
\newblock
\begin{APACrefDOI} \doi{10.1007/s10639-023-12249-8} \end{APACrefDOI}
\PrintBackRefs{\CurrentBib}

\bibitem [\protect \citeauthoryear {%
J.~Li%
, Tang%
, Zhao%
, Nie%
\BCBL {}\ \BBA {} Wen%
}{%
J.~Li%
\ \protect \BOthers {.}}{%
{\protect \APACyear {2024}}%
}]{%
li2024pre}
\APACinsertmetastar {%
li2024pre}%
\begin{APACrefauthors}%
Li, J.%
, Tang, T.%
, Zhao, W\BPBI X.%
, Nie, J\BHBI Y.%
\BCBL {}\ \BBA {} Wen, J\BHBI R.%
\end{APACrefauthors}%
\unskip\
\newblock
\APACrefYearMonthDay{2024}{}{}.
\newblock
{\BBOQ}\APACrefatitle {Pre-trained language models for text generation: A survey} {Pre-trained language models for text generation: A survey}.{\BBCQ}
\newblock
\APACjournalVolNumPages{ACM Computing Surveys}{56}{9}{1--39}.
\newblock
\begin{APACrefDOI} \doi{10.1145/3649449} \end{APACrefDOI}
\PrintBackRefs{\CurrentBib}

\bibitem [\protect \citeauthoryear {%
R.~Li%
}{%
R.~Li%
}{%
{\protect \APACyear {2023}}%
}]{%
li_still_2023}
\APACinsertmetastar {%
li_still_2023}%
\begin{APACrefauthors}%
Li, R.%
\end{APACrefauthors}%
\unskip\
\newblock
\APACrefYearMonthDay{2023}{}{}.
\newblock
{\BBOQ}\APACrefatitle {Still a fallible tool? {Revisiting} effects of automated writing evaluation from activity theory perspective} {Still a fallible tool? {Revisiting} effects of automated writing evaluation from activity theory perspective}.{\BBCQ}
\newblock
\APACjournalVolNumPages{British Journal of Educational Technology}{54}{3}{773--789}.
\newblock
\begin{APACrefDOI} \doi{10.1111/bjet.13294} \end{APACrefDOI}
\PrintBackRefs{\CurrentBib}

\bibitem [\protect \citeauthoryear {%
Lin%
\ \BBA {} Kifer%
}{%
Lin%
\ \BBA {} Kifer%
}{%
{\protect \APACyear {2015}}%
}]{%
lin_information_2015}
\APACinsertmetastar {%
lin_information_2015}%
\begin{APACrefauthors}%
Lin, B\BHBI R.%
\BCBT {}\ \BBA {} Kifer, D.%
\end{APACrefauthors}%
\unskip\
\newblock
\APACrefYearMonthDay{2015}{}{}.
\newblock
{\BBOQ}\APACrefatitle {Information Measures in Statistical Privacy and Data Processing Applications} {Information measures in statistical privacy and data processing applications}.{\BBCQ}
\newblock
\APACjournalVolNumPages{ACM Transactions on Knowledge Discovery from Data}{9}{4}{28:1--28:29}.
\newblock
\begin{APACrefDOI} \doi{10.1145/2700407} \end{APACrefDOI}
\PrintBackRefs{\CurrentBib}

\bibitem [\protect \citeauthoryear {%
Lindner%
, Lüdtke%
\BCBL {}\ \BBA {} Nagy%
}{%
Lindner%
\ \protect \BOthers {.}}{%
{\protect \APACyear {2019}}%
}]{%
lindner_onset_2019}
\APACinsertmetastar {%
lindner_onset_2019}%
\begin{APACrefauthors}%
Lindner, M\BPBI A.%
, Lüdtke, O.%
\BCBL {}\ \BBA {} Nagy, G.%
\end{APACrefauthors}%
\unskip\
\newblock
\APACrefYearMonthDay{2019}{}{}.
\newblock
{\BBOQ}\APACrefatitle {The Onset of Rapid-Guessing Behavior Over the Course of Testing Time: {A} Matter of Motivation and Cognitive Resources} {The onset of rapid-guessing behavior over the course of testing time: {A} matter of motivation and cognitive resources}.{\BBCQ}
\newblock
\APACjournalVolNumPages{Frontiers in Psychology}{10}{}{}.
\newblock
\begin{APACrefDOI} \doi{10.3389/fpsyg.2019.01533} \end{APACrefDOI}
\PrintBackRefs{\CurrentBib}

\bibitem [\protect \citeauthoryear {%
P.~Liu%
\ \protect \BOthers {.}}{%
P.~Liu%
\ \protect \BOthers {.}}{%
{\protect \APACyear {2023}}%
}]{%
liu2023pre}
\APACinsertmetastar {%
liu2023pre}%
\begin{APACrefauthors}%
Liu, P.%
, Yuan, W.%
, Fu, J.%
, Jiang, Z.%
, Hayashi, H.%
\BCBL {}\ \BBA {} Neubig, G.%
\end{APACrefauthors}%
\unskip\
\newblock
\APACrefYearMonthDay{2023}{}{}.
\newblock
{\BBOQ}\APACrefatitle {Pre-train, prompt, and predict: A systematic survey of prompting methods in natural language processing} {Pre-train, prompt, and predict: A systematic survey of prompting methods in natural language processing}.{\BBCQ}
\newblock
\APACjournalVolNumPages{ACM Computing Surveys}{55}{9}{1--35}.
\newblock
\begin{APACrefDOI} \doi{10.1145/3560815} \end{APACrefDOI}
\PrintBackRefs{\CurrentBib}

\bibitem [\protect \citeauthoryear {%
X.~Liu%
\ \BBA {} Fauss%
}{%
X.~Liu%
\ \BBA {} Fauss%
}{%
{\protect \APACyear {2024}}%
}]{%
liu_fauss_2024}
\APACinsertmetastar {%
liu_fauss_2024}%
\begin{APACrefauthors}%
Liu, X.%
\BCBT {}\ \BBA {} Fauss, M.%
\end{APACrefauthors}%
\unskip\
\newblock
\APACrefYearMonthDay{2024}{April}{}.
\newblock
{\BBOQ}\APACrefatitle {A {Bayesian} Nonparametric Model for Flexible Automated Scoring} {A {Bayesian} nonparametric model for flexible automated scoring}.{\BBCQ}
\newblock
\APACjournalVolNumPages{Paper presented at the annual meeting of the National Council for Measurement in Education, Philadelphia, PA}{}{}{}.
\PrintBackRefs{\CurrentBib}

\bibitem [\protect \citeauthoryear {%
Y.~Liu%
, Cheng%
\BCBL {}\ \BBA {} Liu%
}{%
Y.~Liu%
\ \protect \BOthers {.}}{%
{\protect \APACyear {2020}}%
}]{%
liu2020identifying}
\APACinsertmetastar {%
liu2020identifying}%
\begin{APACrefauthors}%
Liu, Y.%
, Cheng, Y.%
\BCBL {}\ \BBA {} Liu, H.%
\end{APACrefauthors}%
\unskip\
\newblock
\APACrefYearMonthDay{2020}{}{}.
\newblock
{\BBOQ}\APACrefatitle {Identifying effortful individuals with mixture modeling response accuracy and response time simultaneously to improve item parameter estimation} {Identifying effortful individuals with mixture modeling response accuracy and response time simultaneously to improve item parameter estimation}.{\BBCQ}
\newblock
\APACjournalVolNumPages{Educational and Psychological Measurement}{80}{4}{775--807}.
\newblock
\begin{APACrefDOI} \doi{10.1177/0013164419895068} \end{APACrefDOI}
\PrintBackRefs{\CurrentBib}

\bibitem [\protect \citeauthoryear {%
Lottridge%
, Ormerod%
\BCBL {}\ \BBA {} Jafari%
}{%
Lottridge%
\ \protect \BOthers {.}}{%
{\protect \APACyear {2023}}%
}]{%
lottridge_psychometric_2023}
\APACinsertmetastar {%
lottridge_psychometric_2023}%
\begin{APACrefauthors}%
Lottridge, S.%
, Ormerod, C.%
\BCBL {}\ \BBA {} Jafari, A.%
\end{APACrefauthors}%
\unskip\
\newblock
\APACrefYearMonthDay{2023}{}{}.
\newblock
{\BBOQ}\APACrefatitle {Psychometric Considerations When Using Deep Learning for Automated Scoring} {Psychometric considerations when using deep learning for automated scoring}.{\BBCQ}
\newblock
\BIn{} V.~Yaneva\ \BBA {} M.~von Davier\ (\BEDS), \APACrefbtitle {Advancing Natural Language Processing in Educational Assessment} {Advancing natural language processing in educational assessment}\ (\BPGS\ 15--30).
\newblock
\APACaddressPublisher{}{Routledge}.
\newblock
\begin{APACrefDOI} \doi{10.4324/9781003278658-3} \end{APACrefDOI}
\PrintBackRefs{\CurrentBib}

\bibitem [\protect \citeauthoryear {%
Lottridge%
\ \BBA {} Young%
}{%
Lottridge%
\ \BBA {} Young%
}{%
{\protect \APACyear {2022}}%
}]{%
lottridge2022examining}
\APACinsertmetastar {%
lottridge2022examining}%
\begin{APACrefauthors}%
Lottridge, S.%
\BCBT {}\ \BBA {} Young, M.%
\end{APACrefauthors}%
\unskip\
\newblock
\APACrefYearMonthDay{2022}{April}{}.
\newblock
{\BBOQ}\APACrefatitle {Examining bias in automated scoring of reading comprehension items.} {Examining bias in automated scoring of reading comprehension items.}{\BBCQ}
\newblock
\BIn{} \APACrefbtitle {the annual meeting of the {National Council on Measurement in Education}.} {the annual meeting of the {National Council on Measurement in Education}.}
\newblock
\APACaddressPublisher{San Diego, CA}{}.
\PrintBackRefs{\CurrentBib}

\bibitem [\protect \citeauthoryear {%
Lyell%
\ \BBA {} Coiera%
}{%
Lyell%
\ \BBA {} Coiera%
}{%
{\protect \APACyear {2017}}%
}]{%
lyell_automation_2017}
\APACinsertmetastar {%
lyell_automation_2017}%
\begin{APACrefauthors}%
Lyell, D.%
\BCBT {}\ \BBA {} Coiera, E.%
\end{APACrefauthors}%
\unskip\
\newblock
\APACrefYearMonthDay{2017}{}{}.
\newblock
{\BBOQ}\APACrefatitle {Automation bias and verification complexity: {A} systematic review} {Automation bias and verification complexity: {A} systematic review}.{\BBCQ}
\newblock
\APACjournalVolNumPages{Journal of the American Medical Informatics Association}{24}{2}{423--431}.
\newblock
\begin{APACrefDOI} \doi{10.1093/jamia/ocw105} \end{APACrefDOI}
\PrintBackRefs{\CurrentBib}

\bibitem [\protect \citeauthoryear {%
Mayfield%
\ \BBA {} Black%
}{%
Mayfield%
\ \BBA {} Black%
}{%
{\protect \APACyear {2020}}%
}]{%
mayfield2020should}
\APACinsertmetastar {%
mayfield2020should}%
\begin{APACrefauthors}%
Mayfield, E.%
\BCBT {}\ \BBA {} Black, A\BPBI W.%
\end{APACrefauthors}%
\unskip\
\newblock
\APACrefYearMonthDay{2020}{}{}.
\newblock
{\BBOQ}\APACrefatitle {Should you fine-tune {BERT} for automated essay scoring?} {Should you fine-tune {BERT} for automated essay scoring?}{\BBCQ}
\newblock
\BIn{} \APACrefbtitle {Proceedings of the Fifteenth Workshop on Innovative Use of {NLP} for Building Educational Applications} {Proceedings of the fifteenth workshop on innovative use of {NLP} for building educational applications}\ (\BPGS\ 151--162).
\PrintBackRefs{\CurrentBib}

\bibitem [\protect \citeauthoryear {%
McCaffrey%
, Casabianca%
, M.%
\BCBL {}\ \BBA {} Johnson%
}{%
McCaffrey%
\ \protect \BOthers {.}}{%
{\protect \APACyear {2024}}%
}]{%
mccaffrey2024}
\APACinsertmetastar {%
mccaffrey2024}%
\begin{APACrefauthors}%
McCaffrey, D\BPBI F.%
, Casabianca%
, M., J.%
\BCBL {}\ \BBA {} Johnson, M\BPBI S.%
\end{APACrefauthors}%
\unskip\
\newblock
\APACrefYearMonthDay{2024}{}{}.
\newblock
\APACrefbtitle {The proportional reduction in mean squared error for use in automated scoring.} {The proportional reduction in mean squared error for use in automated scoring.}
\newblock
\APACrefnote{Unpublished manuscript}
\PrintBackRefs{\CurrentBib}

\bibitem [\protect \citeauthoryear {%
McCaffrey%
, Casabianca%
, Ricker-Pedley%
, Lawless%
\BCBL {}\ \BBA {} Wendler%
}{%
McCaffrey%
\ \protect \BOthers {.}}{%
{\protect \APACyear {2022}}%
}]{%
mccaffrey2022best}
\APACinsertmetastar {%
mccaffrey2022best}%
\begin{APACrefauthors}%
McCaffrey, D\BPBI F.%
, Casabianca, J\BPBI M.%
, Ricker-Pedley, K\BPBI L.%
, Lawless, R\BPBI R.%
\BCBL {}\ \BBA {} Wendler, C.%
\end{APACrefauthors}%
\unskip\
\newblock
\APACrefYearMonthDay{2022}{}{}.
\newblock
{\BBOQ}\APACrefatitle {Best Practices for Constructed-Response Scoring} {Best practices for constructed-response scoring}.{\BBCQ}
\newblock
\APACjournalVolNumPages{ETS Research Report Series}{2022}{1}{1--58}.
\PrintBackRefs{\CurrentBib}

\bibitem [\protect \citeauthoryear {%
Mehrabi%
, Morstatter%
, Saxena%
, Lerman%
\BCBL {}\ \BBA {} Galstyan%
}{%
Mehrabi%
\ \protect \BOthers {.}}{%
{\protect \APACyear {2021}}%
}]{%
mehrabi_survey_2021}
\APACinsertmetastar {%
mehrabi_survey_2021}%
\begin{APACrefauthors}%
Mehrabi, N.%
, Morstatter, F.%
, Saxena, N.%
, Lerman, K.%
\BCBL {}\ \BBA {} Galstyan, A.%
\end{APACrefauthors}%
\unskip\
\newblock
\APACrefYearMonthDay{2021}{}{}.
\newblock
{\BBOQ}\APACrefatitle {A survey on Bias and Fairness in Machine Learning} {A survey on bias and fairness in machine learning}.{\BBCQ}
\newblock
\APACjournalVolNumPages{ACM Computing Surveys}{54}{6}{115:1--115:35}.
\newblock
\begin{APACrefDOI} \doi{10.1145/3457607} \end{APACrefDOI}
\PrintBackRefs{\CurrentBib}

\bibitem [\protect \citeauthoryear {%
Meng%
\ \BBA {} Ma%
}{%
Meng%
\ \BBA {} Ma%
}{%
{\protect \APACyear {2023}}%
}]{%
meng_machine_2023}
\APACinsertmetastar {%
meng_machine_2023}%
\begin{APACrefauthors}%
Meng, H.%
\BCBT {}\ \BBA {} Ma, Y.%
\end{APACrefauthors}%
\unskip\
\newblock
\APACrefYearMonthDay{2023}{}{}.
\newblock
{\BBOQ}\APACrefatitle {Machine Learning–Based Profiling in Test Cheating Detection} {Machine learning–based profiling in test cheating detection}.{\BBCQ}
\newblock
\APACjournalVolNumPages{Educational Measurement: Issues and Practice}{42}{1}{59--75}.
\newblock
\begin{APACrefDOI} \doi{10.1111/emip.12541} \end{APACrefDOI}
\PrintBackRefs{\CurrentBib}

\bibitem [\protect \citeauthoryear {%
Molnar%
}{%
Molnar%
}{%
{\protect \APACyear {2020}}%
}]{%
molnar2020interpretable}
\APACinsertmetastar {%
molnar2020interpretable}%
\begin{APACrefauthors}%
Molnar, C.%
\end{APACrefauthors}%
\unskip\
\newblock
\APACrefYear{2020}.
\newblock
\APACrefbtitle {Interpretable machine learning: A Guide for Making Black Box Models Explainable} {Interpretable machine learning: A guide for making black box models explainable}.
\newblock
\begin{APACrefURL} \url{https://christophm.github.io/interpretable-ml-book/} \end{APACrefURL}
\PrintBackRefs{\CurrentBib}

\bibitem [\protect \citeauthoryear {%
Moore%
\ \BBA {} MacArthur%
}{%
Moore%
\ \BBA {} MacArthur%
}{%
{\protect \APACyear {2016}}%
}]{%
moore_student_2016}
\APACinsertmetastar {%
moore_student_2016}%
\begin{APACrefauthors}%
Moore, N\BPBI S.%
\BCBT {}\ \BBA {} MacArthur, C\BPBI A.%
\end{APACrefauthors}%
\unskip\
\newblock
\APACrefYearMonthDay{2016}{}{}.
\newblock
{\BBOQ}\APACrefatitle {Student use of automated essay evaluation technology during revision} {Student use of automated essay evaluation technology during revision}.{\BBCQ}
\newblock
\APACjournalVolNumPages{Journal of Writing Research}{8}{1}{149--175}.
\newblock
\begin{APACrefDOI} \doi{10.17239/jowr-2016.08.01.05} \end{APACrefDOI}
\PrintBackRefs{\CurrentBib}

\bibitem [\protect \citeauthoryear {%
Moreno-Guerrero%
, Rodríguez-Jiménez%
, Gómez-García%
\BCBL {}\ \BBA {} Ramos Navas-Parejo%
}{%
Moreno-Guerrero%
\ \protect \BOthers {.}}{%
{\protect \APACyear {2020}}%
}]{%
moreno-guerrero_educational_2020}
\APACinsertmetastar {%
moreno-guerrero_educational_2020}%
\begin{APACrefauthors}%
Moreno-Guerrero, A\BHBI J.%
, Rodríguez-Jiménez, C.%
, Gómez-García, G.%
\BCBL {}\ \BBA {} Ramos Navas-Parejo, M.%
\end{APACrefauthors}%
\unskip\
\newblock
\APACrefYearMonthDay{2020}{}{}.
\newblock
{\BBOQ}\APACrefatitle {Educational Innovation in Higher Education: {Use} of Role Playing and Educational Video in Future Teachers’ Training} {Educational innovation in higher education: {Use} of role playing and educational video in future teachers’ training}.{\BBCQ}
\newblock
\APACjournalVolNumPages{Sustainability}{12}{6}{2558}.
\newblock
\begin{APACrefDOI} \doi{10.3390/su12062558} \end{APACrefDOI}
\PrintBackRefs{\CurrentBib}

\bibitem [\protect \citeauthoryear {%
Motwani%
, Nagpal%
, Motwani%
, Nagdev%
\BCBL {}\ \BBA {} Yeole%
}{%
Motwani%
\ \protect \BOthers {.}}{%
{\protect \APACyear {2021}}%
}]{%
motwani_ai-based_2021}
\APACinsertmetastar {%
motwani_ai-based_2021}%
\begin{APACrefauthors}%
Motwani, S.%
, Nagpal, C.%
, Motwani, M.%
, Nagdev, N.%
\BCBL {}\ \BBA {} Yeole, A.%
\end{APACrefauthors}%
\unskip\
\newblock
\APACrefYearMonthDay{2021}{}{}.
\newblock
\APACrefbtitle {{AI}-Based Proctoring System for Online Tests} {{AI}-based proctoring system for online tests}\ [{SSRN} {Scholarly} {Paper}].
\newblock
\APACaddressPublisher{Rochester, NY}{}.
\newblock
\begin{APACrefURL} \url{https://papers.ssrn.com/abstract=3866446} \end{APACrefURL}
\newblock
\begin{APACrefDOI} \doi{10.2139/ssrn.3866446} \end{APACrefDOI}
\PrintBackRefs{\CurrentBib}

\bibitem [\protect \citeauthoryear {%
Mueller%
, Zhang%
\BCBL {}\ \BBA {} Ferrara%
}{%
Mueller%
\ \protect \BOthers {.}}{%
{\protect \APACyear {2016}}%
}]{%
mueller2016have}
\APACinsertmetastar {%
mueller2016have}%
\begin{APACrefauthors}%
Mueller, L.%
, Zhang, Y.%
\BCBL {}\ \BBA {} Ferrara, S.%
\end{APACrefauthors}%
\unskip\
\newblock
\APACrefYearMonthDay{2016}{}{}.
\newblock
{\BBOQ}\APACrefatitle {What Have We Learned?} {What have we learned?}{\BBCQ}
\newblock
\BIn{} G\BPBI J.~Cizek\ \BBA {} J\BPBI A.~Wollack\ (\BEDS), \APACrefbtitle {Handbook of quantitative methods for detecting cheating on tests} {Handbook of quantitative methods for detecting cheating on tests}\ (\BPGS\ 373--389).
\newblock
\APACaddressPublisher{}{Routledge}.
\PrintBackRefs{\CurrentBib}

\bibitem [\protect \citeauthoryear {%
Myers%
\ \BBA {} Wilson%
}{%
Myers%
\ \BBA {} Wilson%
}{%
{\protect \APACyear {2023}}%
}]{%
myers_evaluating_2023}
\APACinsertmetastar {%
myers_evaluating_2023}%
\begin{APACrefauthors}%
Myers, M\BPBI C.%
\BCBT {}\ \BBA {} Wilson, J.%
\end{APACrefauthors}%
\unskip\
\newblock
\APACrefYearMonthDay{2023}{}{}.
\newblock
{\BBOQ}\APACrefatitle {Evaluating the Construct Validity of an Automated Writing Evaluation System with a Randomization Algorithm} {Evaluating the construct validity of an automated writing evaluation system with a randomization algorithm}.{\BBCQ}
\newblock
\APACjournalVolNumPages{International Journal of Artificial Intelligence in Education}{33}{3}{609--634}.
\newblock
\begin{APACrefDOI} \doi{10.1007/s40593-022-00301-6} \end{APACrefDOI}
\PrintBackRefs{\CurrentBib}

\bibitem [\protect \citeauthoryear {%
Ngo%
, Chen%
\BCBL {}\ \BBA {} Lai%
}{%
Ngo%
\ \protect \BOthers {.}}{%
{\protect \APACyear {2024}}%
}]{%
ngo_effectiveness_2024}
\APACinsertmetastar {%
ngo_effectiveness_2024}%
\begin{APACrefauthors}%
Ngo, T\BPBI T\BHBI N.%
, Chen, H\BPBI H\BHBI J.%
\BCBL {}\ \BBA {} Lai, K\BPBI K\BHBI W.%
\end{APACrefauthors}%
\unskip\
\newblock
\APACrefYearMonthDay{2024}{}{}.
\newblock
{\BBOQ}\APACrefatitle {The effectiveness of automated writing evaluation in {EFL}/{ESL} writing: a three-level meta-analysis} {The effectiveness of automated writing evaluation in {EFL}/{ESL} writing: a three-level meta-analysis}.{\BBCQ}
\newblock
\APACjournalVolNumPages{Interactive Learning Environments}{32}{2}{727--744}.
\newblock
\begin{APACrefDOI} \doi{10.1080/10494820.2022.2096642} \end{APACrefDOI}
\PrintBackRefs{\CurrentBib}

\bibitem [\protect \citeauthoryear {%
Nguyen%
, Ngo%
, Hong%
, Dang%
\BCBL {}\ \BBA {} Nguyen%
}{%
Nguyen%
\ \protect \BOthers {.}}{%
{\protect \APACyear {2023}}%
}]{%
nguyen_ethical_2023}
\APACinsertmetastar {%
nguyen_ethical_2023}%
\begin{APACrefauthors}%
Nguyen, A.%
, Ngo, H\BPBI N.%
, Hong, Y.%
, Dang, B.%
\BCBL {}\ \BBA {} Nguyen, B\BHBI P\BPBI T.%
\end{APACrefauthors}%
\unskip\
\newblock
\APACrefYearMonthDay{2023}{}{}.
\newblock
{\BBOQ}\APACrefatitle {Ethical principles for artificial intelligence in education} {Ethical principles for artificial intelligence in education}.{\BBCQ}
\newblock
\APACjournalVolNumPages{Education and Information Technologies}{28}{4}{4221--4241}.
\newblock
\begin{APACrefDOI} \doi{10.1007/s10639-022-11316-w} \end{APACrefDOI}
\PrintBackRefs{\CurrentBib}

\bibitem [\protect \citeauthoryear {%
Nigam%
, Pasricha%
, Singh%
\BCBL {}\ \BBA {} Churi%
}{%
Nigam%
\ \protect \BOthers {.}}{%
{\protect \APACyear {2021}}%
}]{%
nigam_systematic_2021}
\APACinsertmetastar {%
nigam_systematic_2021}%
\begin{APACrefauthors}%
Nigam, A.%
, Pasricha, R.%
, Singh, T.%
\BCBL {}\ \BBA {} Churi, P.%
\end{APACrefauthors}%
\unskip\
\newblock
\APACrefYearMonthDay{2021}{}{}.
\newblock
{\BBOQ}\APACrefatitle {A Systematic Review on {AI}-based Proctoring Systems: {Past}, Present and Future} {A systematic review on {AI}-based proctoring systems: {Past}, present and future}.{\BBCQ}
\newblock
\APACjournalVolNumPages{Education and Information Technologies}{26}{5}{6421--6445}.
\newblock
\begin{APACrefDOI} \doi{10.1007/s10639-021-10597-x} \end{APACrefDOI}
\PrintBackRefs{\CurrentBib}

\bibitem [\protect \citeauthoryear {%
Nunes%
, Cordeiro%
, Limpo%
\BCBL {}\ \BBA {} Castro%
}{%
Nunes%
\ \protect \BOthers {.}}{%
{\protect \APACyear {2022}}%
}]{%
nunes_effectiveness_2022}
\APACinsertmetastar {%
nunes_effectiveness_2022}%
\begin{APACrefauthors}%
Nunes, A.%
, Cordeiro, C.%
, Limpo, T.%
\BCBL {}\ \BBA {} Castro, S\BPBI L.%
\end{APACrefauthors}%
\unskip\
\newblock
\APACrefYearMonthDay{2022}{}{}.
\newblock
{\BBOQ}\APACrefatitle {Effectiveness of automated writing evaluation systems in school settings: {A} systematic review of studies from 2000 to 2020} {Effectiveness of automated writing evaluation systems in school settings: {A} systematic review of studies from 2000 to 2020}.{\BBCQ}
\newblock
\APACjournalVolNumPages{Journal of Computer Assisted Learning}{38}{2}{599--620}.
\newblock
\begin{APACrefDOI} \doi{10.1111/jcal.12635} \end{APACrefDOI}
\PrintBackRefs{\CurrentBib}

\bibitem [\protect \citeauthoryear {%
OECD%
}{%
OECD%
}{%
{\protect \APACyear {2023}}%
}]{%
oecd2023}
\APACinsertmetastar {%
oecd2023}%
\begin{APACrefauthors}%
OECD.%
\end{APACrefauthors}%
\unskip\
\newblock
\APACrefYearMonthDay{2023}{}{}.
\newblock
{\BBOQ}\APACrefatitle {Opportunities, guidelines and guardrails for effective and equitable use of {AI} in education} {Opportunities, guidelines and guardrails for effective and equitable use of {AI} in education}.{\BBCQ}
\newblock
\BIn{} \APACrefbtitle {{OECD} {D}igital {E}ducation {O}utlook 2023: Towards an {E}ffective {D}igital {E}ducation {E}cosystem.} {{OECD} {D}igital {E}ducation {O}utlook 2023: Towards an {E}ffective {D}igital {E}ducation {E}cosystem.}
\newblock
\APACaddressPublisher{}{OECD Publishing}.
\newblock
\begin{APACrefURL} \url{https://doi.org/10.1787/2b39e98b-en.} \end{APACrefURL}
\PrintBackRefs{\CurrentBib}

\bibitem [\protect \citeauthoryear {%
Offerijns%
, Verberne%
\BCBL {}\ \BBA {} Verhoef%
}{%
Offerijns%
\ \protect \BOthers {.}}{%
{\protect \APACyear {2020}}%
}]{%
offerijns2020better}
\APACinsertmetastar {%
offerijns2020better}%
\begin{APACrefauthors}%
Offerijns, J.%
, Verberne, S.%
\BCBL {}\ \BBA {} Verhoef, T.%
\end{APACrefauthors}%
\unskip\
\newblock
\APACrefYearMonthDay{2020}{}{}.
\newblock
{\BBOQ}\APACrefatitle {Better distractions: Transformer-based distractor generation and multiple choice question filtering} {Better distractions: Transformer-based distractor generation and multiple choice question filtering}.{\BBCQ}
\newblock
\APACjournalVolNumPages{arXiv preprint arXiv:2010.09598}{}{}{}.
\PrintBackRefs{\CurrentBib}

\bibitem [\protect \citeauthoryear {%
Ormerod%
}{%
Ormerod%
}{%
{\protect \APACyear {2022}}%
{\protect \APACexlab {{\protect \BCnt {1}}}}}]{%
ormerod2022mapping}
\APACinsertmetastar {%
ormerod2022mapping}%
\begin{APACrefauthors}%
Ormerod, C\BPBI M.%
\end{APACrefauthors}%
\unskip\
\newblock
\APACrefYearMonthDay{2022{\protect \BCnt {1}}}{}{}.
\newblock
{\BBOQ}\APACrefatitle {Mapping between hidden states and features to validate automated essay scoring using {DeBERTa} models} {Mapping between hidden states and features to validate automated essay scoring using {DeBERTa} models}.{\BBCQ}
\newblock
\APACjournalVolNumPages{Psychological Test and Assessment Modeling}{64}{4}{495--526}.
\PrintBackRefs{\CurrentBib}

\bibitem [\protect \citeauthoryear {%
Ormerod%
}{%
Ormerod%
}{%
{\protect \APACyear {2022}}%
{\protect \APACexlab {{\protect \BCnt {2}}}}}]{%
ormerod_short-answer_2022}
\APACinsertmetastar {%
ormerod_short-answer_2022}%
\begin{APACrefauthors}%
Ormerod, C\BPBI M.%
\end{APACrefauthors}%
\unskip\
\newblock
\APACrefYearMonthDay{2022{\protect \BCnt {2}}}{}{}.
\newblock
{\BBOQ}\APACrefatitle {Short-answer scoring with ensembles of pretrained language models} {Short-answer scoring with ensembles of pretrained language models}.{\BBCQ}
\newblock
\APACjournalVolNumPages{arXiv Preprint}{}{}{}.
\newblock
\begin{APACrefDOI} \doi{10.48550/arXiv.2202.11558} \end{APACrefDOI}
\PrintBackRefs{\CurrentBib}

\bibitem [\protect \citeauthoryear {%
Ormerod%
, Malhotra%
\BCBL {}\ \BBA {} Jafari%
}{%
Ormerod%
\ \protect \BOthers {.}}{%
{\protect \APACyear {2021}}%
}]{%
ormerod2021automated}
\APACinsertmetastar {%
ormerod2021automated}%
\begin{APACrefauthors}%
Ormerod, C\BPBI M.%
, Malhotra, A.%
\BCBL {}\ \BBA {} Jafari, A.%
\end{APACrefauthors}%
\unskip\
\newblock
\APACrefYearMonthDay{2021}{}{}.
\newblock
{\BBOQ}\APACrefatitle {Automated essay scoring using efficient transformer-based language models} {Automated essay scoring using efficient transformer-based language models}.{\BBCQ}
\newblock
\APACjournalVolNumPages{arXiv preprint}{}{}{}.
\newblock
\begin{APACrefDOI} \doi{10.48550/arXiv.2102.13136} \end{APACrefDOI}
\PrintBackRefs{\CurrentBib}

\bibitem [\protect \citeauthoryear {%
Page%
}{%
Page%
}{%
{\protect \APACyear {1966}}%
}]{%
page_imminence_1966}
\APACinsertmetastar {%
page_imminence_1966}%
\begin{APACrefauthors}%
Page, E\BPBI B.%
\end{APACrefauthors}%
\unskip\
\newblock
\APACrefYearMonthDay{1966}{}{}.
\newblock
{\BBOQ}\APACrefatitle {The Imminence of... grading Essays by Computer} {The imminence of... grading essays by computer}.{\BBCQ}
\newblock
\APACjournalVolNumPages{The Phi Delta Kappan}{47}{5}{238--243}.
\newblock
\begin{APACrefURL} \url{https://www.jstor.org/stable/20371545} \end{APACrefURL}
\PrintBackRefs{\CurrentBib}

\bibitem [\protect \citeauthoryear {%
Parasuraman%
\ \BBA {} Riley%
}{%
Parasuraman%
\ \BBA {} Riley%
}{%
{\protect \APACyear {1997}}%
}]{%
parasuraman1997humans}
\APACinsertmetastar {%
parasuraman1997humans}%
\begin{APACrefauthors}%
Parasuraman, R.%
\BCBT {}\ \BBA {} Riley, V.%
\end{APACrefauthors}%
\unskip\
\newblock
\APACrefYearMonthDay{1997}{}{}.
\newblock
{\BBOQ}\APACrefatitle {Humans and automation: {U}se, misuse, disuse, abuse} {Humans and automation: {U}se, misuse, disuse, abuse}.{\BBCQ}
\newblock
\APACjournalVolNumPages{Human Factors}{39}{2}{230--253}.
\newblock
\begin{APACrefDOI} \doi{10.1518/001872097778543886} \end{APACrefDOI}
\PrintBackRefs{\CurrentBib}

\bibitem [\protect \citeauthoryear {%
Penfield%
}{%
Penfield%
}{%
{\protect \APACyear {2016}}%
}]{%
penfield2016fairness}
\APACinsertmetastar {%
penfield2016fairness}%
\begin{APACrefauthors}%
Penfield, R\BPBI D.%
\end{APACrefauthors}%
\unskip\
\newblock
\APACrefYearMonthDay{2016}{}{}.
\newblock
{\BBOQ}\APACrefatitle {Fairness in test scoring} {Fairness in test scoring}.{\BBCQ}
\newblock
\BIn{} N\BPBI J.~Dorans\ \BBA {} L\BPBI L.~Cook\ (\BEDS), \APACrefbtitle {Fairness in educational assessment and measurement} {Fairness in educational assessment and measurement}\ (\BPGS\ 55--75).
\newblock
\APACaddressPublisher{}{Routledge}.
\PrintBackRefs{\CurrentBib}

\bibitem [\protect \citeauthoryear {%
Raczynski%
\ \BBA {} Cohen%
}{%
Raczynski%
\ \BBA {} Cohen%
}{%
{\protect \APACyear {2018}}%
}]{%
raczynski_appraising_2018}
\APACinsertmetastar {%
raczynski_appraising_2018}%
\begin{APACrefauthors}%
Raczynski, K.%
\BCBT {}\ \BBA {} Cohen, A.%
\end{APACrefauthors}%
\unskip\
\newblock
\APACrefYearMonthDay{2018}{}{}.
\newblock
{\BBOQ}\APACrefatitle {Appraising the scoring performance of automated essay scoring systems—{Some} additional considerations: {Which} essays? {Which} human raters? {Which} scores?} {Appraising the scoring performance of automated essay scoring systems—{Some} additional considerations: {Which} essays? {Which} human raters? {Which} scores?}{\BBCQ}
\newblock
\APACjournalVolNumPages{Applied Measurement in Education}{31}{}{233--240}.
\newblock
\begin{APACrefDOI} \doi{10.1080/08957347.2018.1464449} \end{APACrefDOI}
\PrintBackRefs{\CurrentBib}

\bibitem [\protect \citeauthoryear {%
Radford%
, Narasimhan%
, Salimans%
\BCBL {}\ \BBA {} Sutskever%
}{%
Radford%
\ \protect \BOthers {.}}{%
{\protect \APACyear {2018}}%
}]{%
radford2018improving}
\APACinsertmetastar {%
radford2018improving}%
\begin{APACrefauthors}%
Radford, A.%
, Narasimhan, K.%
, Salimans, T.%
\BCBL {}\ \BBA {} Sutskever, I.%
\end{APACrefauthors}%
\unskip\
\newblock
\APACrefYearMonthDay{2018}{}{}.
\newblock
\APACrefbtitle {Improving language understanding by generative pre-training.} {Improving language understanding by generative pre-training.}
\newblock
\begin{APACrefURL} \url{https://s3-us-west-2.amazonaws.com/openai-assets/research-covers/language-unsupervised/language_understanding_paper.pdf} \end{APACrefURL}
\PrintBackRefs{\CurrentBib}

\bibitem [\protect \citeauthoryear {%
Rahm%
}{%
Rahm%
}{%
{\protect \APACyear {2023}}%
}]{%
rahm_education_2023}
\APACinsertmetastar {%
rahm_education_2023}%
\begin{APACrefauthors}%
Rahm, L.%
\end{APACrefauthors}%
\unskip\
\newblock
\APACrefYearMonthDay{2023}{}{}.
\newblock
{\BBOQ}\APACrefatitle {Education, automation and {AI}: a genealogy of alternative futures} {Education, automation and {AI}: a genealogy of alternative futures}.{\BBCQ}
\newblock
\APACjournalVolNumPages{Learning, Media and Technology}{48}{1}{6--24}.
\newblock
\begin{APACrefDOI} \doi{10.1080/17439884.2021.1977948} \end{APACrefDOI}
\PrintBackRefs{\CurrentBib}

\bibitem [\protect \citeauthoryear {%
Ramesh%
\ \BBA {} Sanampudi%
}{%
Ramesh%
\ \BBA {} Sanampudi%
}{%
{\protect \APACyear {2022}}%
}]{%
ramesh_automated_2022}
\APACinsertmetastar {%
ramesh_automated_2022}%
\begin{APACrefauthors}%
Ramesh, D.%
\BCBT {}\ \BBA {} Sanampudi, S\BPBI K.%
\end{APACrefauthors}%
\unskip\
\newblock
\APACrefYearMonthDay{2022}{}{}.
\newblock
{\BBOQ}\APACrefatitle {An automated essay scoring systems: A systematic literature review} {An automated essay scoring systems: A systematic literature review}.{\BBCQ}
\newblock
\APACjournalVolNumPages{Artificial Intelligence Review}{55}{3}{2495--2527}.
\newblock
\begin{APACrefDOI} \doi{10.1007/s10462-021-10068-2} \end{APACrefDOI}
\PrintBackRefs{\CurrentBib}

\bibitem [\protect \citeauthoryear {%
Ranger%
, Schmidt%
\BCBL {}\ \BBA {} Wolgast%
}{%
Ranger%
\ \protect \BOthers {.}}{%
{\protect \APACyear {2023}}%
}]{%
ranger2023detecting}
\APACinsertmetastar {%
ranger2023detecting}%
\begin{APACrefauthors}%
Ranger, J.%
, Schmidt, N.%
\BCBL {}\ \BBA {} Wolgast, A.%
\end{APACrefauthors}%
\unskip\
\newblock
\APACrefYearMonthDay{2023}{}{}.
\newblock
{\BBOQ}\APACrefatitle {Detecting cheating in large-scale assessment: The transfer of detectors to new tests} {Detecting cheating in large-scale assessment: The transfer of detectors to new tests}.{\BBCQ}
\newblock
\APACjournalVolNumPages{Educational and Psychological Measurement}{83}{5}{1033--1058}.
\newblock
\begin{APACrefDOI} \doi{10.1177/00131644221132723} \end{APACrefDOI}
\PrintBackRefs{\CurrentBib}

\bibitem [\protect \citeauthoryear {%
Rao%
}{%
Rao%
}{%
{\protect \APACyear {2015}}%
}]{%
rao_universal_2015}
\APACinsertmetastar {%
rao_universal_2015}%
\begin{APACrefauthors}%
Rao, K.%
\end{APACrefauthors}%
\unskip\
\newblock
\APACrefYearMonthDay{2015}{}{}.
\newblock
{\BBOQ}\APACrefatitle {Universal Design for Learning and Multimedia Technology: {Supporting} Culturally and Linguistically Diverse Students} {Universal design for learning and multimedia technology: {Supporting} culturally and linguistically diverse students}.{\BBCQ}
\newblock
\APACjournalVolNumPages{Journal of Educational Multimedia and Hypermedia}{24}{2}{121--137}.
\newblock
\begin{APACrefURL} \url{http://hdl.handle.net/10125/41065} \end{APACrefURL}
\PrintBackRefs{\CurrentBib}

\bibitem [\protect \citeauthoryear {%
Reiss%
}{%
Reiss%
}{%
{\protect \APACyear {2021}}%
}]{%
reiss_use_2021}
\APACinsertmetastar {%
reiss_use_2021}%
\begin{APACrefauthors}%
Reiss, M\BPBI J.%
\end{APACrefauthors}%
\unskip\
\newblock
\APACrefYearMonthDay{2021}{}{}.
\newblock
{\BBOQ}\APACrefatitle {The Use of {Al} in Education: Practicalities and Ethical Considerations} {The use of {Al} in education: Practicalities and ethical considerations}.{\BBCQ}
\newblock
\APACjournalVolNumPages{London Review of Education}{19}{1}{}.
\PrintBackRefs{\CurrentBib}

\bibitem [\protect \citeauthoryear {%
Riordan%
, Bichler%
, Bradford%
, King~Chen%
\BCBL {}\ \protect \BOthers {.}}{%
Riordan%
, Bichler%
, Bradford%
, King~Chen%
\BCBL {}\ \protect \BOthers {.}}{%
{\protect \APACyear {2020}}%
}]{%
riordan_empirical_2020}
\APACinsertmetastar {%
riordan_empirical_2020}%
\begin{APACrefauthors}%
Riordan, B.%
, Bichler, S.%
, Bradford, A.%
, King~Chen, J.%
, Wiley, K.%
, Gerard, L.%
\BCBL {}\ \BBA {} C.~Linn, M.%
\end{APACrefauthors}%
\unskip\
\newblock
\APACrefYearMonthDay{2020}{{\APACmonth{07}}}{}.
\newblock
{\BBOQ}\APACrefatitle {An empirical investigation of neural methods for content scoring of science explanations} {An empirical investigation of neural methods for content scoring of science explanations}.{\BBCQ}
\newblock
\BIn{} J.~Burstein\ \BOthers {.}\ (\BEDS), \APACrefbtitle {Proceedings of the {Fifteenth} {Workshop} on {Innovative} {Use} of {NLP} for {Building} {Educational} {Applications}} {Proceedings of the {Fifteenth} {Workshop} on {Innovative} {Use} of {NLP} for {Building} {Educational} {Applications}}\ (\BPGS\ 135--144).
\newblock
\APACaddressPublisher{Seattle, WA, USA}{Association for Computational Linguistics}.
\newblock
\begin{APACrefDOI} \doi{10.18653/v1/2020.bea-1.13} \end{APACrefDOI}
\PrintBackRefs{\CurrentBib}

\bibitem [\protect \citeauthoryear {%
Riordan%
, Bichler%
, Bradford%
\BCBL {}\ \BBA {} Linn%
}{%
Riordan%
, Bichler%
, Bradford%
\BCBL {}\ \BBA {} Linn%
}{%
{\protect \APACyear {2020}}%
}]{%
riordan_probing_2020}
\APACinsertmetastar {%
riordan_probing_2020}%
\begin{APACrefauthors}%
Riordan, B.%
, Bichler, S.%
, Bradford, A.%
\BCBL {}\ \BBA {} Linn, M\BPBI C.%
\end{APACrefauthors}%
\unskip\
\newblock
\APACrefYearMonthDay{2020}{}{}.
\newblock
{\BBOQ}\APACrefatitle {Probing Saliency in Short Answer Scoring Models for Science Explanations} {Probing saliency in short answer scoring models for science explanations}.{\BBCQ}
\newblock
\APACjournalVolNumPages{New York Academy of Sciences Natural Language, Dialog and Speech Symposium}{}{}{}.
\PrintBackRefs{\CurrentBib}

\bibitem [\protect \citeauthoryear {%
Rodriguez-Torrealba%
, Garcia-Lopez%
\BCBL {}\ \BBA {} Garcia-Cabot%
}{%
Rodriguez-Torrealba%
\ \protect \BOthers {.}}{%
{\protect \APACyear {2022}}%
}]{%
rodriguez2022}
\APACinsertmetastar {%
rodriguez2022}%
\begin{APACrefauthors}%
Rodriguez-Torrealba, R.%
, Garcia-Lopez, E.%
\BCBL {}\ \BBA {} Garcia-Cabot, A.%
\end{APACrefauthors}%
\unskip\
\newblock
\APACrefYearMonthDay{2022}{}{}.
\newblock
{\BBOQ}\APACrefatitle {End-to-End generation of Multiple-Choice questions using Text-to-Text transfer Transformer models} {End-to-end generation of multiple-choice questions using text-to-text transfer transformer models}.{\BBCQ}
\newblock
\APACjournalVolNumPages{Expert Systems with Applications}{208}{}{118258}.
\newblock
\begin{APACrefDOI} \doi{10.1016/j.eswa.2022.118258} \end{APACrefDOI}
\PrintBackRefs{\CurrentBib}

\bibitem [\protect \citeauthoryear {%
Sankey%
, Birch%
\BCBL {}\ \BBA {} Gardiner%
}{%
Sankey%
\ \protect \BOthers {.}}{%
{\protect \APACyear {2010}}%
}]{%
sankey2010engaging}
\APACinsertmetastar {%
sankey2010engaging}%
\begin{APACrefauthors}%
Sankey, M.%
, Birch, D.%
\BCBL {}\ \BBA {} Gardiner, M\BPBI W.%
\end{APACrefauthors}%
\unskip\
\newblock
\APACrefYearMonthDay{2010}{}{}.
\newblock
{\BBOQ}\APACrefatitle {Engaging students through multimodal learning environments: The journey continues} {Engaging students through multimodal learning environments: The journey continues}.{\BBCQ}
\newblock
\APACjournalVolNumPages{Proceedings of the 27th Australasian Society for Computers in Learning in Tertiary Education}{}{}{852--863}.
\PrintBackRefs{\CurrentBib}

\bibitem [\protect \citeauthoryear {%
S{\"a}uberli%
\ \BBA {} Clematide%
}{%
S{\"a}uberli%
\ \BBA {} Clematide%
}{%
{\protect \APACyear {2024}}%
}]{%
sauberli2024automatic}
\APACinsertmetastar {%
sauberli2024automatic}%
\begin{APACrefauthors}%
S{\"a}uberli, A.%
\BCBT {}\ \BBA {} Clematide, S.%
\end{APACrefauthors}%
\unskip\
\newblock
\APACrefYearMonthDay{2024}{}{}.
\newblock
{\BBOQ}\APACrefatitle {Automatic Generation and Evaluation of Reading Comprehension Test Items with Large Language Models} {Automatic generation and evaluation of reading comprehension test items with large language models}.{\BBCQ}
\newblock
\APACjournalVolNumPages{arXiv preprint arXiv:2404.07720}{}{}{}.
\PrintBackRefs{\CurrentBib}

\bibitem [\protect \citeauthoryear {%
Sayin%
\ \BBA {} Gierl%
}{%
Sayin%
\ \BBA {} Gierl%
}{%
{\protect \APACyear {2024}}%
}]{%
sayin2024using}
\APACinsertmetastar {%
sayin2024using}%
\begin{APACrefauthors}%
Sayin, A.%
\BCBT {}\ \BBA {} Gierl, M.%
\end{APACrefauthors}%
\unskip\
\newblock
\APACrefYearMonthDay{2024}{}{}.
\newblock
{\BBOQ}\APACrefatitle {Using {OpenAI GPT} to Generate Reading Comprehension Items} {Using {OpenAI GPT} to generate reading comprehension items}.{\BBCQ}
\newblock
\APACjournalVolNumPages{Educational Measurement: Issues and Practice}{43}{1}{5--18}.
\newblock
\begin{APACrefDOI} \doi{10.1111/emip.12590} \end{APACrefDOI}
\PrintBackRefs{\CurrentBib}

\bibitem [\protect \citeauthoryear {%
Seger%
, Ovadya%
, Siddarth%
, Garfinkel%
\BCBL {}\ \BBA {} Dafoe%
}{%
Seger%
\ \protect \BOthers {.}}{%
{\protect \APACyear {2023}}%
}]{%
seger2023democratising}
\APACinsertmetastar {%
seger2023democratising}%
\begin{APACrefauthors}%
Seger, E.%
, Ovadya, A.%
, Siddarth, D.%
, Garfinkel, B.%
\BCBL {}\ \BBA {} Dafoe, A.%
\end{APACrefauthors}%
\unskip\
\newblock
\APACrefYearMonthDay{2023}{}{}.
\newblock
{\BBOQ}\APACrefatitle {Democratising AI: Multiple meanings, goals, and methods} {Democratising ai: Multiple meanings, goals, and methods}.{\BBCQ}
\newblock
\BIn{} \APACrefbtitle {{Proceedings of the 2023 {AAAI/ACM} Conference on AI, Ethics, and Society}} {{Proceedings of the 2023 {AAAI/ACM} Conference on AI, Ethics, and Society}}\ (\BPGS\ 715--722).
\PrintBackRefs{\CurrentBib}

\bibitem [\protect \citeauthoryear {%
Sharma%
\ \BBA {} Giannakos%
}{%
Sharma%
\ \BBA {} Giannakos%
}{%
{\protect \APACyear {2020}}%
}]{%
sharma2020multimodal}
\APACinsertmetastar {%
sharma2020multimodal}%
\begin{APACrefauthors}%
Sharma, K.%
\BCBT {}\ \BBA {} Giannakos, M.%
\end{APACrefauthors}%
\unskip\
\newblock
\APACrefYearMonthDay{2020}{}{}.
\newblock
{\BBOQ}\APACrefatitle {Multimodal data capabilities for learning: What can multimodal data tell us about learning?} {Multimodal data capabilities for learning: What can multimodal data tell us about learning?}{\BBCQ}
\newblock
\APACjournalVolNumPages{British Journal of Educational Technology}{51}{5}{1450-1484}.
\newblock
\begin{APACrefDOI} \doi{https://doi.org/10.1111/bjet.12993} \end{APACrefDOI}
\PrintBackRefs{\CurrentBib}

\bibitem [\protect \citeauthoryear {%
Shermis%
}{%
Shermis%
}{%
{\protect \APACyear {2024}}%
}]{%
shermis2024ai}
\APACinsertmetastar {%
shermis2024ai}%
\begin{APACrefauthors}%
Shermis, M\BPBI D.%
\end{APACrefauthors}%
\unskip\
\newblock
\APACrefYearMonthDay{2024}{}{}.
\newblock
{\BBOQ}\APACrefatitle {{AI} Scoring and Writing Fairness} {{AI} scoring and writing fairness}.{\BBCQ}
\newblock
\BIn{} \APACrefbtitle {The {Routledge} International Handbook of Automated Essay Evaluation} {The {Routledge} international handbook of automated essay evaluation}\ (\BPGS\ 386--420).
\newblock
\APACaddressPublisher{}{Routledge}.
\PrintBackRefs{\CurrentBib}

\bibitem [\protect \citeauthoryear {%
Shermis%
, Mao%
, Mulholland%
\BCBL {}\ \BBA {} Kieftenbeld%
}{%
Shermis%
\ \protect \BOthers {.}}{%
{\protect \APACyear {2017}}%
}]{%
shermis2017use}
\APACinsertmetastar {%
shermis2017use}%
\begin{APACrefauthors}%
Shermis, M\BPBI D.%
, Mao, L.%
, Mulholland, M.%
\BCBL {}\ \BBA {} Kieftenbeld, V.%
\end{APACrefauthors}%
\unskip\
\newblock
\APACrefYearMonthDay{2017}{}{}.
\newblock
{\BBOQ}\APACrefatitle {Use of automated scoring features to generate hypotheses regarding language-based {DIF}} {Use of automated scoring features to generate hypotheses regarding language-based {DIF}}.{\BBCQ}
\newblock
\APACjournalVolNumPages{International Journal of Testing}{17}{4}{351--371}.
\newblock
\begin{APACrefDOI} \doi{10.1080/15305058.2017.1308949} \end{APACrefDOI}
\PrintBackRefs{\CurrentBib}

\bibitem [\protect \citeauthoryear {%
Shi%
, Liu%
, Lai%
\BCBL {}\ \BBA {} Jin%
}{%
Shi%
\ \protect \BOthers {.}}{%
{\protect \APACyear {2022}}%
}]{%
shi_enhancing_2022}
\APACinsertmetastar {%
shi_enhancing_2022}%
\begin{APACrefauthors}%
Shi, Z.%
, Liu, F.%
, Lai, C.%
\BCBL {}\ \BBA {} Jin, T.%
\end{APACrefauthors}%
\unskip\
\newblock
\APACrefYearMonthDay{2022}{}{}.
\newblock
{\BBOQ}\APACrefatitle {Enhancing the use of evidence in argumentative writing through collaborative processing of content-based automated writing evaluation feedback} {Enhancing the use of evidence in argumentative writing through collaborative processing of content-based automated writing evaluation feedback}.{\BBCQ}
\newblock
\APACjournalVolNumPages{Language Learning \& Technology}{}{}{}.
\newblock
\APACrefnote{Publisher: University of Hawaii, National Foreign Language Resource Center}
\PrintBackRefs{\CurrentBib}

\bibitem [\protect \citeauthoryear {%
Shneiderman%
}{%
Shneiderman%
}{%
{\protect \APACyear {2022}}%
}]{%
shneiderman2022human}
\APACinsertmetastar {%
shneiderman2022human}%
\begin{APACrefauthors}%
Shneiderman, B.%
\end{APACrefauthors}%
\unskip\
\newblock
\APACrefYear{2022}.
\newblock
\APACrefbtitle {Human-centered {AI}} {Human-centered {AI}}.
\newblock
\APACaddressPublisher{}{Oxford University Press}.
\PrintBackRefs{\CurrentBib}

\bibitem [\protect \citeauthoryear {%
Singh%
, Aggarwal%
, Tiwari%
\BCBL {}\ \BBA {} Joshi%
}{%
Singh%
\ \protect \BOthers {.}}{%
{\protect \APACyear {2022}}%
}]{%
singh_exam_2022}
\APACinsertmetastar {%
singh_exam_2022}%
\begin{APACrefauthors}%
Singh, J.%
, Aggarwal, R.%
, Tiwari, S.%
\BCBL {}\ \BBA {} Joshi, V.%
\end{APACrefauthors}%
\unskip\
\newblock
\APACrefYearMonthDay{2022}{{\APACmonth{10}}}{}.
\newblock
{\BBOQ}\APACrefatitle {Exam Proctoring Classification Using Eye Gaze Detection} {Exam proctoring classification using eye gaze detection}.{\BBCQ}
\newblock
\BIn{} \APACrefbtitle {the 3rd {International} {Conference} on {Smart} {Electronics} and {Communication}} {the 3rd {International} {Conference} on {Smart} {Electronics} and {Communication}}\ (\BPGS\ 371--376).
\newblock
\begin{APACrefDOI} \doi{10.1109/ICOSEC54921.2022.9951987} \end{APACrefDOI}
\PrintBackRefs{\CurrentBib}

\bibitem [\protect \citeauthoryear {%
Slusky%
}{%
Slusky%
}{%
{\protect \APACyear {2020}}%
}]{%
slusky_cybersecurity_2020}
\APACinsertmetastar {%
slusky_cybersecurity_2020}%
\begin{APACrefauthors}%
Slusky, L.%
\end{APACrefauthors}%
\unskip\
\newblock
\APACrefYearMonthDay{2020}{}{}.
\newblock
{\BBOQ}\APACrefatitle {Cybersecurity of Online Proctoring Systems} {Cybersecurity of online proctoring systems}.{\BBCQ}
\newblock
\APACjournalVolNumPages{Journal of International Technology and Information Management}{29}{1}{56--83}.
\newblock
\begin{APACrefDOI} \doi{10.58729/1941-6679.1445} \end{APACrefDOI}
\PrintBackRefs{\CurrentBib}

\bibitem [\protect \citeauthoryear {%
Smith%
, Pacheco%
\BCBL {}\ \BBA {} Khorosheva%
}{%
Smith%
\ \protect \BOthers {.}}{%
{\protect \APACyear {2021}}%
}]{%
smith_emergent_2021}
\APACinsertmetastar {%
smith_emergent_2021}%
\begin{APACrefauthors}%
Smith, B\BPBI E.%
, Pacheco, M\BPBI B.%
\BCBL {}\ \BBA {} Khorosheva, M.%
\end{APACrefauthors}%
\unskip\
\newblock
\APACrefYearMonthDay{2021}{}{}.
\newblock
{\BBOQ}\APACrefatitle {Emergent Bilingual Students and Digital Multimodal Composition: {A} Systematic Review of Research in Secondary Classrooms} {Emergent bilingual students and digital multimodal composition: {A} systematic review of research in secondary classrooms}.{\BBCQ}
\newblock
\APACjournalVolNumPages{Reading Research Quarterly}{56}{1}{33--52}.
\newblock
\begin{APACrefDOI} \doi{10.1002/rrq.298} \end{APACrefDOI}
\PrintBackRefs{\CurrentBib}

\bibitem [\protect \citeauthoryear {%
Soland%
, Kuhfeld%
\BCBL {}\ \BBA {} Rios%
}{%
Soland%
\ \protect \BOthers {.}}{%
{\protect \APACyear {2021}}%
}]{%
soland_comparing_2021}
\APACinsertmetastar {%
soland_comparing_2021}%
\begin{APACrefauthors}%
Soland, J.%
, Kuhfeld, M.%
\BCBL {}\ \BBA {} Rios, J.%
\end{APACrefauthors}%
\unskip\
\newblock
\APACrefYearMonthDay{2021}{}{}.
\newblock
{\BBOQ}\APACrefatitle {Comparing different response time threshold setting methods to detect low effort on a large-scale assessment} {Comparing different response time threshold setting methods to detect low effort on a large-scale assessment}.{\BBCQ}
\newblock
\APACjournalVolNumPages{Large-scale Assessments in Education}{9}{1}{8}.
\newblock
\begin{APACrefDOI} \doi{10.1186/s40536-021-00100-w} \end{APACrefDOI}
\PrintBackRefs{\CurrentBib}

\bibitem [\protect \citeauthoryear {%
Stahl%
\ \BBA {} Karger%
}{%
Stahl%
\ \BBA {} Karger%
}{%
{\protect \APACyear {2016}}%
}]{%
stahl_student_2016}
\APACinsertmetastar {%
stahl_student_2016}%
\begin{APACrefauthors}%
Stahl, W\BPBI M.%
\BCBT {}\ \BBA {} Karger, J.%
\end{APACrefauthors}%
\unskip\
\newblock
\APACrefYearMonthDay{2016}{}{}.
\newblock
{\BBOQ}\APACrefatitle {Student Data Privacy, Digital Learning, and Special Education: {C}hallenges at the Intersection of Policy and Practice} {Student data privacy, digital learning, and special education: {C}hallenges at the intersection of policy and practice}.{\BBCQ}
\newblock
\APACjournalVolNumPages{Journal of Special Education Leadership}{29}{2}{79--88}.
\newblock
\begin{APACrefURL} \url{https://www.learntechlib.org/p/192627} \end{APACrefURL}
\PrintBackRefs{\CurrentBib}

\bibitem [\protect \citeauthoryear {%
Strubell%
, Ganesh%
\BCBL {}\ \BBA {} McCallum%
}{%
Strubell%
\ \protect \BOthers {.}}{%
{\protect \APACyear {2020}}%
}]{%
strubell2020energy}
\APACinsertmetastar {%
strubell2020energy}%
\begin{APACrefauthors}%
Strubell, E.%
, Ganesh, A.%
\BCBL {}\ \BBA {} McCallum, A.%
\end{APACrefauthors}%
\unskip\
\newblock
\APACrefYearMonthDay{2020}{}{}.
\newblock
{\BBOQ}\APACrefatitle {Energy and policy considerations for modern deep learning research} {Energy and policy considerations for modern deep learning research}.{\BBCQ}
\newblock
\BIn{} \APACrefbtitle {{Proceedings of the AAAI conference on artificial intelligence}} {{Proceedings of the AAAI conference on artificial intelligence}}\ (\BVOL~34, \BPGS\ 13693--13696).
\PrintBackRefs{\CurrentBib}

\bibitem [\protect \citeauthoryear {%
Sun%
\ \protect \BOthers {.}}{%
Sun%
\ \protect \BOthers {.}}{%
{\protect \APACyear {2020}}%
}]{%
sun2020mobilebert}
\APACinsertmetastar {%
sun2020mobilebert}%
\begin{APACrefauthors}%
Sun, Z.%
, Yu, H.%
, Song, X.%
, Liu, R.%
, Yang, Y.%
\BCBL {}\ \BBA {} Zhou, D.%
\end{APACrefauthors}%
\unskip\
\newblock
\APACrefYearMonthDay{2020}{}{}.
\newblock
{\BBOQ}\APACrefatitle {MobileBERT: A compact task-agnostic bert for resource-limited devices} {Mobilebert: A compact task-agnostic bert for resource-limited devices}.{\BBCQ}
\newblock
\APACjournalVolNumPages{arXiv preprint arXiv:2004.02984}{}{}{}.
\PrintBackRefs{\CurrentBib}

\bibitem [\protect \citeauthoryear {%
Suresh%
\ \BBA {} Guttag%
}{%
Suresh%
\ \BBA {} Guttag%
}{%
{\protect \APACyear {2021}}%
{\protect \APACexlab {{\protect \BCnt {1}}}}}]{%
suresh_framework_2021}
\APACinsertmetastar {%
suresh_framework_2021}%
\begin{APACrefauthors}%
Suresh, H.%
\BCBT {}\ \BBA {} Guttag, J.%
\end{APACrefauthors}%
\unskip\
\newblock
\APACrefYearMonthDay{2021{\protect \BCnt {1}}}{{\APACmonth{10}}}{}.
\newblock
{\BBOQ}\APACrefatitle {A Framework for Understanding Sources of Harm throughout the Machine Learning Life Cycle} {A framework for understanding sources of harm throughout the machine learning life cycle}.{\BBCQ}
\newblock
\BIn{} \APACrefbtitle {Equity and Access in Algorithms, Mechanisms, and Optimization} {Equity and access in algorithms, mechanisms, and optimization}\ (\BPGS\ 1--9).
\newblock
\APACaddressPublisher{NY, USA}{ACM}.
\newblock
\begin{APACrefDOI} \doi{10.1145/3465416.3483305} \end{APACrefDOI}
\PrintBackRefs{\CurrentBib}

\bibitem [\protect \citeauthoryear {%
Suresh%
\ \BBA {} Guttag%
}{%
Suresh%
\ \BBA {} Guttag%
}{%
{\protect \APACyear {2021}}%
{\protect \APACexlab {{\protect \BCnt {2}}}}}]{%
suresh_understanding_2021}
\APACinsertmetastar {%
suresh_understanding_2021}%
\begin{APACrefauthors}%
Suresh, H.%
\BCBT {}\ \BBA {} Guttag, J.%
\end{APACrefauthors}%
\unskip\
\newblock
\APACrefYearMonthDay{2021{\protect \BCnt {2}}}{}{}.
\newblock
{\BBOQ}\APACrefatitle {Understanding Potential Sources of Harm throughout the Machine Learning Life Cycle} {Understanding potential sources of harm throughout the machine learning life cycle}.{\BBCQ}
\newblock
\APACjournalVolNumPages{MIT Case Studies in Social and Ethical Responsibilities of Computing}{}{Summer 2021}{}.
\newblock
\begin{APACrefDOI} \doi{10.21428/2c646de5.c16a07bb} \end{APACrefDOI}
\PrintBackRefs{\CurrentBib}

\bibitem [\protect \citeauthoryear {%
Taiwo%
, Akinsola%
, Tella%
, Makinde%
\BCBL {}\ \BBA {} Akinwande%
}{%
Taiwo%
\ \protect \BOthers {.}}{%
{\protect \APACyear {2023}}%
}]{%
taiwo_review_2023}
\APACinsertmetastar {%
taiwo_review_2023}%
\begin{APACrefauthors}%
Taiwo, E.%
, Akinsola, A.%
, Tella, E.%
, Makinde, K.%
\BCBL {}\ \BBA {} Akinwande, M.%
\end{APACrefauthors}%
\unskip\
\newblock
\APACrefYearMonthDay{2023}{}{}.
\newblock
\APACrefbtitle {A Review of the Ethics of Artificial Intelligence and its Applications in the {United States}.} {A review of the ethics of artificial intelligence and its applications in the {United States}.}
\newblock
\APACrefnote{arXiv:2310.05751 [cs]}
\newblock
\begin{APACrefDOI} \doi{10.5121/ijci.2023.1206010} \end{APACrefDOI}
\PrintBackRefs{\CurrentBib}

\bibitem [\protect \citeauthoryear {%
Tan%
, Armoush%
, Mazzullo%
, Bulut%
\BCBL {}\ \BBA {} Gierl%
}{%
Tan%
\ \protect \BOthers {.}}{%
{\protect \APACyear {2024}}%
}]{%
tan2024review}
\APACinsertmetastar {%
tan2024review}%
\begin{APACrefauthors}%
Tan, B.%
, Armoush, N.%
, Mazzullo, E.%
, Bulut, O.%
\BCBL {}\ \BBA {} Gierl, M.%
\end{APACrefauthors}%
\unskip\
\newblock
\APACrefYearMonthDay{2024}{}{}.
\newblock
{\BBOQ}\APACrefatitle {A Review of Automatic Item Generation Techniques Leveraging Large Language Models} {A review of automatic item generation techniques leveraging large language models}.{\BBCQ}
\newblock
\APACjournalVolNumPages{EdArXiv Preprints}{}{}{}.
\newblock
\begin{APACrefDOI} \doi{10.35542/osf.io/6d8tj} \end{APACrefDOI}
\PrintBackRefs{\CurrentBib}

\bibitem [\protect \citeauthoryear {%
Tang%
}{%
Tang%
}{%
{\protect \APACyear {2023}}%
}]{%
tang2023latent}
\APACinsertmetastar {%
tang2023latent}%
\begin{APACrefauthors}%
Tang, X.%
\end{APACrefauthors}%
\unskip\
\newblock
\APACrefYearMonthDay{2023}{}{}.
\newblock
{\BBOQ}\APACrefatitle {A Latent Hidden {M}arkov Model for Process Data} {A latent hidden {M}arkov model for process data}.{\BBCQ}
\newblock
\APACjournalVolNumPages{Psychometrika}{89}{1}{1--36}.
\newblock
\begin{APACrefDOI} \doi{10.1007/s11336-023-09938-1} \end{APACrefDOI}
\PrintBackRefs{\CurrentBib}

\bibitem [\protect \citeauthoryear {%
Trivedi%
}{%
Trivedi%
}{%
{\protect \APACyear {2022}}%
}]{%
trivedi_improving_2022}
\APACinsertmetastar {%
trivedi_improving_2022}%
\begin{APACrefauthors}%
Trivedi, S.%
\end{APACrefauthors}%
\unskip\
\newblock
\APACrefYearMonthDay{2022}{}{}.
\newblock
{\BBOQ}\APACrefatitle {Improving Students' Retention Using Machine Learning: Impacts and Implications} {Improving students' retention using machine learning: Impacts and implications}.{\BBCQ}
\newblock
\APACjournalVolNumPages{ScienceOpen Preprints}{}{}{}.
\newblock
\begin{APACrefDOI} \doi{10.14293/S2199-1006.1.SOR-.PPZMB0B.v2} \end{APACrefDOI}
\PrintBackRefs{\CurrentBib}

\bibitem [\protect \citeauthoryear {%
Uto%
, Xie%
\BCBL {}\ \BBA {} Ueno%
}{%
Uto%
\ \protect \BOthers {.}}{%
{\protect \APACyear {2020}}%
}]{%
uto_neural_2020}
\APACinsertmetastar {%
uto_neural_2020}%
\begin{APACrefauthors}%
Uto, M.%
, Xie, Y.%
\BCBL {}\ \BBA {} Ueno, M.%
\end{APACrefauthors}%
\unskip\
\newblock
\APACrefYearMonthDay{2020}{}{}.
\newblock
{\BBOQ}\APACrefatitle {Neural Automated Essay Scoring Incorporating Handcrafted Features} {Neural automated essay scoring incorporating handcrafted features}.{\BBCQ}
\newblock
\BIn{} D.~Scott, N.~Bel\BCBL {}\ \BBA {} C.~Zong\ (\BEDS), \APACrefbtitle {Proceedings of the 28th {International} {Conference} on {Computational} {Linguistics}} {Proceedings of the 28th {International} {Conference} on {Computational} {Linguistics}}\ (\BPGS\ 6077--6088).
\newblock
\APACaddressPublisher{Barcelona, Spain (Online)}{International Committee on Computational Linguistics}.
\newblock
\begin{APACrefDOI} \doi{10.18653/v1/2020.coling-main.535} \end{APACrefDOI}
\PrintBackRefs{\CurrentBib}

\bibitem [\protect \citeauthoryear {%
Verdegem%
}{%
Verdegem%
}{%
{\protect \APACyear {2021}}%
}]{%
verdegem2021ai}
\APACinsertmetastar {%
verdegem2021ai}%
\begin{APACrefauthors}%
Verdegem, P.%
\end{APACrefauthors}%
\unskip\
\newblock
\APACrefYear{2021}.
\newblock
\APACrefbtitle {{AI} for Everyone?: Critical Perspectives} {{AI} for everyone?: Critical perspectives}.
\newblock
\APACaddressPublisher{}{University of Westminster Press}.
\PrintBackRefs{\CurrentBib}

\bibitem [\protect \citeauthoryear {%
Vincent-Lancrin%
\ \BBA {} Vlies%
}{%
Vincent-Lancrin%
\ \BBA {} Vlies%
}{%
{\protect \APACyear {2020}}%
}]{%
vincent-lancrin_trustworthy_2020}
\APACinsertmetastar {%
vincent-lancrin_trustworthy_2020}%
\begin{APACrefauthors}%
Vincent-Lancrin, S.%
\BCBT {}\ \BBA {} Vlies, R\BPBI v\BPBI d.%
\end{APACrefauthors}%
\unskip\
\newblock
\APACrefYearMonthDay{2020}{}{}.
\newblock
\APACrefbtitle {Trustworthy artificial intelligence ({AI}) in education: {Promises} and challenges} {Trustworthy artificial intelligence ({AI}) in education: {Promises} and challenges}\ \APACbVolEdTR{}{\BTR{}}.
\newblock
\APACaddressInstitution{Paris}{OECD}.
\newblock
\begin{APACrefDOI} \doi{10.1787/a6c90fa9-en} \end{APACrefDOI}
\PrintBackRefs{\CurrentBib}

\bibitem [\protect \citeauthoryear {%
Vo%
, Rickels%
, Welch%
\BCBL {}\ \BBA {} Dunbar%
}{%
Vo%
\ \protect \BOthers {.}}{%
{\protect \APACyear {2023}}%
}]{%
vo2023human}
\APACinsertmetastar {%
vo2023human}%
\begin{APACrefauthors}%
Vo, Y.%
, Rickels, H.%
, Welch, C.%
\BCBL {}\ \BBA {} Dunbar, S.%
\end{APACrefauthors}%
\unskip\
\newblock
\APACrefYearMonthDay{2023}{}{}.
\newblock
{\BBOQ}\APACrefatitle {Human scoring versus automated scoring for {E}nglish learners in a statewide evidence-based writing assessment} {Human scoring versus automated scoring for {E}nglish learners in a statewide evidence-based writing assessment}.{\BBCQ}
\newblock
\APACjournalVolNumPages{Assessing Writing}{56}{}{100719}.
\newblock
\begin{APACrefDOI} \doi{10.1016/j.asw.2023.100719} \end{APACrefDOI}
\PrintBackRefs{\CurrentBib}

\bibitem [\protect \citeauthoryear {%
Wan%
\ \BBA {} Keller%
}{%
Wan%
\ \BBA {} Keller%
}{%
{\protect \APACyear {2023}}%
}]{%
wan_using_2023}
\APACinsertmetastar {%
wan_using_2023}%
\begin{APACrefauthors}%
Wan, S.%
\BCBT {}\ \BBA {} Keller, L\BPBI A.%
\end{APACrefauthors}%
\unskip\
\newblock
\APACrefYearMonthDay{2023}{}{}.
\newblock
{\BBOQ}\APACrefatitle {Using Cumulative Sum Control Chart to Detect Aberrant Responses in Educational Assessments} {Using cumulative sum control chart to detect aberrant responses in educational assessments}.{\BBCQ}
\newblock
\APACjournalVolNumPages{Practical Assessment, Research \& Evaluation}{28}{}{}.
\newblock
\begin{APACrefURL} [{2024-05-22}]\url{https://eric.ed.gov/?id=EJ1380532} \end{APACrefURL}
\PrintBackRefs{\CurrentBib}

\bibitem [\protect \citeauthoryear {%
Wang%
, Xu%
, Shang%
\BCBL {}\ \BBA {} Kuncel%
}{%
Wang%
\ \protect \BOthers {.}}{%
{\protect \APACyear {2018}}%
}]{%
wang2018detecting}
\APACinsertmetastar {%
wang2018detecting}%
\begin{APACrefauthors}%
Wang, C.%
, Xu, G.%
, Shang, Z.%
\BCBL {}\ \BBA {} Kuncel, N.%
\end{APACrefauthors}%
\unskip\
\newblock
\APACrefYearMonthDay{2018}{}{}.
\newblock
{\BBOQ}\APACrefatitle {Detecting aberrant behavior and item preknowledge: A comparison of mixture modeling method and residual method} {Detecting aberrant behavior and item preknowledge: A comparison of mixture modeling method and residual method}.{\BBCQ}
\newblock
\APACjournalVolNumPages{Journal of Educational and Behavioral Statistics}{43}{4}{469--501}.
\newblock
\begin{APACrefDOI} \doi{10.3102/1076998618767123} \end{APACrefDOI}
\PrintBackRefs{\CurrentBib}

\bibitem [\protect \citeauthoryear {%
Ware%
}{%
Ware%
}{%
{\protect \APACyear {2014}}%
}]{%
ware_feedback_2014}
\APACinsertmetastar {%
ware_feedback_2014}%
\begin{APACrefauthors}%
Ware, P.%
\end{APACrefauthors}%
\unskip\
\newblock
\APACrefYearMonthDay{2014}{}{}.
\newblock
{\BBOQ}\APACrefatitle {Feedback for Adolescent Writers in the English Classroom} {Feedback for adolescent writers in the english classroom}.{\BBCQ}
\newblock
\APACjournalVolNumPages{Writing \& Pedagogy}{6}{2}{223--249}.
\newblock
\begin{APACrefDOI} \doi{10.1558/wap.v6i2.223} \end{APACrefDOI}
\PrintBackRefs{\CurrentBib}

\bibitem [\protect \citeauthoryear {%
Warschauer%
\ \BBA {} Grimes%
}{%
Warschauer%
\ \BBA {} Grimes%
}{%
{\protect \APACyear {2008}}%
}]{%
warschauer_automated_2008}
\APACinsertmetastar {%
warschauer_automated_2008}%
\begin{APACrefauthors}%
Warschauer, M.%
\BCBT {}\ \BBA {} Grimes, D.%
\end{APACrefauthors}%
\unskip\
\newblock
\APACrefYearMonthDay{2008}{}{}.
\newblock
{\BBOQ}\APACrefatitle {Automated Writing Assessment in the Classroom} {Automated writing assessment in the classroom}.{\BBCQ}
\newblock
\APACjournalVolNumPages{Pedagogies: An International Journal}{3}{1}{22--36}.
\newblock
\begin{APACrefDOI} \doi{10.1080/15544800701771580} \end{APACrefDOI}
\PrintBackRefs{\CurrentBib}

\bibitem [\protect \citeauthoryear {%
Warschauer%
\ \BBA {} Ware%
}{%
Warschauer%
\ \BBA {} Ware%
}{%
{\protect \APACyear {2006}}%
}]{%
warschauer2006automated}
\APACinsertmetastar {%
warschauer2006automated}%
\begin{APACrefauthors}%
Warschauer, M.%
\BCBT {}\ \BBA {} Ware, P.%
\end{APACrefauthors}%
\unskip\
\newblock
\APACrefYearMonthDay{2006}{}{}.
\newblock
{\BBOQ}\APACrefatitle {Automated writing evaluation: Defining the classroom research agenda} {Automated writing evaluation: Defining the classroom research agenda}.{\BBCQ}
\newblock
\APACjournalVolNumPages{Language Teaching Research}{10}{2}{157--180}.
\newblock
\begin{APACrefDOI} \doi{10.1191/1362168806lr190oa} \end{APACrefDOI}
\PrintBackRefs{\CurrentBib}

\bibitem [\protect \citeauthoryear {%
Webber%
}{%
Webber%
}{%
{\protect \APACyear {2019}}%
}]{%
webber_use_2019}
\APACinsertmetastar {%
webber_use_2019}%
\begin{APACrefauthors}%
Webber, K\BPBI L.%
\end{APACrefauthors}%
\unskip\
\newblock
\APACrefYearMonthDay{2019}{}{}.
\newblock
{\BBOQ}\APACrefatitle {The Use and Potential Misuse of Data in Higher Education: {A} Compilation of Examples} {The use and potential misuse of data in higher education: {A} compilation of examples}.{\BBCQ}
\newblock
\APACjournalVolNumPages{IHE Research in Progress Series 2019-001}{}{}{}.
\newblock
\begin{APACrefURL} \url{https://ihe.uga.edu/sites/default/files/inline-files/Webber_2019001_paper.pdf} \end{APACrefURL}
\PrintBackRefs{\CurrentBib}

\bibitem [\protect \citeauthoryear {%
Weiss%
\ \BBA {} Kingsbury%
}{%
Weiss%
\ \BBA {} Kingsbury%
}{%
{\protect \APACyear {1984}}%
}]{%
weiss1984application}
\APACinsertmetastar {%
weiss1984application}%
\begin{APACrefauthors}%
Weiss, D\BPBI J.%
\BCBT {}\ \BBA {} Kingsbury, G\BPBI G.%
\end{APACrefauthors}%
\unskip\
\newblock
\APACrefYearMonthDay{1984}{}{}.
\newblock
{\BBOQ}\APACrefatitle {Application of computerized adaptive testing to educational problems} {Application of computerized adaptive testing to educational problems}.{\BBCQ}
\newblock
\APACjournalVolNumPages{Journal of educational measurement}{21}{4}{361--375}.
\newblock
\begin{APACrefDOI} \doi{10.1111/j.1745-3984.1984.tb01040.x} \end{APACrefDOI}
\PrintBackRefs{\CurrentBib}

\bibitem [\protect \citeauthoryear {%
Whitmer%
\ \protect \BOthers {.}}{%
Whitmer%
\ \protect \BOthers {.}}{%
{\protect \APACyear {2023}}%
}]{%
whitmer_results_2023}
\APACinsertmetastar {%
whitmer_results_2023}%
\begin{APACrefauthors}%
Whitmer, J.%
, Deng, E\BPBI Y.%
, Blankenship, C.%
, Beiting-Parrish, M.%
, Zhang, T.%
\BCBL {}\ \BBA {} Bailey, P.%
\end{APACrefauthors}%
\unskip\
\newblock
\APACrefYearMonthDay{2023}{}{}.
\newblock
\APACrefbtitle {Results of {NAEP} Reading Item Automated Scoring Data Challenge ({Fall} 2021).} {Results of {NAEP} reading item automated scoring data challenge ({Fall} 2021).}
\newblock
\APACaddressPublisher{}{OSF}.
\newblock
\begin{APACrefURL} \url{https://osf.io/2hevq} \end{APACrefURL}
\PrintBackRefs{\CurrentBib}

\bibitem [\protect \citeauthoryear {%
B.~Williamson%
, Macgilchrist%
\BCBL {}\ \BBA {} Potter%
}{%
B.~Williamson%
\ \protect \BOthers {.}}{%
{\protect \APACyear {2023}}%
}]{%
williamson_re-examining_2023}
\APACinsertmetastar {%
williamson_re-examining_2023}%
\begin{APACrefauthors}%
Williamson, B.%
, Macgilchrist, F.%
\BCBL {}\ \BBA {} Potter, J.%
\end{APACrefauthors}%
\unskip\
\newblock
\APACrefYearMonthDay{2023}{}{}.
\newblock
{\BBOQ}\APACrefatitle {Re-examining {AI}, automation and datafication in education} {Re-examining {AI}, automation and datafication in education}.{\BBCQ}
\newblock
\APACjournalVolNumPages{Learning, Media and Technology}{48}{1}{1--5}.
\newblock
\begin{APACrefDOI} \doi{10.1080/17439884.2023.2167830} \end{APACrefDOI}
\PrintBackRefs{\CurrentBib}

\bibitem [\protect \citeauthoryear {%
D\BPBI M.~Williamson%
, Xi%
\BCBL {}\ \BBA {} Breyer%
}{%
D\BPBI M.~Williamson%
\ \protect \BOthers {.}}{%
{\protect \APACyear {2012}}%
}]{%
williamson_framework_2012}
\APACinsertmetastar {%
williamson_framework_2012}%
\begin{APACrefauthors}%
Williamson, D\BPBI M.%
, Xi, X.%
\BCBL {}\ \BBA {} Breyer, F\BPBI J.%
\end{APACrefauthors}%
\unskip\
\newblock
\APACrefYearMonthDay{2012}{}{}.
\newblock
{\BBOQ}\APACrefatitle {A Framework for Evaluation and Use of Automated Scoring} {A framework for evaluation and use of automated scoring}.{\BBCQ}
\newblock
\APACjournalVolNumPages{Educational Measurement: Issues and Practice}{31}{1}{2--13}.
\newblock
\begin{APACrefURL} [{2024-05-22}]\url{https://onlinelibrary.wiley.com/doi/abs/10.1111/j.1745-3992.2011.00223.x} \end{APACrefURL}
\newblock
\begin{APACrefDOI} \doi{10.1111/j.1745-3992.2011.00223.x} \end{APACrefDOI}
\PrintBackRefs{\CurrentBib}

\bibitem [\protect \citeauthoryear {%
Wilson%
\ \protect \BOthers {.}}{%
Wilson%
\ \protect \BOthers {.}}{%
{\protect \APACyear {2021}}%
}]{%
wilson_elementary_2021}
\APACinsertmetastar {%
wilson_elementary_2021}%
\begin{APACrefauthors}%
Wilson, J.%
, Ahrendt, C.%
, Fudge, E\BPBI A.%
, Raiche, A.%
, Beard, G.%
\BCBL {}\ \BBA {} MacArthur, C.%
\end{APACrefauthors}%
\unskip\
\newblock
\APACrefYearMonthDay{2021}{}{}.
\newblock
{\BBOQ}\APACrefatitle {Elementary teachers’ perceptions of automated feedback and automated scoring: {Transforming} the teaching and learning of writing using automated writing evaluation} {Elementary teachers’ perceptions of automated feedback and automated scoring: {Transforming} the teaching and learning of writing using automated writing evaluation}.{\BBCQ}
\newblock
\APACjournalVolNumPages{Computers \& Education}{168}{}{104208}.
\newblock
\begin{APACrefDOI} \doi{10.1016/j.compedu.2021.104208} \end{APACrefDOI}
\PrintBackRefs{\CurrentBib}

\bibitem [\protect \citeauthoryear {%
Wilson%
, Chen%
, Sandbank%
\BCBL {}\ \BBA {} Hebert%
}{%
Wilson%
\ \protect \BOthers {.}}{%
{\protect \APACyear {2019}}%
}]{%
wilson2019generalizability}
\APACinsertmetastar {%
wilson2019generalizability}%
\begin{APACrefauthors}%
Wilson, J.%
, Chen, D.%
, Sandbank, M\BPBI P.%
\BCBL {}\ \BBA {} Hebert, M.%
\end{APACrefauthors}%
\unskip\
\newblock
\APACrefYearMonthDay{2019}{}{}.
\newblock
{\BBOQ}\APACrefatitle {Generalizability of automated scores of writing quality in Grades 3--5.} {Generalizability of automated scores of writing quality in grades 3--5.}{\BBCQ}
\newblock
\APACjournalVolNumPages{Journal of Educational Psychology}{111}{4}{619--640}.
\newblock
\begin{APACrefDOI} \doi{10.1037/edu0000311} \end{APACrefDOI}
\PrintBackRefs{\CurrentBib}

\bibitem [\protect \citeauthoryear {%
Wilson%
\ \BBA {} Czik%
}{%
Wilson%
\ \BBA {} Czik%
}{%
{\protect \APACyear {2016}}%
}]{%
wilson_automated_2016}
\APACinsertmetastar {%
wilson_automated_2016}%
\begin{APACrefauthors}%
Wilson, J.%
\BCBT {}\ \BBA {} Czik, A.%
\end{APACrefauthors}%
\unskip\
\newblock
\APACrefYearMonthDay{2016}{}{}.
\newblock
{\BBOQ}\APACrefatitle {Automated essay evaluation software in {English} Language Arts classrooms: {Effects} on teacher feedback, student motivation, and writing quality} {Automated essay evaluation software in {English} language arts classrooms: {Effects} on teacher feedback, student motivation, and writing quality}.{\BBCQ}
\newblock
\APACjournalVolNumPages{Computers \& Education}{100}{}{94--109}.
\newblock
\begin{APACrefDOI} \doi{10.1016/j.compedu.2016.05.004} \end{APACrefDOI}
\PrintBackRefs{\CurrentBib}

\bibitem [\protect \citeauthoryear {%
Wilson%
, Palermo%
\BCBL {}\ \BBA {} Wibowo%
}{%
Wilson%
, Palermo%
\BCBL {}\ \BBA {} Wibowo%
}{%
{\protect \APACyear {2024}}%
}]{%
wilson_elementary_2024}
\APACinsertmetastar {%
wilson_elementary_2024}%
\begin{APACrefauthors}%
Wilson, J.%
, Palermo, C.%
\BCBL {}\ \BBA {} Wibowo, A.%
\end{APACrefauthors}%
\unskip\
\newblock
\APACrefYearMonthDay{2024}{}{}.
\newblock
{\BBOQ}\APACrefatitle {Elementary {English} learners’ engagement with automated feedback} {Elementary {English} learners’ engagement with automated feedback}.{\BBCQ}
\newblock
\APACjournalVolNumPages{Learning and Instruction}{91}{}{101890}.
\newblock
\begin{APACrefDOI} \doi{10.1016/j.learninstruc.2024.101890} \end{APACrefDOI}
\PrintBackRefs{\CurrentBib}

\bibitem [\protect \citeauthoryear {%
Wilson%
\ \BBA {} Roscoe%
}{%
Wilson%
\ \BBA {} Roscoe%
}{%
{\protect \APACyear {2020}}%
}]{%
wilson2020automated}
\APACinsertmetastar {%
wilson2020automated}%
\begin{APACrefauthors}%
Wilson, J.%
\BCBT {}\ \BBA {} Roscoe, R\BPBI D.%
\end{APACrefauthors}%
\unskip\
\newblock
\APACrefYearMonthDay{2020}{}{}.
\newblock
{\BBOQ}\APACrefatitle {Automated writing evaluation and feedback: Multiple metrics of efficacy} {Automated writing evaluation and feedback: Multiple metrics of efficacy}.{\BBCQ}
\newblock
\APACjournalVolNumPages{Journal of Educational Computing Research}{58}{1}{87--125}.
\newblock
\begin{APACrefDOI} \doi{10.1177/0735633119830764} \end{APACrefDOI}
\PrintBackRefs{\CurrentBib}

\bibitem [\protect \citeauthoryear {%
Wilson%
, Zhang%
\BCBL {}\ \protect \BOthers {.}}{%
Wilson%
, Zhang%
\BCBL {}\ \protect \BOthers {.}}{%
{\protect \APACyear {2024}}%
}]{%
wilson_predictors_2024}
\APACinsertmetastar {%
wilson_predictors_2024}%
\begin{APACrefauthors}%
Wilson, J.%
, Zhang, F.%
, Palermo, C.%
, Cordero, T\BPBI C.%
, Myers, M\BPBI C.%
, Eacker, H.%
\BDBL {}Coles, J.%
\end{APACrefauthors}%
\unskip\
\newblock
\APACrefYearMonthDay{2024}{}{}.
\newblock
{\BBOQ}\APACrefatitle {Predictors of middle school students’ perceptions of automated writing evaluation} {Predictors of middle school students’ perceptions of automated writing evaluation}.{\BBCQ}
\newblock
\APACjournalVolNumPages{Computers \& Education}{211}{}{104985}.
\newblock
\begin{APACrefDOI} \doi{10.1016/j.compedu.2023.104985} \end{APACrefDOI}
\PrintBackRefs{\CurrentBib}

\bibitem [\protect \citeauthoryear {%
Wind%
, Wolfe%
, Engelhard~Jr.%
, Foltz%
\BCBL {}\ \BBA {} Rosenstein%
}{%
Wind%
\ \protect \BOthers {.}}{%
{\protect \APACyear {2018}}%
}]{%
wind_influence_2018}
\APACinsertmetastar {%
wind_influence_2018}%
\begin{APACrefauthors}%
Wind, S\BPBI A.%
, Wolfe, E\BPBI W.%
, Engelhard~Jr., G.%
, Foltz, P.%
\BCBL {}\ \BBA {} Rosenstein, M.%
\end{APACrefauthors}%
\unskip\
\newblock
\APACrefYearMonthDay{2018}{}{}.
\newblock
{\BBOQ}\APACrefatitle {The Influence of Rater Effects in Training Sets on the Psychometric Quality of Automated Scoring for Writing Assessments} {The influence of rater effects in training sets on the psychometric quality of automated scoring for writing assessments}.{\BBCQ}
\newblock
\APACjournalVolNumPages{International Journal of Testing}{18}{1}{27--49}.
\newblock
\begin{APACrefDOI} \doi{10.1080/15305058.2017.1361426} \end{APACrefDOI}
\PrintBackRefs{\CurrentBib}

\bibitem [\protect \citeauthoryear {%
Wise%
\ \BBA {} Kong%
}{%
Wise%
\ \BBA {} Kong%
}{%
{\protect \APACyear {2005}}%
}]{%
wise_response_2005}
\APACinsertmetastar {%
wise_response_2005}%
\begin{APACrefauthors}%
Wise, S\BPBI L.%
\BCBT {}\ \BBA {} Kong, X.%
\end{APACrefauthors}%
\unskip\
\newblock
\APACrefYearMonthDay{2005}{}{}.
\newblock
{\BBOQ}\APACrefatitle {Response Time Effort: {A} New Measure of Examinee Motivation in Computer-Based Tests} {Response time effort: {A} new measure of examinee motivation in computer-based tests}.{\BBCQ}
\newblock
\APACjournalVolNumPages{Applied Measurement in Education}{18}{2}{163--183}.
\newblock
\begin{APACrefDOI} \doi{10.1207/s15324818ame1802_2} \end{APACrefDOI}
\PrintBackRefs{\CurrentBib}

\bibitem [\protect \citeauthoryear {%
Wolf%
\ \protect \BOthers {.}}{%
Wolf%
\ \protect \BOthers {.}}{%
{\protect \APACyear {2019}}%
}]{%
wolf2019huggingface}
\APACinsertmetastar {%
wolf2019huggingface}%
\begin{APACrefauthors}%
Wolf, T.%
, Debut, L.%
, Sanh, V.%
, Chaumond, J.%
, Delangue, C.%
, Moi, A.%
\BDBL {}others%
\end{APACrefauthors}%
\unskip\
\newblock
\APACrefYearMonthDay{2019}{}{}.
\newblock
{\BBOQ}\APACrefatitle {Huggingface's transformers: State-of-the-art natural language processing} {Huggingface's transformers: State-of-the-art natural language processing}.{\BBCQ}
\newblock
\APACjournalVolNumPages{arXiv Preprint}{}{}{}.
\newblock
\begin{APACrefDOI} \doi{10.48550/arXiv.1910.03771} \end{APACrefDOI}
\PrintBackRefs{\CurrentBib}

\bibitem [\protect \citeauthoryear {%
Wongvorachan%
, Lai%
, Bulut%
, Tsai%
\BCBL {}\ \BBA {} Chen%
}{%
Wongvorachan%
\ \protect \BOthers {.}}{%
{\protect \APACyear {2022}}%
}]{%
wongvorachan_artificial_2022}
\APACinsertmetastar {%
wongvorachan_artificial_2022}%
\begin{APACrefauthors}%
Wongvorachan, T.%
, Lai, K\BPBI W.%
, Bulut, O.%
, Tsai, Y\BHBI S.%
\BCBL {}\ \BBA {} Chen, G.%
\end{APACrefauthors}%
\unskip\
\newblock
\APACrefYearMonthDay{2022}{}{}.
\newblock
{\BBOQ}\APACrefatitle {Artificial Intelligence: Transforming the Future of Feedback in Education} {Artificial intelligence: Transforming the future of feedback in education}.{\BBCQ}
\newblock
\APACjournalVolNumPages{Journal of Applied Testing Technology}{}{}{95--116}.
\newblock
\begin{APACrefURL} \url{http://jattjournal.net/index.php/atp/article/view/170387} \end{APACrefURL}
\PrintBackRefs{\CurrentBib}

\bibitem [\protect \citeauthoryear {%
Yildirim-Erbasli%
\ \BBA {} Bulut%
}{%
Yildirim-Erbasli%
\ \BBA {} Bulut%
}{%
{\protect \APACyear {2022}}%
}]{%
yildirim-erbasli_designing_2022}
\APACinsertmetastar {%
yildirim-erbasli_designing_2022}%
\begin{APACrefauthors}%
Yildirim-Erbasli, S\BPBI N.%
\BCBT {}\ \BBA {} Bulut, O.%
\end{APACrefauthors}%
\unskip\
\newblock
\APACrefYearMonthDay{2022}{}{}.
\newblock
{\BBOQ}\APACrefatitle {Designing Predictive Models for Early Prediction of Students' Test-taking Engagement in Computerized Formative Assessments} {Designing predictive models for early prediction of students' test-taking engagement in computerized formative assessments}.{\BBCQ}
\newblock
\APACjournalVolNumPages{Journal of Applied Testing Technology}{}{}{}.
\newblock
\begin{APACrefURL} [{2024-05-22}]\url{http://jattjournal.net/index.php/atp/article/view/167548} \end{APACrefURL}
\PrintBackRefs{\CurrentBib}

\bibitem [\protect \citeauthoryear {%
Yildirim-Erbasli%
\ \BBA {} Bulut%
}{%
Yildirim-Erbasli%
\ \BBA {} Bulut%
}{%
{\protect \APACyear {2023}}%
}]{%
yildirim-erbasli_conversation-based_2023}
\APACinsertmetastar {%
yildirim-erbasli_conversation-based_2023}%
\begin{APACrefauthors}%
Yildirim-Erbasli, S\BPBI N.%
\BCBT {}\ \BBA {} Bulut, O.%
\end{APACrefauthors}%
\unskip\
\newblock
\APACrefYearMonthDay{2023}{}{}.
\newblock
{\BBOQ}\APACrefatitle {Conversation-based assessment: {A} novel approach to boosting test-taking effort in digital formative assessment} {Conversation-based assessment: {A} novel approach to boosting test-taking effort in digital formative assessment}.{\BBCQ}
\newblock
\APACjournalVolNumPages{Computers and Education: Artificial Intelligence}{4}{}{100135}.
\newblock
\begin{APACrefDOI} \doi{10.1016/j.caeai.2023.100135} \end{APACrefDOI}
\PrintBackRefs{\CurrentBib}

\bibitem [\protect \citeauthoryear {%
Yildirim-Erbasli%
, Bulut%
, Epp%
\BCBL {}\ \BBA {} Cui%
}{%
Yildirim-Erbasli%
, Bulut%
\BCBL {}\ \protect \BOthers {.}}{%
{\protect \APACyear {2023}}%
}]{%
conversation2023}
\APACinsertmetastar {%
conversation2023}%
\begin{APACrefauthors}%
Yildirim-Erbasli, S\BPBI N.%
, Bulut, O.%
, Epp, C\BPBI D.%
\BCBL {}\ \BBA {} Cui, Y.%
\end{APACrefauthors}%
\unskip\
\newblock
\APACrefYearMonthDay{2023}{}{}.
\newblock
{\BBOQ}\APACrefatitle {Conversation-Based Assessments in Education: Design, Implementation, and Cognitive Walkthroughs for Usability Testing} {Conversation-based assessments in education: Design, implementation, and cognitive walkthroughs for usability testing}.{\BBCQ}
\newblock
\APACjournalVolNumPages{Journal of Educational Technology Systems}{52}{1}{27-51}.
\newblock
\begin{APACrefDOI} \doi{10.1177/00472395231178943} \end{APACrefDOI}
\PrintBackRefs{\CurrentBib}

\bibitem [\protect \citeauthoryear {%
Yildirim-Erbasli%
, Gorgun%
\BCBL {}\ \BBA {} Bulut%
}{%
Yildirim-Erbasli%
, Gorgun%
\BCBL {}\ \BBA {} Bulut%
}{%
{\protect \APACyear {2023}}%
}]{%
selfregulated2024}
\APACinsertmetastar {%
selfregulated2024}%
\begin{APACrefauthors}%
Yildirim-Erbasli, S\BPBI N.%
, Gorgun, G.%
\BCBL {}\ \BBA {} Bulut, O.%
\end{APACrefauthors}%
\unskip\
\newblock
\APACrefYearMonthDay{2023}{}{}.
\newblock
{\BBOQ}\APACrefatitle {Enhancing self-regulated learning with artificial intelligence-powered learning analytics} {Enhancing self-regulated learning with artificial intelligence-powered learning analytics}.{\BBCQ}
\newblock
\BIn{} N.~Kavaklı~Ulutaş\ \BBA {} D.~Höl\ (\BEDS), \APACrefbtitle {Advances in Early Childhood and {K-12} Education} {Advances in early childhood and {K-12} education}\ (\BPGS\ 57--83).
\newblock
\APACaddressPublisher{}{IGI Global}.
\newblock
\begin{APACrefDOI} \doi{10.4018/979-8-3693-0066-4.ch004} \end{APACrefDOI}
\PrintBackRefs{\CurrentBib}

\bibitem [\protect \citeauthoryear {%
Yoder-Himes%
\ \protect \BOthers {.}}{%
Yoder-Himes%
\ \protect \BOthers {.}}{%
{\protect \APACyear {2022}}%
}]{%
yoder2022racial}
\APACinsertmetastar {%
yoder2022racial}%
\begin{APACrefauthors}%
Yoder-Himes, D\BPBI R.%
, Asif, A.%
, Kinney, K.%
, Brandt, T\BPBI J.%
, Cecil, R\BPBI E.%
, Himes, P\BPBI R.%
\BDBL {}Ross, E.%
\end{APACrefauthors}%
\unskip\
\newblock
\APACrefYearMonthDay{2022}{}{}.
\newblock
{\BBOQ}\APACrefatitle {Racial, skin tone, and sex disparities in automated proctoring software} {Racial, skin tone, and sex disparities in automated proctoring software}.{\BBCQ}
\newblock
\APACjournalVolNumPages{Frontiers in Education}{7}{881449}{}.
\newblock
\begin{APACrefDOI} \doi{10.3389/feduc.2022.881449} \end{APACrefDOI}
\PrintBackRefs{\CurrentBib}

\bibitem [\protect \citeauthoryear {%
Zenisky%
, Hambleton%
\BCBL {}\ \BBA {} Luecht%
}{%
Zenisky%
\ \protect \BOthers {.}}{%
{\protect \APACyear {2009}}%
}]{%
zenisky2009multistage}
\APACinsertmetastar {%
zenisky2009multistage}%
\begin{APACrefauthors}%
Zenisky, A.%
, Hambleton, R\BPBI K.%
\BCBL {}\ \BBA {} Luecht, R\BPBI M.%
\end{APACrefauthors}%
\unskip\
\newblock
\APACrefYearMonthDay{2009}{}{}.
\newblock
{\BBOQ}\APACrefatitle {Multistage testing: Issues, designs, and research} {Multistage testing: Issues, designs, and research}.{\BBCQ}
\newblock
\BIn{} \APACrefbtitle {Elements of adaptive testing} {Elements of adaptive testing}\ (\BPGS\ 355--372).
\newblock
\APACaddressPublisher{}{Springer}.
\PrintBackRefs{\CurrentBib}

\bibitem [\protect \citeauthoryear {%
M.~Zhang%
}{%
M.~Zhang%
}{%
{\protect \APACyear {2013}}%
}]{%
zhang_contrasting_2013}
\APACinsertmetastar {%
zhang_contrasting_2013}%
\begin{APACrefauthors}%
Zhang, M.%
\end{APACrefauthors}%
\unskip\
\newblock
\APACrefYearMonthDay{2013}{}{}.
\newblock
{\BBOQ}\APACrefatitle {Contrasting automated and human scoring of essays} {Contrasting automated and human scoring of essays}.{\BBCQ}
\newblock
\APACjournalVolNumPages{R \& D Connections}{21}{2}{1--11}.
\PrintBackRefs{\CurrentBib}

\bibitem [\protect \citeauthoryear {%
M.~Zhang%
, Dorans%
, Li%
\BCBL {}\ \BBA {} Rupp%
}{%
M.~Zhang%
\ \protect \BOthers {.}}{%
{\protect \APACyear {2017}}%
}]{%
zhang2017differential}
\APACinsertmetastar {%
zhang2017differential}%
\begin{APACrefauthors}%
Zhang, M.%
, Dorans, N.%
, Li, C.%
\BCBL {}\ \BBA {} Rupp, A.%
\end{APACrefauthors}%
\unskip\
\newblock
\APACrefYearMonthDay{2017}{}{}.
\newblock
{\BBOQ}\APACrefatitle {Differential feature functioning in automated essay scoring} {Differential feature functioning in automated essay scoring}.{\BBCQ}
\newblock
\BIn{} H.~Jiao\ \BBA {} R\BPBI W.~Lissitz\ (\BEDS), \APACrefbtitle {Test fairness in the new generation of large-scale assessment} {Test fairness in the new generation of large-scale assessment}\ (\BPGS\ 185--208).
\newblock
\APACaddressPublisher{}{Information Age Publishing}.
\PrintBackRefs{\CurrentBib}

\bibitem [\protect \citeauthoryear {%
M.~Zhang%
, Ruan%
\BCBL {}\ \BBA {} Johnson%
}{%
M.~Zhang%
\ \protect \BOthers {.}}{%
{\protect \APACyear {2024}}%
}]{%
zhang_fauss_2024}
\APACinsertmetastar {%
zhang_fauss_2024}%
\begin{APACrefauthors}%
Zhang, M.%
, Ruan, C.%
\BCBL {}\ \BBA {} Johnson, M\BPBI S.%
\end{APACrefauthors}%
\unskip\
\newblock
\APACrefYearMonthDay{2024}{}{}.
\newblock
{\BBOQ}\APACrefatitle {Explainable {AI}: Exploring Subgroup Differences in Short-Response Scoring} {Explainable {AI}: Exploring subgroup differences in short-response scoring}.{\BBCQ}
\newblock
\APACjournalVolNumPages{Paper presented at the annual meeting of the National Council for Measurement in Education, Philadelphia, PA}{}{}{}.
\PrintBackRefs{\CurrentBib}

\bibitem [\protect \citeauthoryear {%
N.~Zhang%
\ \protect \BOthers {.}}{%
N.~Zhang%
\ \protect \BOthers {.}}{%
{\protect \APACyear {2021}}%
}]{%
zhang2021differentiable}
\APACinsertmetastar {%
zhang2021differentiable}%
\begin{APACrefauthors}%
Zhang, N.%
, Li, L.%
, Chen, X.%
, Deng, S.%
, Bi, Z.%
, Tan, C.%
\BDBL {}Chen, H.%
\end{APACrefauthors}%
\unskip\
\newblock
\APACrefYearMonthDay{2021}{}{}.
\newblock
{\BBOQ}\APACrefatitle {Differentiable prompt makes pre-trained language models better few-shot learners} {Differentiable prompt makes pre-trained language models better few-shot learners}.{\BBCQ}
\newblock
\APACjournalVolNumPages{arXiv preprint}{}{}{}.
\newblock
\begin{APACrefDOI} \doi{10.48550/arXiv.2108.13161} \end{APACrefDOI}
\PrintBackRefs{\CurrentBib}

\bibitem [\protect \citeauthoryear {%
J.~Zhou%
\ \protect \BOthers {.}}{%
J.~Zhou%
\ \protect \BOthers {.}}{%
{\protect \APACyear {2020}}%
}]{%
zhou_survey_2020}
\APACinsertmetastar {%
zhou_survey_2020}%
\begin{APACrefauthors}%
Zhou, J.%
, Chen, F.%
, Berry, A.%
, Reed, M.%
, Zhang, S.%
\BCBL {}\ \BBA {} Savage, S.%
\end{APACrefauthors}%
\unskip\
\newblock
\APACrefYearMonthDay{2020}{}{}.
\newblock
{\BBOQ}\APACrefatitle {A Survey on Ethical Principles of {AI} and Implementations} {A survey on ethical principles of {AI} and implementations}.{\BBCQ}
\newblock
\BIn{} \APACrefbtitle {2020 {IEEE} {Symposium} {Series} on {Computational} {Intelligence} ({SSCI})} {2020 {IEEE} {Symposium} {Series} on {Computational} {Intelligence} ({SSCI})}\ (\BPGS\ 3010--3017).
\newblock
\begin{APACrefDOI} \doi{10.1109/SSCI47803.2020.9308437} \end{APACrefDOI}
\PrintBackRefs{\CurrentBib}

\bibitem [\protect \citeauthoryear {%
T.~Zhou%
\ \BBA {} Jiao%
}{%
T.~Zhou%
\ \BBA {} Jiao%
}{%
{\protect \APACyear {2023}}%
}]{%
zhou2023exploration}
\APACinsertmetastar {%
zhou2023exploration}%
\begin{APACrefauthors}%
Zhou, T.%
\BCBT {}\ \BBA {} Jiao, H.%
\end{APACrefauthors}%
\unskip\
\newblock
\APACrefYearMonthDay{2023}{}{}.
\newblock
{\BBOQ}\APACrefatitle {Exploration of the stacking ensemble machine learning algorithm for cheating detection in large-scale assessment} {Exploration of the stacking ensemble machine learning algorithm for cheating detection in large-scale assessment}.{\BBCQ}
\newblock
\APACjournalVolNumPages{Educational and Psychological Measurement}{83}{4}{831--854}.
\newblock
\begin{APACrefDOI} \doi{10.1177/00131644221117193} \end{APACrefDOI}
\PrintBackRefs{\CurrentBib}

\end{thebibliography}

\end{document}